\documentclass[12pt]{article}
\usepackage{amsfonts}
\usepackage{youngtab}
\usepackage{tikz-cd}
\usepackage{amsmath,amssymb,cite}
\usepackage{amsthm}
 \usepackage{newtxtext,newtxmath}

\usepackage{cite}
\usepackage[]{hyperref}
\usepackage{xcolor}
\input xy
\xyoption{all}

\def\sqr#1#2{{\vcenter{\vbox{\hrule height.#2pt
            \hbox{\vrule width.#2pt height#1pt \kern#1pt
                  \vrule width.#2pt}\hrule height.#2pt}}}}

\def\square
 {\mathop{\mathchoice{\sqr{12}{15}}{\sqr{9}{12}}
{\sqr{6.3}{9}}{\sqr{4.5}{9}}}}

\def\sqra#1#2#3{{\vcenter{\vbox{\hrule height.#2pt
            \hbox{\vrule width.#2pt height#1pt \kern5pt 
#3
                  \vrule width.#2pt}\hrule height.#2pt}}}}

\def\Z{\mathbb{Z}}

\numberwithin{equation}{section}

\numberwithin{table}{section}

\begin{document}

\vspace*{0.5in}

\begin{center}

 {\large\bf Total instanton restriction via multiverse interference:\\ Noncompact gauge theories and \texorpdfstring{$(-1)$}{(-1)}-form symmetries}

Alonso Perez-Lona, Eric Sharpe, Xingyang Yu, Hao Zhang

\begin{tabular}{l}
                Department of Physics MC 0435\\
                                850 West Campus Drive\\
                                Virginia Tech\\
                                Blacksburg, VA  24061 \end{tabular}

{\tt aperezl@vt.edu}, {\tt ersharpe@vt.edu}, {\tt xingyangy@vt.edu}, 
{\tt hzhang96@vt.edu}

\end{center}

In this note we consider examples of decomposition (in which a local QFT
is equivalent to a disjoint union of multiple independent theories, known as
universes) where
there is a continuous family of universes,
rather than a finite or countably infinite collection.  
In particular, this allows us to consistently eliminate all instantons in a local QFT via
a suitable topological gauging of the $(-1)$-form symmetry.
In two dimensional $U(1)$ gauge theories, this is equivalent to changing the gauge group to ${\mathbb R}$. This makes both locality as well as the instanton restriction explicit.
We apply this to clarify the Gross-Taylor string interpretation of the decomposition of
two-dimensional pure Yang-Mills. 
We also apply decomposition to study two-dimensional ${\mathbb R}$ gauge theories, such as the pure ${\mathbb R}$ Maxwell theory, and
two-dimensional supersymmetric gauged linear sigma models whose gauge groups have factors of ${\mathbb R}$.  In that context, we find that
analogues of the Witten effect for dyons, here rotating between universes, play a role in relating anomalies of the individual universes to (different) anomalies in the disjoint union.
Finally, we discuss limits of the Tanizaki-\"Unsal construction, which accomplish instanton restriction by topologically gauging a ${\mathbb Q}/{\mathbb Z}$ $(-1)$-form symmetry,
and speculate in two-dimensional theories on possible interpretations of those limits in terms of the adelic solenoid.

\begin{flushleft}
July 2025
\end{flushleft}

\newpage

\tableofcontents

\newpage

\section{Introduction}

Restrictions on instantons, as explored in e.g.~\cite{Pantev:2005rh,Pantev:2005wj}, \cite[section 3.1]{Pantev:2005zs}, are a common feature of decomposition \cite{Hellerman:2006zs,Hellerman:2010fv,Sharpe:2022ene}, a phenomenon in which a quantum field theory in $D$ spacetime dimensions, endowed with a global $(D-1)$-form symmetry, is equivalent to a disjoint union of distinct local QFTs, often referred to as universes.  (These subtheories, the universes, are analogous to, but distinct from, superselection sectors.)

While the notion of restricting instantons may seem problematic --- indeed, as emphasized long ago by Weinberg ~\cite[section 23.6]{Weinberg:1996kr}, such restrictions violate cluster decomposition --- it is precisely this violation that decomposition renders tractable. A theory that decomposes naturally violates cluster decomposition, not due to pathology, but because it is not a single theory in the traditional sense. The restrictions on instantons we study here are of this type and hence are under control.

Most previous studies have focused on cases where:
\begin{itemize}
    \item instantons are constrained to lie in sublattices defined by divisibility conditions, and/or
    \item the number of universes is finite or at most countably infinite.
\end{itemize}

In this paper, we shall pursue two generalizations, extending analyses of 
\cite{Yu:2024jtk,Lin:2025oml,Robbins:2025apg}:
\begin{itemize}
    \item instantons are entirely removed --- the allowed topological sectors are restricted to degree zero, as anticipated in e.g.~\cite{Pantev:2023dim},
    \item the decomposition involves a continuous family of universes.
\end{itemize}

In particular, such complete instanton elimination arises from interference among universes --- a sort of destructive quantum superposition across a continuum of vacua. This mechanism, which we refer to as `multiverse interference,' provides a continuous analogue of the familiar cancellation effects seen in discrete decompositions. In two-dimensional examples, this structure appears when the gauge group is extended from a compact group $U(1)$ to the noncompact group $\mathbb{R}$. The absence of instantons in the latter is then not mysterious but reflects the property of a decomposed theory.

Working with noncompact gauge groups introduces subtleties. Two issues in particular deserve mention:
\begin{itemize}
    \item In principle, one typically normalizes partition functions of gauge theories by the volume of the gauge group.  For a noncompact gauge group, that volume is infinite, suggesting at least naively that partition functions should vanish.
    \item For a general noncompact gauge group, the gauge kinetic term
    \begin{equation}
        \int d^D x \, {\rm Tr} F_{\mu \nu} F^{\mu \nu}
    \end{equation}
    need not be positive-definite in Euclidean signature (see e.g.~\cite{Julia:1979pv}).
\end{itemize}

In this paper, we restrict attention primarily to abelian noncompact groups $\mathbb{R}$, $\mathbb{R}^\times$, and $\mathbb{C}^\times$, where the second issue can be addressed --- we can identify the perturbative Lagrangian of an ${\mathbb R}$ gauge theory with that of a corresponding $U(1)$ gauge theory\footnote{We leave more general cases for future work.}. As we shall see, the first issue --- the infinite volume of the gauge group --- may play a meaningful role in understanding quantum cohomology computations in this context.

The subtle features of noncompact gauge theories, while initially puzzling, become more transparent when viewed from a different perspective. Rather than treating such theories as fundamental, we will regard them as emerging from the topological gauging\footnote{As we shall explain in Section~\ref{sect:gauge-1}, `topological' gauging involves summing over flat connections --- in the case of $(-1)$-form symmetries, over constant $\theta$-parameters on each connected component of spacetime. We use the term `topological' to distinguish this from an ordinary gauging,
where $\theta$ is promoted to a dynamical axion. This difference becomes crucial for continuous symmetry groups, and we emphasize it can be accomplished with a local action, as we will discuss in Section~\ref{sect:gauge-1}.} of $(-1)$-form symmetries.
This point of view will guide much of what follows.

A $(-1)$-form symmetry corresponds to background parameters labeling distinct theories, and couples to space-filling topological defects. Gauging such a symmetry amounts to summing over such defects, which is equivalent to forming a disjoint union of universes labeled by parameters in the language of multiverse decomposition.

In the special case that the spacetime has dimension $D=2$, we will discuss how topologically gauging a $(-1)$-form symmetry in a gauge theory can be equivalent to changing the gauge group. In particular, this allows us to relate, for example, gauge theories with gauge group $U(1)$ to gauge theories with gauge group ${\mathbb R}$. This realization provides a physical interpretation of the total instanton restriction via multiverse interference.

To be clear, the idea of topologically gauging $(-1)$-form symmetries was explicitly proposed in~\cite{Yu:2024jtk}, drawing on various guises already present in the literature. Related constructions have appeared both in general discussions and in the specific context of relating two-dimensional $U(1)$ and $\mathbb{R}$ gauge theories~\cite{Tanizaki:2019rbk,Brennan:2024fgj,Antinucci:2024zjp}, as well as in~\cite[section 4.3]{Aloni:2024jpb}\footnote{Note that~\cite{Aloni:2024jpb} largely uses the phrase “gauging the $(-1)$-form symmetry” to refer to an ordinary, non-topological gauging—a somewhat different operation from the topological gauging we employ here.} (see also~\cite{Banks:1991mb} for earlier hints, which considered both two-dimensional $\mathbb{R}$ gauge theories and, separately, integration over the zero modes of an axion). Subsequently, the idea has been further developed in~\cite[section 2.3]{Lin:2025oml}, \cite{Robbins:2025apg,Najjar:2024vmm, Najjar:2025htp}, and~\cite[section 7]{Oguz:2025ftx}. We do not claim novelty in the operation itself. Our focus lies instead in how such gaugings realize decomposition, and in particular, how total instanton restriction naturally emerges in this framework. As this mechanism is both subtle and central to our discussion, we will provide multiple consistency checks across a range of examples.

\subsubsection*{Organization of the paper}

We begin in section~\ref{sect:rev} with a short review of decomposition, listing a few examples we will revisit several times later.

In section~\ref{sect:rev:-1form} we explain how gauging $(-1)$-form symmetries --- finite or continuous --- produces disjoint unions of theories (universes). We review how existing examples of decomposition can be easily understood in terms of gauging a finite $(-1)$-form symmetry. We then give an overview of topologically gauging continuous $(-1)$-form symmetries, explaining what this means concretely, and clarify how it leads to a continuous family of universes and (typically) total instanton restriction.

In section~\ref{sect:change} we specialize to two-dimensional theories, and explore the relation between topological gauging and enlargement of the gauge group. 
After reviewing some examples involving finite $(-1)$-form symmetries, we turn to continuous $(-1)$-form symmetries. In particular, we discuss the key example of topologically gauging continuous $(-1)$-form symmetries in two-dimensional $U(1)$ gauge theories to create an ${\mathbb R}$ gauge theory which decomposes into a continuous family of $U(1)$ gauge theories.  This exhibits total instanton restriction, which can be understood both as a multiverse interference effect from the continuous family of universes, and also as a consequence of working with an ${\mathbb R}$ gauge theory.  We discuss how gauging symmetries relates the $U(1)$ and ${\mathbb R}$ theories.
We also discuss ${\mathbb R}^{\times}$ and ${\mathbb C}^{\times}$ gauge theories.

In section~\ref{sect:sigma} we extend these ideas to sigma models and their gauged linear sigma model (GLSM) realizations. We discuss how topologically gauging $(-1)$-form symmetries can be in general understood in that context. As an application, we  revisit the Gross–Taylor string interpretation of $2D$ Yang–Mills decomposition \cite{Pantev:2023dim}, where it was noted that a total instanton restriction is required. We also analyze sigma models on ${\mathbb Z}$-gerbes, and study their GLSM realizations in terms of abelian gauge theories with factors of ${\mathbb R}$ in the gauge group. We outline analogues of quantum cohomology rings, mirrors, and quantum K theory computations in such theories. We also examine the fate of axial anomalies in decomposed theories, where analogues of the Witten effect lead to transitions between universes. This explains how the constituent universes of the ${\mathbb R}$ gauge theory can have axial anomalies while the ${\mathbb R}$ gauge theory itself does not. We briefly comment on how supersymmetric localization expressions for partition functions are consistent with our general framework.  We also give a general discussion of symmetries in these theories.

In section~\ref{sect:tu} we turn to the Tanizaki-\"Unsal construction \cite{Tanizaki:2019rbk}, which restricts instantons to degrees divisible by an integer $p$ by coupling the QFT of interest to a TQFT \cite{Kapustin:2014gua}. We study its limit as $p \rightarrow \infty$, interpreted as gauging a ${\mathbb Q}/{\mathbb Z}$ $(-1)$-form symmetry, which results in a countably infinite family of universes, parametrized by ${\mathbb Q}/{\mathbb Z}$, and discuss various interpretations.  
We also discuss supersymmetric generalizations and quantum cohomology computations in such theories, both for finite $p$ and in the limit $p \rightarrow \infty$. 

In several appendices we collect some supporting material.
In appendix~\ref{app:nonabel}, we briefly outline the analogue of higher-form symmetries for nonabelian groups.
In appendix~\ref{sect:r-gauge} we discuss some basic facts about ${\mathbb R}$ gauge theories, including expressions for partition functions and states of pure ${\mathbb R}$ gauge theories in two dimensions.
In appendix~\ref{app:gauge-1orb} we demonstrate how (topologically) gauging $(-1)$-form symmetries in two-dimensional orbifolds can be understood as gauging a larger group with explicit construction.

We close this introduction with a brief remark on potential applications to the strong CP problem. In the spirit of \cite{Banks:1991mb}, it is tempting to speculate on applications to the strong CP problem, say by replacing a topological gauging of $(-1)$-form symmetries with an ensemble in which contributions from different $\theta$ are weighted slightly differently.  However, the spirit of the constructions in this paper is to integrate out $\theta$, so in the resulting gauge theory, $\theta$ is no longer defined a priori.  As a result, the models we discuss here are not really relevant to questions about symmetry properties that could make $\theta$ small. We leave this direction to future work.

\section{Short review of decomposition}  \label{sect:rev}

Briefly, decomposition \cite{Hellerman:2006zs,Sharpe:2022ene} is the statement that a local QFT in $D$ spacetime dimensions with a global $(D-1)$-form symmetry is equivalent to a disjoint union of local quantum field theories, known as universes.
Decomposition is frequently (though not always) related to instanton restriction,
via ``multiverse interference effects'' in which the contributions of instantons
from different universes cancel out
due to theta angle phases.

We list here some simple prototypical\footnote{
As this paper is focused on gauging $(-1)$-form symmetries, we have chosen examples of a relatively special form, in which the universes are all copies of the same underlying theory, albeit with different discrete theta angles.  However, more complicated decompositions also exist.  For example, let ${\mathbb H}$ denote the eight-element group of unit quaternions,
and consider a Calabi-Yau $X$ with an action of ${\mathbb H}$ such that the noncentral subgroup
$\langle i \rangle \cong {\mathbb Z}_4 \subset {\mathbb H}$ acts trivially.
Then, as discussed in \cite[section 5.4]{Hellerman:2006zs},
\begin{equation}
    {\rm QFT}\left( [X/{\mathbb H}] \right) \: = \:
    {\rm QFT}\left( X \, \coprod \, [X/{\mathbb Z}_2  \times {\mathbb Z}_2]
    \, \coprod \, [X/{\mathbb Z}_2 \times {\mathbb Z}_2] \right),
\end{equation}
which is not of the same form, because of the mismatched universes.
} examples:
\begin{itemize}
\item Two-dimensional orbifolds $[X/G]$ with trivially-acting
$K$ in the center of $G$ are equivalent to disjoint unions
\cite{Hellerman:2006zs}
\begin{equation}  \label{eq:orbifold-decomp}
{\rm QFT}\left( [X/G] \right) \: = \: 
{\rm QFT}\left( \coprod_{\rho \in \hat{K}} [X/ (G/K) ]_{\omega(\rho)} 
\right),
\end{equation}
where $\omega(\rho)$ denotes discrete torsion, realized mathematically
as the
image of the characteristic class of $[X/G]$ viewed as
a $K$-gerbe on $[X/(G/K)]$:
\begin{equation}
\rho: \: H^2([X/(G/K)], K) \: \longrightarrow \: H^2([X/(G/K)], U(1)).
\end{equation}
(To be clear, decomposition is understood much more generally in orbifolds,
one does not need to restrict to $K$ central to have a decomposition,
but for simplicity, we choose $K$ to be central in this example.)

In an orbifold, the instantons are precisely the `partial traces,'
the contributions to the partition function from nontrivial boundary
conditions, as those correspond to nontrivial $G$ bundles on the
worldsheet.  However, not all the partial traces of a $(G/K)$ orbifold
appear in the $G$ orbifold in general -- because commuting elements of $G/K$ need not lift to commuting elements of $G$.
In the presentation as a disjoint union, such presentations arise because contributions from the same partial trace are weighted differently by discrete torsion, and their contributions cancel out.

For example, if $G = D_4$, $K= {\mathbb Z}_2$, then $G/K = 
{\mathbb Z}_2 \times {\mathbb Z}_2$, which is abelian.
In general, only commuting pairs of group elements $g, h \in G$ define
a partial trace contribution to a $T^2$ partition function
\begin{equation}
{\scriptstyle g} \square_h ,
\end{equation}
so as $G = D_4$ is nonabelian, only commuting group elements appear above,
whereas since $G/K = {\mathbb Z}_2 \times {\mathbb Z}_2$ is abelian,
all pairs define partial traces.  There exist some pairs of elements
in $G/K$ which do not lift to commuting pairs in $G = D_4$, which ultimately
is the mechanism reason why $D_4$ sees fewer instanton sectors (partial
traces) than $G/K$.  Those same partial traces are weighted differently by discrete torsion, and so their contributions to the partition function cancel out, as has been extensively reviewed elsewhere.

\item Two-dimensional $G$ gauge theories with trivially-acting $K$ in the center of $G$
are equivalent to \cite[section 2.3]{Sharpe:2014tca} disjoint unions of $G/K$ gauge theories with the same matter content:
\begin{equation}  \label{eq:decomp:2dg}
    \mbox{$G$ gauge theory} \: = \: \coprod_{\rho \in \hat{K}}
    \left( \mbox{$G/K$ gauge theory} \right)_{\rho},
\end{equation}
where the $\rho$ indicates a discrete theta angle.  
A $G/K$ gauge theory has more bundles / nonperturbative effects than a $G$ gauge theory, measured by a characteristic class, the same characteristic class to which the discrete theta angles couple.  By summing over contributions from $G/K$ gauge theories with different discrete theta angles, we can cancel out contributions from bundles for which that characteristic class is nonzero.

For example, schematically,
\begin{equation}
    SU(2) \: = \: SO(3)_+ \, \coprod \, SO(3)_-,
\end{equation}
where the $\pm$ indicate discrete theta angles.  
Now, an $SO(3)$ gauge theory has more bundles / nonperturbative effects than an $SU(2)$ theory, specifically, bundles with nontrivial Stiefel-Whitney class $w_2$.  However, the ${\mathbb Z}_2$ discrete theta angle couples to $w_2$, so that the effect of summing over $SO(3)$ theories with both choices of discrete theta angle is to cancel out contributions from bundles with $w_2 \neq 0$.
\item Four-dimensional Yang-Mills theory with a restriction to instanton
number divisible by $k$ is equivalent to a disjoint union of
$k$ universes \cite{Tanizaki:2019rbk},
each of which is an ordinary Yang-Mills theory
(with all instantons), but in which each has a slightly rotated
theta angle.  The effect is that in the sum over universes,
contributions from instantons of instanton number not divisible by $k$
cancel out (because of mismatched phases), and only contributions
from instanton numbers divislble by $k$ survive.
\end{itemize}

We will see that when we consider continua of universes,
we can produce restrictions to single instanton sectors, not just to a
subset.

Such a restriction is not unanticipated.
Consider for example two-dimensional pure Yang-Mills theory.
This theory has a decomposition indexed by the irreducible representations
of the gauge group \cite{Nguyen:2021yld,Nguyen:2021naa}.
In the context of the Gross-Taylor \cite{Gross:1993hu} description of
two-dimensional pure Yang-Mills as a string theory,
it was argued in \cite{Pantev:2023dim}
that the decomposition of two-dimensional pure
Yang-Mills corresponds to a restriction of the Gross-Taylor sigma model
to maps of a single fixed degree -- a restriction on sigma model instantons
to a single degree, in other words.
The paper \cite{Pantev:2023dim} had some suggestions but did not make a completely clean statement about the interpretation of this restriction; now, with hindsight, we can.
We will return to this matter in section~\ref{sect:app:gt}.

Now, as discussed in  \cite[section 4.1]{Pantev:2023dim}, it is not
always possible to restrict to a single instanton sector in any theory.
For example, consider an orbifold $[X/G]$ where $G$ is a finite group.
A single instanton sector corresponds to a single `partial trace'
in the computation of the partition function -- a single set of boundary
conditions, corresponding to a single (equivalence class of)
principal $G$-bundles.  Such a partial trace is not, by itself,
modular invariant, and so we do not expect it to define a meaningful
physical theory (except of course when the boundary conditions are all
trivial, and the partial trace is the partition function of the unorbifolded
theory).

However, even in the case of orbifolds, where we cannot restrict to a single
instanton sector, we can still sum over the analogue of theta angles --
namely, choices of discrete torsion in the case of orbifolds.
The result will not be a single instanton sector, but rather a
modular-invariant sum of twisted sectors, defining a decomposing theory,
as in the prototypical example~(\ref{eq:orbifold-decomp}).

\section{Gauging $(-1)$-form symmetries}
\label{sect:rev:-1form}

Many examples of decomposition can be naturally described via gauging $(-1)$-form symmetries. Formally, this is because (topological) $(-1)$-form symmetries are quantum symmetries of gauged $(D-1)$-form symmetries in $D$ spacetime dimensions.
Abstractly, given a theory
$T$ that decomposes into a collection of universes $U_i$,
\begin{equation}
    T \: = \: \coprod_i U_i
\end{equation}
by gauging a $(D-1)$-form symmetry in $D$ spacetime dimensions, and turning on a suitable
theta angle, one can select out any one universe $U_i$.  The quantum symmetry for this operation is a $(-1)$-form symmetry, so gauging the $(-1)$-form symmetry returns the original theory.
Schematically,
\begin{equation}
    U_i \: 
     \mathrel{ \mathop{\rightleftarrows}^{ (-1)-{\rm form}  }_{ (D-1)-{\rm form} } }
     \:
     T.
\end{equation}

Theta angles serve as background fields (connections) for $(-1)$-form symmetries,
so gauging a $(-1)$-form symmetry means to sum over theta angles (for discrete theta angles)
or integrate over (constant) values of $\theta$ (for continuous theta angles),
in effect, integrating over flat connections on the $(-1)$-form symmetry.
Throughout this paper, we restrict attention to topological gaugings of continuous $(-1)$-form symmetries, which means that we only integrate over (analogues of) flat
connections --- locally constant $\theta$, in other words.  Morally, a (full) dynamical gauging of a $(-1)$-form symmetry would correspond to promoting $\theta$ to an axion, which we are not doing here.  To distinguish these cases, we speak of integrating over locally constant $\theta$ angles as a topological gauging.  We will elaborate on this matter further 
in section~\ref{sect:gauge-1}.

In the next two subsections, we will quickly revisit examples of gauging finite $(-1)$-form symmetries (summing over discrete theta angles)
in existing decompositions, and then discuss topologically gauging continuous $(-1)$-form symmetries
(integrating over theta angles).

\subsection{Gauging finite $(-1)$-form symmetries}  \label{sect:finite-1}

Let us revisit the decomposition examples of section~\ref{sect:rev},
in the language of gauged $(-1)$-form symmetries.

\begin{itemize}
    \item Two-dimensional orbifolds.  In the decomposition~(\ref{eq:orbifold-decomp}), namely
    \begin{equation} 
    {\rm QFT}\left( [X/G] \right) \: = \: 
    {\rm QFT}\left( \coprod_{\rho \in \hat{K}} [X/ (G/K) ]_{\omega(\rho)} 
    \right),
    \end{equation}
    we can interpret the right-hand-side as a gauged Rep$(K) = \hat{K}$ $(-1)$-form symmetry, in which one sums over (finitely-many) values of the discrete theta angles $\omega(\rho)$.
    (In an orbifold, a `discrete theta angle' can be identified with a choice of discrete torsion, both because it assigns a phase weighting nonperturbative sectors, and also because it can be understood as a lift of the orbifold group action to $B$ fields \cite{Sharpe:2000ki,Sharpe:2000wu}, and $B$ fields act as theta angles in sigma models.)

For example, summing over the two choices of discrete torsion in 
${\mathbb Z}_2 \times {\mathbb Z}_2$ orbifolds is equivalent to working with a $D_4$ orbifold
\cite[section 5.2]{Hellerman:2006zs}:
\begin{equation}
    [X/D_4] \: = \: [X/{\mathbb Z}_2 \times {\mathbb Z}_2] \, \coprod \, [X/{\mathbb Z}_2 \times {\mathbb Z}_2]_{\omega},
\end{equation}
where $\omega$ denotes the nontrivial choice of discrete torsion.
(Note we are
using the fact that $H^2( {\mathbb Z}_2 \times {\mathbb Z}_2, U(1)) = {\mathbb Z}_2$, so that there are two choices of discrete torsion in this example -- one trivial, one nontrivial.)

Now, in the example above, one summed over all choices of discrete torsion,
but in the decomposition~(\ref{eq:orbifold-decomp}), one might only sum over some special choices of discrete torsion, possibly with multiplicity.  Nevertheless, in general one can choose to sum over all choices, and if one does so, the result is equivalent to an orbifold by a larger group, as we will discuss in section~\ref{sect:gauge-1:finite}.

\item Two-dimensional $G$ gauge theories with trivially-acting $K$ in the center of $G$.
In the decomposition~(\ref{eq:decomp:2dg}), namely
\begin{equation}  
    \mbox{$G$ gauge theory} \: = \: \coprod_{\rho \in \hat{K}}
    \left( \mbox{$G/K$ gauge theory} \right)_{\rho},
\end{equation}
we can interpret the right-hand-side as a gauged Rep$(K) = \hat{K}$ $(-1)$-form symmetry, in which one sums over (finitely-many) values of the discrete theta angles.

For example, the decomposition
\begin{equation}
    SU(2) \: = \: SO(3)_+ \, \coprod \, SO(3)_-
\end{equation}
can be understood as gauging the ${\mathbb Z}_2^{[-1]}$ symmetry by summing over discrete theta angles in $SO(3)$ theories.  In section~\ref{sect:gauge-1:finite} we will briefly discuss how gauging a $(-1)$-form symmetry in two dimensions, enlarges the gauge group.

\end{itemize}

\subsection{Topologically gauging continuous $(-1)$-form symmetries}
\label{sect:gauge-1}

So far, we have considered finite $(-1)$-form symmetries --- corresponding to finite sums over discrete $\theta$ angles --- which naturally lead to instanton number constraints via divisibility conditions.

We now turn to continuous $(-1)$-form symmetries. In this case, the gauging procedure imposes a much stronger restriction: it entirely eliminates nonzero-instanton sectors. This formal step plays a central role in realizing total instanton suppression and decomposition, as previewed in the Introduction.

In this section, we outline the basic framework, and postpone detailed examples and consistency checks to later sections.

\subsubsection{Definition of topological gauging}  \label{sect:top-gauge-defn}

We begin by clarifying what we mean by \emph{topological} gauging, and how it differs from more conventional notions of gauging.
This operation is sometimes referred to as \emph{flat} gauging. For an abelian $p$-form symmetry, it involves a $(p+1)$-form gauge field constrained to be flat.
In the special case of a $(-1)$-form symmetry, the relevant “background gauge field” is a scalar field $\phi$ with $d\phi = 0$, i.e., a locally constant parameter over each connected component of spacetime. The path integral then integrates over such $\phi$.
See, for instance, \cite[section 3.2]{Brennan:2024fgj} and \cite{Antinucci:2024zjp}, where this idea is explored from the perspective of SymTFTs with continuous $U(1)$ $p$-form symmetries.

Topological gauging becomes particularly relevant in our context because we are interested in gauging continuous symmetries that are dual to (or in some cases, themselves) quantum symmetries. A classic setting for quantum symmetries arises in orbifolds (see e.g.~\cite[section 8.5]{Ginsparg:1988ui}): For finite $G$, the orbifold $[X/G]$ possesses a quantum symmetry described by $\text{Rep}(G)$, such that orbifolding $[X/G]$ again by $\text{Rep}(G)$ recovers the original theory.\footnote{In fact, it is also possible to gauge symmetries beyond $\text{Rep}(\mathcal{H})$, where $\mathcal{H}$ is a group or a Hopf algebra. See, e.g., \cite{Yu:2025iqf} for explicit computations.}

More generally, gauging a finite $G$ $p$-form symmetry in $D$ spacetime dimensions produces a quantum Rep$(G)$ $(D - p - 2)$-form symmetry.
(For example, when $D = 2$ and $p = 0$, the quantum symmetry is a standard $0$-form symmetry.) For finite groups $G$, both the notion of Rep$(G)$ and its gauging are well understood. In this work, however, we are primarily interested in cases where Rep$(G)$ is continuous—a setting that requires more care.

These topological gaugings have been described in the language of SymTFTs (see e.g.,~\cite[section 3.2]{Brennan:2024fgj}, \cite{Antinucci:2024zjp,Yu:2024jtk}), but they can also be understood more traditionally, within the framework of coupling a local QFT to a TQFT \cite{Kapustin:2014gua}. More precisely, to topologically gauge a continuous $p$-form symmetry in $D$ spacetime dimensions, we begin with a background $(p+1)$-form gauge field $A_{p+1}$ and promote it to a dynamical field $a_{p+1}$. We then introduce a dynamical $(D - p - 2)$-form field $c_{D-p-2}$, and couple them via the interaction:
\begin{equation} \label{eq:top:int}
\int \left( a_{p+1} \wedge B_{D - p - 1} + d a_{p+1} \wedge c_{D - p - 2} \right),
\end{equation}
where $B_{D - p - 1}$ is a fixed background field to which $a_{p+1}$ couples.

The field $c_{D - p - 2}$ enters solely as a Lagrange multiplier. Its equation of motion imposes
\begin{equation}
d a_{p+1} = 0,
\end{equation}
thereby enforcing the desired flatness condition on the gauge field.

Stated alternatively, the procedure can be schematically summarized at the level of the path integral as follows:
\begin{eqnarray}
    \int [D \phi] \exp(- S[\phi, A]) & \mapsto & \int [D \phi] [D a] [D c] \exp\left( - S[\phi, a] + \int \left( a \wedge B + d a \wedge  c \right) \right),
    \nonumber \\
    & & = \int [D \phi] [d a] \left. \exp\left( - S[\phi, a] + \int a \wedge B \right)  \right|_{d a = 0},
\end{eqnarray}
where $\phi$ collectively denotes all matter fields, and $S[\phi, A]$ is the original action before gauging.

Note\footnote{E.S. would like to thank Y.~Tanizaki for this argument.} that gauge invariance of the $a \wedge B$ term under gauge transformations of $B$ requires that $a$ be flat. This flatness is enforced by the Lagrange multiplier $c_{D - p - 2}$ introduced above. If we interpret $B_{D - p - 1}$ as the gauge field for the quantum symmetry dual to $a_{p + 1}$ (or vice versa), this provides yet another perspective on why gauging a continuous quantum symmetry necessitates a topological rather than ordinary gauging.

In the remainder of this paper, when referring to `topological gauging,' for simplicity we will omit the explicit coupling to background fields $B$ or Lagrange multipliers $c$. Instead, we will simply indicate that the dynamical gauge field is constrained to be flat. We emphasize that the resulting theory is still defined by a local action, but we will usually omit the details above for simplicity of exposition.

In the present context, we are interested in $(-1)$-form symmetries, coupling to the theta angle, which plays the role of a background connection.  Topologically gauging this symmetry amounts to promoting the constant parameter $\theta$ to a dynamical scalar field $\phi$ with $d\phi = 0$, so that $\phi$ is locally constant across spacetime. In practice, this amounts to integrating over theta angles, independently 
on each connected component of the spacetime manifold. This should be contrasted with an ordinary gauging of the $U(1)^{[-1]}$ symmetry, in which $\theta$ would be promoted to a dynamical axion field, which is not a procedure we consider here.

For simplicity of notation, we will uniformly refer to this as `topological gauging' for both finite and continuous $G$. In the case of finite $G$, this distinction is immaterial: topological gauging coincides with ordinary gauging.
It is only in the continuous case that the difference becomes conceptually and technically meaningful.

\subsubsection{Prototypical example and instanton restriction}

As a concrete illustration, consider a Yang-Mills theory in spacetime dimension $D = 2k$ with gauge
group $G$ and a theta angle $\theta$.
As is well-known, the partition function can be written schematically
in the form
\begin{equation}
Z(\theta) \: = \: \sum_n \int [D A_{\mu}]_n \exp\left( - i S_{YM} \right)
\exp\left( i \theta n \right),
\end{equation}
where $n$ is the instanton number, and $\theta$ is the theta angle.

For simplicity, assume the spacetime is connected.
After topologically gauging the $(-1)$-form symmetry involving the theta angle,
the partition function is\footnote{
As a technical note, we are implicitly working at a fixed scale, so that we can disregard any renormalization of $\theta$ across different scales.
}
\begin{eqnarray}
Z & = & \int d \theta \, Z(\theta) \: = \:
\int d \theta \sum_n \int [D A_{\mu}]_n \exp\left( - i S_{YM} \right)
\exp\left( i \theta n \right),    \label{eq:theta-int}
\\
& \propto &
\sum_n \int [D A_{\mu}]_n \exp\left( - i S_{YM} \right)
\delta_{n,0},    \label{eq:inst-res}
\end{eqnarray}
where $[D A_{\mu}]_n$ denotes the path integral in the sector of
instanton number $n$, and we integrate over $\theta$ with an ordinary integral --- $\theta$ is treated as an ordinary number integrated over $\mathbb{R}$ instead of an axion\footnote{We are implicitly assuming that the spacetime has a single connected component, for simplicity.}.
From~(\ref{eq:inst-res}), the delta function clearly restricts\footnote{
Similar restrictions, arising as limits of the Tanizaki-\"Unsal mechanism \cite{Tanizaki:2019rbk}, 
have been discussed in \cite[section 3.1]{Lin:2025oml}, \cite[appendix D.1]{Pantev:2023dim}.  We will discuss limits of the Tanizaki-\"Unsal mechanism
in section~\ref{sect:tu}.
} to the
zero-instanton sector.

In most of this paper, we focus on the restriction to instanton number zero.
However, it is also possible to project onto a fixed nonzero instanton sector.
For example, consider
\begin{eqnarray}
Z(k) & = & \int d \theta \, Z(\theta) \exp(-i \theta k),   \label{eq:rest:k:1}
\\
& = & 
\sum_n \int [D A_{\mu}]_n \exp\left(-i S_{YM}  \right)
\delta_{n,k},
\end{eqnarray}
which projects onto the contribution from instanton number $k$.

Phrased another way, we can extract $Z(k)$ (the contribution to the partition function from a fixed instanton number $k$) from the partition function by a Fourier transform, which is the integration over $\theta$ angles, or equivalently, gauging the $(-1)$-form symmetry.
In terms of symmetries, the exponential phase factor encodes a coupling to a background field. More generally, for a $p$-form symmetry with background connection $b_{p+1}$ and dual $(D - p - 1)$-form field $C$, one has the well-known schematic form 
\begin{equation}
    Z_{\rm gauged}[C] \: = \: \int d b_{p+1} Z[b_{p+1}] \exp\left( i \int b_{p+1} \wedge C \right),
\end{equation}
as a direct generalization of~(\ref{eq:rest:k:1}). This viewpoint will become important in our discussion of the Gross–Taylor model in section~\ref{sect:app:gt}.

\subsubsection{Interpretation as decomposition}

So far, we have discussed the mechanics of topological gauging and the resulting restrictions on instantons. We now interpret these topologically-gauged theories in terms of decomposition, relying on the duality between $(D-1)$-form and $(-1)$-form symmetries in $D$ spacetime dimensions.

Concretely, given a theory with a $G^{[-1]}$-form symmetry for abelian $G$, topological gauging produces a new theory with a Rep$(G)^{[D-1]}$-form symmetry, which then decomposes. Conversely, starting from a theory with a $G^{[D-1]}$-form symmetry, topologically gauging it yields a theory with Rep$(G)^{[-1]}$-form symmetry.

In our setting, gauging the $U(1)^{[-1]}$-form symmetry results in a theory that decomposes into components carrying a
Rep$(U(1))^{[D-1]} \cong \mathbb{Z}^{[D-1]}$-form symmetry. Such a theory contains (infinitely-heavy) domain walls labeled by elements of $U(1)$, and topological local operators labeled by $\mathbb{Z} \cong \text{Rep}(U(1))$.
(See also \cite{Yu:2024jtk}, \cite[section 2.3]{Lin:2025oml}, \cite[section 2]{Robbins:2025apg} where this is discussed as well.)

This decomposition is reflected in the structure of the partition function~(\ref{eq:theta-int}), where the ordinary integral over $\theta$ amounts to summing over a disjoint union of uncountably many universes, each distinguished by the value of the theta angle. The cancellation of contributions from nonzero-instanton sectors by interference between different $\theta$-sectors is an example of what is known as `multiverse interference.' We will check and study this mechanism in more detail through explicit examples later in the paper.

Occasionally, we will encounter cases with no $\theta$ angle. Even in such situations, one can still topologically gauge a trivially-acting $(-1)$-form symmetry (see, e.g., \cite{Gu:2025gtb}). The resulting theory exhibits not only a $0$-form symmetry arising from gauging a trivially-acting symmetry, but also a quantum $(D-1)$-form symmetry signaling decomposition.

In passing, one could also consider integrals over more general parameters than just theta angles.\footnote{More generally, the parameter space may not even be a group manifold. In such cases, referring to these symmetries as “$(-1)$-form symmetries” is somewhat of an abuse of notation. See, for example, the discussion in \cite[Section 1.5]{Cordova:2019jnf}.}
For example, in \cite{Yu:2024jtk}, a multiverse decomposition was constructed by gauging the $(-1)$-form symmetry associated with the rank of the gauge group in ABJ(M) theories, leading to a formal sum over ABJM theories with different numbers of fractional branes.
Such more general constructions are interesting, but we will not consider them in this paper.

\subsubsection{The role of chiral anomalies}

A technical point we have not addressed involves the well-definedness of the theta angle in the presence of anomalous global symmetries. As is well-known, acting by such global symmetries can effectively shift the theta angle (as one interpretation of the anomaly).

In our construction, however, we integrate over all values of $\theta$, regardless of whether such anomalies are present. Acting with an anomalous global symmetry maps $\theta \mapsto \theta + \alpha$ for some $\alpha$, but this shift leaves the $\theta$-integral invariant.

In fact, since the theta angles parametrize universes, we see that a chiral rotation has the effect of rotating universes into one another. This perspective aligns naturally with the description in terms of topologically gauging a $(-1)$-form symmetry:
because chiral symmetry transformations shift $\theta$, the topological gauging renders such shifts redundant. In this sense, the chiral anomaly has been absorbed --- or perhaps better, sublimated --- into the multiverse structure\footnote{In the case of ordinary gauging, i.e., promoting $\theta$ to a dynamical axion field $\phi$, the chiral anomaly is also canceled.
In more detail, let $C_1$ denote a background field, let $\phi$ denote the axion, and write the action as
\begin{equation}
    S[C_1] \: = \: S_{\rm QFT} \: + \: \frac{1}{2\pi} \int \phi \wedge F \wedge \cdots \wedge F \: + \: \int C_1 \wedge * J_A,
\end{equation}
where $d(* J_A) = (1/2\pi) F \wedge \cdots \wedge F$.  Then, it is straightforward to verify that the action is invariant under 
\begin{equation}
    C_1 \: \mapsto \: C_1 + d \alpha, \: \: \: \phi \: \mapsto \: \phi - \alpha.
\end{equation}
}
A chiral symmetry rotation that would ordinarily be anomalous within a single universe now simply rotates between distinct universes in the disjoint union.

We will see this phenomenon explicitly in the supersymmetric ${\mathbb P}^n$ model, discussed in section~\ref{sect:qc:r}.

\subsubsection{Comparison to ensemble averaging}

Integrating over theta angles is strongly reminiscent of ensemble averaging (see e.g.~\cite{syk,Maldacena:2016hyu,Saad:2019lba,Witten:2020bvl,Maloney:2020nni,Afkhami-Jeddi:2020ezh,Cotler:2020hgz,Schlenker:2022dyo}), with dilaton shift factors playing the role of probability distributions, as has been previously noted in e.g.~\cite[Section IV]{Yu:2024jtk}, \cite{Sharpe:2023lfk}. However, the two notions are conceptually distinct.
\begin{itemize}
    \item If the spacetime has multiple
connected components, 
\begin{itemize}
\item In an ensemble, there is a single sum (or integral) over element theories of the ensemble, independent of the spacetime topology.
\item In a decomposition, each connected component of the spacetime carries its own independent sum over universes.
\end{itemize}
See e.g.~\cite[section 6]{Sharpe:2023lfk} for a more detailed discussion.
\item If the spacetime is connected, then another difference lies in the normalization of the partition functions:
\begin{itemize}
    \item In an ensemble, the probability distribution is normalized to integrate to one.
    \item In a decomposition, the partition function is canonically normalized so that, after projecting to any finite number of universes, the resulting partition function recovers integer state counts (modulo dilaton shift factors\footnote{
    These are Euler-number-type, universal counterterms.  For example, in the partition function of a two-dimensional theory, they contribute factors multiplying partition functions of the individual universes, of the form
    \begin{equation}
        \exp\left(a \chi(\Sigma) + b A(\Sigma) \right),
    \end{equation}
    where $a, b \in {\mathbb R}$, $\chi$ is the Euler characteristic, and $A$ the area.
    Their presence in decomposition was observed in the original paper \cite{Hellerman:2006zs}.
    See \cite{Sharpe:2023lfk} for a more detailed discussion.
    }).
\end{itemize}
\end{itemize}
In particular, on a connected spacetime, the difference between ensemble averaging and decomposition reduces to a matter of normalization
of the partition functions, which reflects different interpretations of the theory\footnote{See also \cite{Heckman:2021vzx} and \cite[Section 8]{Baume:2023kkf} for a string-theoretic construction of ensemble averages, where the element theories are defined on multiple copies of spacetime manifolds. The random coupling constants arise from different vevs of dynamical moduli fields, making the construction feel, up to a normalization by probability distributions, like a reversed process of (topological) gauging.}. In decomposition, the universes are treated as physically real sectors with well-defined state counts; in an ensemble, they are integrated over with probabilistic weights.

It should also be said that the main purposes of decompositions and ensembles are different.  For example, typically what makes a decomposition interesting is that the starting point is a manifestly local QFT, which (unexpectedly) is equivalent to a disjoint union.

Having outlined the general framework for topologically gauging continuous $(-1)$-form symmetries and its implications, in the next section we turn to concrete examples and perform several consistency checks.

 \section{Changing the gauge group in $D=2$}
 \label{sect:change}

\subsection{General statement}  \label{sect:genlclaim}

In this section, we review how, in two-dimensional gauge theories, topologically gauging $(-1)$-form symmetries and the resulting instanton restriction can often be understood in terms of changing the gauge group. We motivate this interpretation by comparing two kinds of constructions related by gauging symmetries:
\begin{itemize}
    \item Topologically gauging a $(-1)$-form symmetry in $D$ spacetime dimensions.

As discussed in earlier sections, in $D$ spacetime dimensions, topologically
gauging a $(-1)$-form symmetry leads to a QFT with a global
$(D-1)$-form symmetry (as the quantum symmetry), which therefore decomposes
(see e.g.~\cite{Hellerman:2006zs,Sharpe:2022ene}).
For example, suppose one has 
a $G/H$ gauge theory in $D$ spacetime dimensions, with theta angles labeled by Rep$(H)$. Then we have the schematic relation:
\begin{equation} \label{eq:yoga-decomp}
    \mbox{$G/H$ gauge theory} \: 
    \mathrel{ \mathop{\rightleftarrows}^{ {\rm Rep}(H)^{[-1]} }_{ \mathrm{{H}^{[D-1]}} } }
\:
\mbox{sum of universes}.
\end{equation}

Due to multiverse interference, the path integral typically restricts the contributing instantons to a subset of those in the $G/H$ gauge theory.

\item Topologically gauging a $(D-3)$-form symmetry in $D$ spacetime dimensions.

Suppose one has a $G$ gauge theory in $D$ spacetime dimensions
in which a subgroup $H \subset G$ acts trivially on the
matter fields.  For simplicity, we shall assume that $H$ is central in $G$ (i.e.~commutes with all elements
of $G$), but variations of this story will exist more generally.

This $G$ gauge theory has a one-form symmetry $BH = H^{[1]}$.  Topologically gauging this symmetry yields a $G/H$ gauge theory with the same matter content (for example, the characteristic classes of $H$ gerbes provide the characteristic classes of $G/H$ bundles not present for $G$ bundles), and induces a quantum Rep$(H)^{[D-3]}$-form symmetry \cite{Gaiotto:2014kfa}. Conversely, topologically gauging the quantum symmetry Rep$(H)^{[D-3]}$ then returns the original $G$ gauge theory. Schematically:
\begin{equation}   \label{eq:yogaD}
\mbox{$G/H$ gauge theory} \:
\mathrel{ \mathop{\rightleftarrows}^{ {\rm Rep}(H)^{[D-3]} }_{ \mathrm{{H}^{[1]}} } }
\:
\mbox{$G$ gauge theory},
\end{equation}
where
\begin{equation}
    1 \: \longrightarrow \: H \: \longrightarrow \: G \: \longrightarrow \: G/H \: \longrightarrow \: 1.
\end{equation}

That is to say, topologically gauging a $(D-3)$-form symmetry in a $G/H$ theory effectively lifts the gauge group to a cover.

\end{itemize}

Now, specializing to $D = 2$, we compare~(\ref{eq:yogaD}) with~(\ref{eq:yoga-decomp}) to see that the sum of universes should be the $G$ gauge theory. In this case, Rep$(H)^{[D-3]} = \text{Rep}(H)^{[-1]}$ becomes a $(-1)$-form symmetry, and topological gauging it typically\footnote{
We will also see examples in which the $G/H$ gauge theory has no
Rep$(H)$ theta angles.  One can still gauge the Rep$(H)^{[-1]}$ as a trivially-acting symmetry,
and we will see that this still results in a $G$ gauge theory.
} amounts to summing or integrating over theta angles:
\begin{itemize}
    \item If Rep$(H)$ is continuous, this corresponds to integration over  continuous theta angles.
    \item If Rep$(H)$ is discrete, this corresponds to summing over discrete theta angles or a subset of the continuous theta angle depending upon the circumstances.
\end{itemize}

The general statements above reduce, for $D=2$, to the diagram 
\begin{equation}  \label{eq:yoga2}
\mbox{$G/H$ gauge theory} \:
\mathrel{ \mathop{\rightleftarrows}^{ {\rm Rep}(H)^{[-1]} }_{ \mathrm{{H}^{[1]}} } }
\:
\mbox{$G$ gauge theory}.
\end{equation}

Furthermore, because the $G$ gauge theory in $D=2$ has a one-form symmetry, it decomposes.
Contributions from the different universes of the decomposition can have the effect of cancelling out instanton contributions, a phenomenon now referred to as a ``multiverse interference effect,'' resulting in a restriction on instantons.  

After briefly reviewing familiar examples of decomposition exhibiting this phenomenon, we will explore more exotic cases, where we relate compact and noncompact gauge groups.
Gauge theories with noncompact groups often lack instantons altogether, and we will see that the structure above implements a \emph{total} restriction: not merely eliminating a subset, but removing all instantons altogether

\subsection{Finite examples}  \label{sect:gauge-1:finite}

In this section we will review some basic examples of decomposition and (topologically\footnote{All gaugings of $(-1)$-form symmetries in this paper are topological gaugings, but for finite groups, `topological' is redundant.}) gauged finite $(-1)$-form symmetries discussed in sections~\ref{sect:rev}, \ref{sect:finite-1}, and review how gauging the $(-1)$-form symmetry in two-dimensional cases can lead to an extension of the gauge group in a gauge theory.

\begin{itemize}
    \item Two-dimensional orbifolds.  We have reviewed how an orbifold by a finite group with a trivially-acting subgroup is equivalent to a disjoint union of orbifolds, and how that can often be interpreted as topologically gauging a finite $(-1)$-form symmetry.  Next, we consider the converse:
    start with an ordinary orbifold, and topologically gauge a $(-1)$-form symmetry.

    Consider any two-dimensional orbifold $[X/G]$, where $G$ need not have a trivially-acting subgroup.  By topologically gauging the $(-1)$-form symmetry coupling to discrete torsion in this orbifold, we construct a disjoint union
    \begin{equation}
         \coprod_{\omega \in H^2(G,U(1))}
    {\rm QFT}\left( [X/G]_{\omega} \right).
    \end{equation}
    The yoga we have presented so far would suggest that this disjoint union can be interpreted in terms of an orbifold by a larger orbifold group, with a trivially-acting subgroup.
    
    In appendix~\ref{app:gauge-1orb} we outline a construction of a covering group $\widetilde{G}$, with trivially-acting subset, so that
\begin{equation}
    {\rm QFT}\left( [X/\widetilde{G}] \right) \: = \: \coprod_{\omega \in H^2(G,U(1))}
    {\rm QFT}\left( [X/G]_{\omega} \right),
\end{equation}
where the sum is over all possible choices of discrete torsion in $G$.
We interpret this to mean that for any finite group $G$, if we gauge the $(-1)$-form symmetry corresponding to choices of discrete torsion, the result is not just a disjoint union, but is also a $\widetilde{G}$ gauge theory for some larger $\widetilde{G} \twoheadrightarrow G$.  We will see this same pattern,
that topologically gauging a $(-1)$-form symmetry can increase the size of the gauge group in two-dimensional theories, later in this paper.
\item Two-dimensional gauge theories.  We have reviewed how a gauge theory in which a subgroup of the center acts trivially on the matter is equivalent to a disjoint union of gauge theories, with a smaller gauge group and discrete theta angles.

Now, we consider the converse.
Start with an ordinary two-dimensional gauge theory with a non-simply-connected gauge group $G$,
and topologically gauge discrete theta angles in $\hat{K}$ for $K$ some subgroup of $\pi_1(G)$,
to form the disjoint union
\begin{equation}
    \coprod_{\rho \in \hat{K}} (\mbox{$G$ gauge theory})_{\rho},
\end{equation}
where the notation indicates a $G$-gauge theory with discrete theta angle $\rho$.

From equation~(\ref{eq:decomp:2dg}), one immediately recognizes that this disjoint union is equivalent to a decomposition of a $\widetilde{G}$ gauge theory, where
\begin{equation}
    1 \: \longrightarrow \: K \: \longrightarrow \: \widetilde{G} \: \longrightarrow \: G \: \longrightarrow \: 1,
\end{equation}
and where $K$ acts trivially on the matter (the same matter as the $G$ gauge theory).

Again, we see in two dimensions that topologically gauging a $(-1)$-form symmetry in a gauge theory can be interpreted as enlarging the gauge group.
\end{itemize}

So far we have reviewed some common examples involving finite $(-1)$-form symmetry topological gaugings.
In the remainder of this section, we will focus on continuous $(-1)$-form symmetry topological gaugings in two dimensions, and resulting gauge group enlargements and total instanton restrictions.

\subsection{Total instanton restriction in $U(1)$ theory as ${\mathbb R}$ theory}   \label{sect:int}  \label{sect:ex:u1-r}

\subsubsection{Generalities} \label{sect:u1-r:gen}

Next, we relate (ordinary) $U(1)$ gauge theories to ${\mathbb R}$ gauge theories in spacetime dimension $D=2$, using
the relation~(\ref{eq:yoga2}), 
which in this case says
\begin{equation}
\mbox{$U(1)$ gauge theory} \: \mathrel{\mathop{\rightleftarrows}^{\mathrm{U(1)^{[-1]}}}_{\mathrm{{\mathbb Z}^{[1]}}}}
\:
\mbox{${\mathbb R}$ gauge theory}.
\end{equation}
(Note that the map from $U(1)$ theories to ${\mathbb R}$ theories is not surjective:
only ${\mathbb R}$ gauge theories with a $B {\mathbb Z} = {\mathbb Z}^{[1]}$ symmetry can be
constructed in this way from $U(1)$ gauge theories.)

As mentioned in the introduction, topologically
gauging $(-1)$-form symmetries to relate $U(1)$ and ${\mathbb R}$ gauge theories has been previously studied in e.g.~\cite[section 2.3]{Lin:2025oml}, \cite{Robbins:2025apg,Antinucci:2024zjp}, \cite[section 4.3]{Aloni:2024jpb}, \cite[section 7]{Oguz:2025ftx}.
(See also, e.g., \cite{Margolin:1992rg,Chiodaroli:2023tvo,Bonati:2023ybq} for studies of gauge theories with noncompact gauge groups, albeit unrelated to topological gauging of $(-1)$-form symmetries.)
However, as this perspective has not been thoroughly investigated and plays a central role in our paper, we will carry out a series of consistency checks to better understand and substantiate this correspondence.

As a first check, note that gauging the $B{\mathbb Z} = {\mathbb Z}^{[1]}$ symmetry of the $\mathbb{R}$ gauge theory to obtain the $U(1)$ theory introduces $U(1)$ vortices or instantons\footnote{In the context of two-dimensional $U(1)$ gauge theories, we will use the terms `vortex' and `instanton' interchangeably, referring to principal $U(1)$ bundles with nontrivial first Chern class.}, which are absent in the original ${\mathbb R}$ gauge theory. This arises because gauging $B {\mathbb Z}$ amounts to summing over ${\mathbb Z}$-gerbes in the path integral, which are classified on a spacetime $X$ by $H^2(X, {\mathbb Z})$, the same group that classifies first Chern classes. In other words, the Chern classes of $U(1)$ bundles in the resulting $U(1)$ theory are precisely the characteristic classes of the ${\mathbb Z}$-gerbes summed over in the gauging procedure.

Furthermore, one can recover the theta angle of the $U(1)$ theory by turning on a theta term in the $B {\mathbb Z}$ gauging, as discussed in similar contexts in, e.g., \cite{Sharpe:2019ddn}.
Here, the idea is that one adds a term to the action to weight the sum over ${\mathbb Z}$-gerbes in the path integral.  Schematically, this means that the path integral is modified as
\begin{equation}
    \int [D \phi] \cdots \: \mapsto \: 
    \sum_{n \in H^2(X,{\mathbb Z})} \int [D \phi] \exp\left( i \theta \int n \right) \cdots,
\end{equation}
where the sum is over gerbe characteristic classes $n \in H^2(X, {\mathbb Z})$, and the exponential term assigns a $\theta$-dependent weight to each sector. As we will see explicitly in later examples, these classes become the vortex numbers of the $U(1)$ theory, and the $\theta$-angle from the gauging becomes the usual $\theta$ parameter in the $U(1)$ gauge theory.

Conversely, in two dimensions, topologically gauging the $U(1)^{[-1]}$ magnetic symmetry corresponds to integrating over the $\theta$ angle, as discussed in Section~\ref{sect:gauge-1}. This produces a continuous family of universes, each described by a $U(1)$ gauge theory with a different value of the $\theta$ angle. Multiverse interference effects imply a complete cancellation between vortices/instantons of nonzero degree.

We will perform other consistency checks to reinforce the statements above in the following subsections.

\subsubsection{Example:  pure gauge theories}   

In this section we will explore the relation between pure $U(1)$ and ${\mathbb R}$ gauge theories in $D=2$, and check that partition functions match.

 A $D$-dimensional pure $U(1)$ Maxwell theory has the symmetry
\begin{equation}
U(1)_e^{[1]} \times U(1)^{[D-3]}_m,
\end{equation}
with a mixed anomaly.  If we topologically gauge the $U(1)^{[D-3]}_m$ symmetry,
then taking into account the mixed anomaly, the resulting theory has
a
${\mathbb R}^{[1]}$ symmetry, arising as the extension
\begin{equation}
1 \: \longrightarrow \: {\mathbb Z}^{[1]} \: \longrightarrow \:
{\mathbb R}^{[1]} \: \longrightarrow \:
U(1)^{[1]}_e \: \longrightarrow \: 1.
\end{equation}
The resulting theory is the pure $\mathbb{R}$ gauge theory with this 1-form symmetry. 
Topologically gauging the ${\mathbb Z}^{[1]} \subset {\mathbb R}^{[1]}$ is performing a quotient $\mathbb{R}/\mathbb{Z}\cong U(1)$, thus returning to the original theory.

Now, consider the special case of $D=2$.  Topologically
gauging the $(-1)$-form symmetry of the pure $U(1)$ Maxwell theory (i.e.~integrating over theta angles, as discussed in section~\ref{sect:rev:-1form}) then results in the ${\mathbb R}$ Maxwell theory with
an ${\mathbb R}^{[1]}$ symmetry.

Conversely, in $D=2$, if we start with the two-dimensional pure ${\mathbb R}$ gauge theory and
gauge ${\mathbb Z}^{[1]} \subset {\mathbb R}^{[1]}$, we recover
the pure $U(1)$ Maxwell theory.

Schematically, in two dimensions,
\begin{equation}
\mbox{pure $U(1)$ Maxwell} \: \mathrel{\mathop{\rightleftarrows}^{\mathrm{U(1)^{[-1]}}}_{\mathrm{{\mathbb Z}^{[1]}}}}
\:
\mbox{pure ${\mathbb R}$ Maxwell}.
\end{equation}

Each of these two-dimensional theories has a one-form symmetry, and so decomposes:
\begin{itemize}
    \item The pure $U(1)$ Maxwell theory has a $U(1)_e^{[1]}$ symmetry and
    decomposes (see e.g.~\cite{Cherman:2020cvw,Nguyen:2021yld,Nguyen:2021naa}, \cite[section 2]{Komargodski:2020mxz})
into Rep$(U(1)) \cong {\mathbb Z}$ universes, each an invertible field theory.
\item The pure ${\mathbb R}$ Maxwell theory has a ${\mathbb R}^{[1]}$ symmetry
and decomposes (see appendix~\ref{sect:r-gauge}) into Rep$({\mathbb R}) \cong {\mathbb R}$
universes, each an invertible field theory.
\end{itemize}

Now, let us reconcile the counting of universes in the ${\mathbb R}$ gauge theory,
viewed from two perspectives:
\begin{itemize}
\item Integrating the pure $U(1)$ theory over theta angles is equivalent to working with a family of universes, parametrized by $U(1)$, each a copy of the $U(1)$ gauge theory, hence each has a decomposition into
${\mathbb Z}$ invertible field theories.  Now, naively one might think that the decomposition into invertible field theories would be parametrized by $U(1) \times {\mathbb Z}$; however, because of the mixed anomaly between $U(1)_e^{[1]}$ and $U(1)_m^{[-1]}$, the correct one-form symmetry is
${\mathbb R}^{[1]}$, essentially because of the extension
\begin{equation}
    1 \: \longrightarrow \: {\mathbb Z} \: \longrightarrow \: {\mathbb R} \: \longrightarrow \: U(1) \: \longrightarrow \: 1.
\end{equation}
Alternatively, it was noted in \cite[section 4.1.2]{Pantev:2023dim} that there exists an analogue of the Witten effect in the decomposition of pure $U(1)$ Maxwell theory:  rotating the theta angle $\theta \mapsto \theta + 2 \pi$ also shifts universes.  This is another way to understand why the decomposition is into invertible field theories parametrized by ${\mathbb R}$ rather than
$U(1) \times {\mathbb Z}$ -- rotations of the $U(1)$ also shift ${\mathbb Z}$.

\item To compare, the pure ${\mathbb R}$ gauge theory was argued in appendix~\ref{sect:r-gauge} to decompose into Rep$({\mathbb R}) \cong {\mathbb R}$ universes, each of which is an invertible field theory.
\end{itemize}
Thus, we get the same decomposition, which is a consistency check of our claims.

We can also perform the same consistency check on the $U(1)$ gauge theory:
\begin{itemize}
    \item The pure $U(1)$ gauge theory in spacetime dimension $D=2$ is known 
    (see e.g.~\cite{Cherman:2020cvw,Nguyen:2021yld,Nguyen:2021naa},  \cite[section 2]{Komargodski:2020mxz})  to decompose into ${\mathbb Z}$-many universes, each of which is an invertible field theory.
    \item The pure ${\mathbb R}$ gauge theory has Rep$({\mathbb R}) \cong {\mathbb R}$ many universes, each an invertible field theory, but after gauging the ${\mathbb Z}^{[1]}$, the remaining symmetry will be ${\mathbb R}^{[1]} / {\mathbb Z}^{[1]} = U(1)^{[1]}$,
    hence the total number of universes will be
    Rep$(U(1)) \cong {\mathbb Z}$, each again an invertible field theory. 
\end{itemize}
Again, we see that the counts of universes match in these two descriptions of each theory.

Finally, as a consistency check on topologically gauging the $(-1)$-form symmetry, we will explicitly compute the partition function of the $(-1)$-form-symmetry-gauged $U(1)$ gauge theory, and demonstrate that it matches that of the ${\mathbb R}$ gauge theory, and also check the converse.

Elaborating on the partition function expressions given earlier, and following e.g.~\cite{Caporaso:2006kk,Paniak:2002fi,Paniak:2003gn}, \cite[section 4.1.2]{Pantev:2023dim},
the partition function of the pure $U(1)$ gauge theory, with theta angle $\theta$, can be written in the form
\begin{equation}
    Z_{U(1)}(\theta) \: \propto \: \sum_{n=-\infty}^{\infty} \exp\left( - \frac{ \pi^2}{ A}{n^2} \: + \: i \theta n\right),
\end{equation}
where $n$ is vortex number (the $c_1$ of a principal $U(1)$ bundle) and
$A$ is the area of the worldsheet.  (The first term, proportional to $1/A$, is due to the Maxwell kinetic term.)  As has already been noted, 
\begin{equation}
    \int d\theta\, Z_{U(1)}(\theta) \: \propto \: \sum_{n=-\infty}^{\infty} \exp\left( - \frac{ \pi^2}{ A}{n^2} \right) \, \delta_{n,0},
\end{equation}
reflecting the earlier statement that integrating over $\theta$ restricts the allowed vortex number to vanish, consistent with the interpretation of the result as an ${\mathbb R}$ gauge theory.

Following the analysis of e.g.~\cite{Caporaso:2006kk,Paniak:2002fi,Paniak:2003gn}, \cite[equations (4.15), (4.19)]{Pantev:2023dim},
\cite[section 2]{Komargodski:2020mxz},
we can Poisson resum the expression above to write
the partition function of the pure $U(1)$ Maxwell theory with theta angle $\theta$ in $D=2$ in the form
\begin{equation}
    Z_{U(1)}(\theta) \: = \: \sum_{q \in {\mathbb Z}} \exp\left( - A \left( q + \frac{\theta}{2\pi} \right)^2 \right).
\end{equation}
We note
\begin{itemize}
    \item For $\theta=0$, this is precisely the same form as the exact partition function of a $U(1)$ theory \cite[section 3.7]{Cordes:1994fc},
\cite{Migdal:1975zg,Drouffe:1978py,Lang:1981rj,Menotti:1981ry,Rusakov:1990rs,Witten:1991we,Witten:1992xu,Blau:1993hj}.
    \item The $q$ can be understood as total charge sectors.  In $D=2$, these are the universes \cite{Cherman:2020cvw,Nguyen:2021yld,Nguyen:2021naa}, \cite[section 2]{Komargodski:2020mxz}; in $D>2$, these correspond to superselection sectors.  (The $D=2$ case is a stronger version of the usual lore (see e.g.~\cite{Wick:1970bd}) concerning total charge sectors being superselection sectors, essentially because in $1+1$ dimensions, a point charge acts as a domain wall in spacetime.)
    See appendix~\ref{app:r:decomp} for a more detailed discussion.
    \item The expression above was derived in e.g.~\cite[equations (4.15), (4.19)]{Pantev:2023dim} on genus zero Riemann surfaces, but given the known form of the exact result, the only modifications at other genus would be factors of the dimension of the representation, which for $U(1)$ is always $1$, hence the expression above holds on any Riemann surface, modulo overall normalizations.
\end{itemize}

Now, when we integrate over $\theta$ (topologically gauge the $U(1)^{[-1]}$), this becomes
\begin{eqnarray}
\int_0^{2 \pi} d \theta \, Z_{U(1)}(\theta) 
&= &
    \int_0^{2 \pi} d \theta \,  \sum_{q \in {\mathbb Z}} \exp\left( - A \left( q + \frac{\theta}{2\pi} \right)^2 \right)
    \\
    & = &
    \int_{{\mathbb R}} dr \, \exp\left( - A \frac{r^2}{4 \pi^2} \right),
\end{eqnarray}
where we identify $r \in {\mathbb R}$ with $2 \pi q + \theta$ (using e.g.~the Witten effect, under which $\theta \mapsto \theta+2\pi$ induces $q \mapsto q + 1$ to get ${\mathbb R}$ instead of
$U(1) \times {\mathbb Z}$).
Now, this is precisely the partition function $Z_{\mathbb R}$ of the pure ${\mathbb R}$ gauge theory~(\ref{eq:2d:partfn-R}), hence
\begin{equation}
    Z_{\mathbb R} \: = \: \int_0^{2\pi} d\theta \, Z_{U(1)}(\theta),
\end{equation}
consistent with expectations.

We should note that ordinarily, the partition function of a gauge theory is normalized by the volume of the gauge group, which for a ${\mathbb R}$ gauge theory is infinite.
Such a factor obviously does not appear in the expression above, or in the expressions in
appendix~\ref{sect:r-gauge}, and so presumably
reflects a suitable renormalization in the definition of $Z_{\mathbb R}$,
which turns out to be consistent with the $U(1)$ partition function.
Later we will see other circumstances, however, (namely correlation functions in supersymmetric theories with gauge group ${\mathbb R}$) where that volume may play a role.

So far we have checked that the partition function
for the ${\mathbb R}$ gauge theory can be computed from that of the $U(1)$ gauge theory by 
integrating over $\theta$ angles.
We can also do the reverse, namely, start with the partition function for the ${\mathbb R}$ gauge theory and gauge the $B {\mathbb Z} = {\mathbb Z}^{[1]}$ to explicitly recover the partition function of the $U(1)$ theory, following closely analogous computations for compact groups
in \cite[section 7]{Sharpe:2019ddn}, \cite{Santilli:2024dyz}.  

First, in general terms, gauging the $B {\mathbb Z}$ adds a path integral sum over ${\mathbb Z}$-gerbes, of characteristic class $H^2(\Sigma, {\mathbb Z})$, so the gauging will contribute a sum
\begin{equation}
    \sum_{n \in H^2(\Sigma,{\mathbb Z})} \exp( - i n \theta) \, \cdots,
\end{equation}
where $\theta$ is the analogue of a theta angle for the $B {\mathbb Z}$ gauging.
(Specifying a theta angle in the $B{\mathbb Z}$ gauging will determine a theta angle in the resulting gauge theory, just as in the compact Lie group examples in \cite[section 7]{Sharpe:2019ddn}.)
(In addition, we have assumed for simplicity that the worldsheet $\Sigma$ is connected; if there are multiple components, then $n$ will be a vector of integers with multiple components, one for each connected component of $\Sigma$.)
Furthermore, each ${\mathbb R}$ gauge theory will be twisted by the ${\mathbb Z}$ gerbes.
We define a twisted cap partition function as
\begin{equation}
    Z_{\rm cap, twisted}(s) \: = \: \int_{ {\rm Rep}({\mathbb R})} dr \,
    \chi_r( s + 2 \pi n) \, \exp\left( -  A \frac{r^2}{4\pi^2} \right),
\end{equation}
where $s \in {\mathbb R}$, and
\begin{equation}
    \chi_r(s + 2 \pi n) \: = \: \exp(i n r) \chi_r(s).
\end{equation}
The resulting twisted partition function, summing over ${\mathbb Z}$-gerbes of
characteristic class $n$, and again assuming a connected worldsheet, is
\begin{eqnarray}
    Z(\theta) & = & \sum_{n \in H^2(\Sigma,{\mathbb Z})} \exp( - i n \theta)
    \int_{ {\rm Rep}({\mathbb R})} dr \,
    \exp(i n r) \exp\left( -  A \frac{r^2}{4 \pi^2} \right),
    \\
    & = & 
    \sum_{n \in {\mathbb Z}} \exp( - i n \theta) \int_{-\infty}^{+\infty} dr \,
    \exp\left( - \frac{A}{4\pi^2} \left( r - \frac{i n}{2} \frac{4 \pi^2}{A}  \right)^2 - \frac{n^2 \pi^2}{A} \right),
    \\
    & \propto & \sum_{n \in {\mathbb Z}} \exp( - i n \theta) \exp\left( - \frac{n^2 \pi^2}{A} \right),
    \\
    & \propto & \sum_{q \in {\mathbb Z}} \exp\left( - A \left( q + \frac{\theta}{2\pi} \right)^2 \right).
\end{eqnarray}
This is the partition function $Z_{U(1)}(\theta)$ of two-dimensional pure $U(1)$ gauge theory with theta angle $\theta$.  So, we see that gauging the $B {\mathbb Z}$ recovers pure $U(1)$ Maxwell theory.

\subsubsection{Example:  $U(1)$ theories with minimally-charged matter}  
\label{sect:ru1:mincharge}

Consider a two-dimensional $U(1)$ gauge theory with minimally-charged matter,
say complex scalars $\phi^i$, which are transformed under gauge transformations as
\begin{equation}
    \phi^i \: \mapsto \: \exp(i \alpha) \, \phi^i,
\end{equation}
for $\exp(i \alpha) \in U(1)$.
As discussed above, the theta angle can be described as a connection for a $(-1)$-form symmetry, and performing an (ordinary) integral over $\theta$ angles results
in a new theory which decomposes into a continuous family of universes, parametrized by $U(1)$. 

That theory described by a continuous family of universes is an ${\mathbb R}$
gauge theory with charged complex scalars $\phi^i$,
on which the ${\mathbb R}$ acts as
\begin{equation}
    \phi^i \: \mapsto \: \exp(i a) \, \phi^i,
\end{equation}
for $a \in {\mathbb R}$.
More formally, this can be understood as composing the $U(1)$ action with the projection map
${\mathbb R} \rightarrow U(1)$, and in general this is how ${\mathbb R}$ actions on the matter
are defined if one starts with a $U(1)$ gauge theory.

In this example, a $2 \pi {\mathbb Z} \subset {\mathbb R}$ subgroup of the gauge group acts
trivially, hence the ${\mathbb R}$ gauge theory has a $B {\mathbb Z} = {\mathbb Z}^{[1]}$ symmetry,
and on general principles \cite{Hellerman:2006zs} we expect this gauge theory to decompose into Rep$({\mathbb Z})\cong U(1)$-many universes, each of which is a ${\mathbb R} / 2 \pi {\mathbb Z} \cong U(1)$ gauge theory, consistent with the description derived from integrating the $U(1)$ theory over theta angles.

Now, the ${\mathbb R}$ gauge theory has no instantons, but a $U(1)$ gauge theory certainly can admit instantons.

Since we have a decomposition into a continuous family ($U(1)$) of universes, we expect that this is a special case of the instanton annihilation of sections~\ref{sect:gauge-1}, \ref{sect:int}, in which one integrates over $\theta$ to remove instantons.

To give a completely concrete example, consider a two-dimensional gauged linear sigma model
for ${\mathbb P}^N$, with $N+1$ chiral superfields of charge $1$, but with gauge group
${\mathbb R}$ instead of $U(1)$.  This is an example of the form above.
The gauge group has a trivially-acting $2 \pi {\mathbb Z}$ subgroup, hence a $B {\mathbb Z} = {\mathbb Z}^{[1]}$ symmetry, hence decomposes into a Rep$({\mathbb Z}) \cong U(1)$-family of universes,
each a copy of the ordinary supersymmetric ${\mathbb P}^N$ model with gauge group
$U(1)$, but with a theta angle determined by the position of the universe on
Rep(${\mathbb Z}$), so that contributions from all instantons cancel out.
Geometrically, this gauge theory describes a ${\mathbb Z}$ gerbe over ${\mathbb P}^N$, as we will discuss later in section~\ref{sect:zgerbe}.

Furthermore, we will study properties of gauged linear sigma models with gauge group ${\mathbb R}$ and matter of the form above in
more detail in section~\ref{sect:qc:r}.

\subsubsection{Example:  $U(1)$ theories with non-minimally-charged matter}

Consider a two-dimensional $U(1)$ gauge theory in which all matter has charges divisible by
some integer $p$, or equivalently all instanton/vortex numbers are divisible by $p$,
as in \cite{Pantev:2005rh,Pantev:2005wj,Pantev:2005zs}.

Now, topologically gauge the $U(1)^{[-1]}$ symmetry to get a ${\mathbb R}$ gauge theory.
If one further gauges a $(q {\mathbb Z})^{[1]}$ symmetry, the result will be a $U(1)$ gauge theory
with instanton/vortex numbers divisible by $p q$.  We can summarize this as follows:
\begin{equation}
    \xymatrix{
    T[U(1),p] \ar[rr]^{ U(1)^{[-1]} } & & T[ {\mathbb R}, p] \ar[dl]^{ (q {\mathbb Z})^{[1]} } 
    \\
    & T[U(1), pq] \ar[ul]^{ {\mathbb Z}_q^{[1]} } &
    }
\end{equation}
where $T[G,p]$ denotes an abelian gauge theory with gauge group $G$ and matter of charge divisible by $p$.

Similarly, if we start with a ${\mathbb R}$ gauge theory with a $B {\mathbb Z}$ symmetry, such that gauging the $B {\mathbb Z}$ would result in
a $U(1)$ gauge theory with minimally-charged matter,
and then in the ${\mathbb R}$ gauge theory, gauge instead a $(q {\mathbb Z})^{1]} \subset {\mathbb Z}^{[1]}$ symmetry, the result is a $U(1)$ theory with fields of charge divisible by $q$.  We can describe this more formally as follows:
\begin{equation}
T[U(1), q]  \: \mathrel{\mathop{\rightleftarrows}^{\mathrm{ \left( \frac{ U(1) }{ {\mathbb Z}_q } \right)^{[-1]}}}_{\mathrm{(q{\mathbb Z})^{[1]}}}}
\:
T[{\mathbb R}, 1],
\end{equation} 
where 
\begin{equation}
    {\rm Rep}(q {\mathbb Z}) \: = \: \frac{ U(1) }{ {\mathbb Z}_q }.
\end{equation}
Of course, $U(1)/{\mathbb Z}_q \cong U(1)$, but we keep track of this form to emphasize that the
radius of the circle has changed, which is reflected in the nonminimal charges and instanton restriction.

Note that this means that even for higher-form symmetries, one must take the global form of the group into account, not just for zero-form symmetries.

\subsubsection{Examples: ${\mathbb R}$ gauge theories with more general matter}

So far in this section we have discussed how topologically gauging $(-1)$-form symmetries in $U(1)$ gauge theories in two dimensions leads to ${\mathbb R}$ gauge theories, with the same matter.
In this section we consider, in some sense, the converse:  interpreting an ${\mathbb R}$ gauge theory with matter of somewhat more general charges.

Specifically, consider a two-dimensional ${\mathbb R}$ gauge theory with $n$ matter fields $\phi_i$ of nonzero charges $q_i \in {\mathbb R}$, $i \in \{1, \cdots, n\}$, by which we mean that $a \in {\mathbb R}$ acts as
\begin{equation}
    \phi_i \: \mapsto \: \exp\left( 2 \pi i q_i a \right) \, \phi_i.
\end{equation}
(For convenience in this section, we extracted a factor of $2 \pi$.)

Now, suppose the charges $q_i \in \mathbb{R}$ of the matter fields are all rationally related, in the sense that there exist integers $k_i$ such that
\begin{equation}
    \frac{k_1}{q_1} \: = \: \frac{k_2}{q_2} \: = \: \cdots \: = \: \frac{k_n}{q_n}\in \mathbb{Q},
\end{equation}
or equivalently,
\begin{equation}
    \frac{q_i}{q_j} \: = \: \frac{k_i}{k_j} \: \in \: {\mathbb Q},
\end{equation}
for every pair $i, j$ for which $q_j, k_j \neq 0$.  In this case, there is a trivially-acting subgroup ${\mathbb Z}$ of the gauge group ${\mathbb R}$, given by $a \in {\mathbb R}$ of the form 
\begin{equation}
    a \: = \: n \frac{k_1}{q_1}
\end{equation}
for $n \in {\mathbb Z}$.  Explicitly, for such $a$, $\phi_i$ picks up the phase
\begin{equation}
    \exp\left( 2 \pi i q_i n \frac{k_1}{q_1} \right) \: = \: \exp\left( 2 \pi i q_i n \frac{k_i}{q_i} \right)
    \: = \: \exp\left( 2 \pi i n k_i \right) \: = \: 1.
\end{equation}
This demonstrates that, in this case, there is a trivially-acting ${\mathbb Z} \subset {\mathbb R}$,
hence we expect the ${\mathbb R}$ gauge theory to decompose into Rep$({\mathbb Z}) = U(1)$ universes,
each a ${\mathbb R}/{\mathbb Z} \cong U(1)$ gauge theory.

Next, we turn to the constituent universes.  We claim that each universe is a $U(1)$ gauge theory with
$n$ matter fields of charges $k_1, \cdots, k_n \in {\mathbb Z}$.

Now, that description is not unique.  For example, the integers could all be divisible by an integer.
However, in this case, each $U(1)$ gauge theory of charges $\{ k_i \}$ would itself further decompose into $s$ universes, where $s = {\rm gcd}(k_1, \cdots, k_n)$, each a $U(1)$ gauge theory with matter of charges
$\{ k_1/s, k_2/s, \cdots, k_n/s \}$.  That said, regardless, the ${\mathbb R}$ gauge theory would still decompose into a family of universes parametrized by $U(1)$.  The decompositions into universes with charges $\{ k_i \}$ and into universes into universes with charges $\{ k_i/s \}$ would both be parametrized by circles, albeit of different radii:
\begin{equation}
    1 \: \longrightarrow \: {\mathbb Z}_s \: \longrightarrow \: U(1)_{ \{ k_i/s \} } \: \longrightarrow \:
    U(1)_{ \{ k_i \} } \: \longrightarrow \: 1.
\end{equation}

\subsubsection{Example:  An ${\mathbb R}$ gauge theory not relatable to $U(1)$}

Consider a two-dimensional ${\mathbb R}$ gauge theory with a real scalar field $\phi$ on which
${\mathbb R}$ acts by translations:  $\phi \mapsto \phi + a$.

This theory does not have a $B {\mathbb Z} = {\mathbb Z}^{[1]}$ symmetry, or any other one-form
symmetry, as no nontrivial subgroup of the gauge group acts trivially.

As a result, this theory is an example of an ${\mathbb R}$ gauge theory that cannot be related to a $U(1)$ gauge theory:  there is no $B {\mathbb Z}$ to gauge to turn it into a $U(1)$ gauge
theory, and conversely, we believe there is no $(-1)$-form symmetry topological gauging of a $U(1)$ gauge theory that will produce this ${\mathbb R}$ gauge theory.  

Phrased another way, the operation of topologically gauging $U(1)^{[-1]}$ in a $U(1)$ gauge theory is not surjective:  it will generate ${\mathbb R}$ gauge theories, but not every ${\mathbb R}$ gauge theory is of that form.

\subsection{$G$ gauge theory versus trivial theory}

Let $G$ be an abelian group.
($G$ can be finite or nonfinite, compact or noncompact; we only require that $G$ be abelian, for simplicity.)

As another special case of the yoga above, that informs other examples,
consider a $G$ gauge theory with a $BG = G^{[1]}$ symmetry, which we topologically gauge to completely remove the gauge symmetry.  
In terms of the yoga~(\ref{eq:yogaD}),
since $G/G = 1$, the predicted structure is
\begin{equation}  
\mbox{$G/G =$no gauge theory} \:
\mathrel{ \mathop{\rightleftarrows}^{ {\rm Rep}(G)^{[D-3]} }_{ \mathrm{{G}^{[1]}} } }
\:
\mbox{$G$ gauge theory},
\end{equation}

Starting with the $G$ gauge theory, if we topologically gauge $BG = G^{[1]}$, we remove the gauging altogether.  Mathematically, gauging $BG$ adds $G$ gerbes to the path integral, which twist the $G$ bundles.

In physics, if the $G$ gauge theory has a $BG = G^{[1]}$ symmetry, then $G$ must act trivially on all the matter.  The quantum field theory may have a sector that decouples from the gauge theory; for simplicity, we set such a factor aside and focus on the $G$ gauge theory itself.  If $G$ acts trivially on all the matter, then the gauge theory decomposes into Rep$(G)$ universes:
\begin{itemize}
    \item If $G$ is finite, then the gauge theory is a Dijkgraaf-Witten theory, or equivalently an orbifold of a point, and its decomposition In two dimensions was studied in 
    \cite{Hellerman:2006zs}. It decomposes to copies of sigma models with target space a point, meaning, invertible field theories.
    \item If $G$ is continuous and abelian, it also decomposes to invertible field theories, as discussed in e.g.~\cite{Cherman:2020cvw},  \cite[section 2]{Komargodski:2020mxz}.
    \item If $G$ is nonabelian, the theory still decomposes to invertible field theories, but the result can no longer be understood in terms of existence of a $BG = G^{[1]}$ symmetry, as that language requires $G$ be abelian.
    See instead appendix~\ref{app:nonabel} for a discussion of the analogue of a one-form symmetry in such cases.  For $G$ finite and nonabelian, the gauge theory is again Dijkgraaf-Witten theory, and the decomposition to invertible field theories was discussed in \cite{Hellerman:2006zs}; for $G$ continuous and nonabelian, the decomposition to invertible field theories was discussed in \cite{Nguyen:2021yld,Nguyen:2021naa}.
\end{itemize}

We can also understand this mathematically.
A pure $G$-gauge theory (for $G$ not necessarily abelian) can be regarded as a $\sigma$-model whose target space is a stack $[*/G]$, a smooth groupoid (known as the classifying stack, denoted $BG$) with a single object and morphisms labelled by $G$. Now, for $H\leq G$ a central subgroup of $G$, there is an action on this groupoid that sends a morphism $g$ to $h\cdot g$, the left multiplication by $h\in H$. Gauging this action requires considering as a target space a higher stack $[[*/G]/[*/H]]$ that introduces 2-isomorphisms between the 1-isomorphisms $g, \tilde{g}\in G$ for $\tilde{g}=h\cdot g$, labeled by $h\in H$.  But this quotient stack is equivalent to the stack with a single object and $G/H$ worth of 1-isomorphisms, and trivial $2$-isomorphisms. In other words, it is $[ * / (G/H) ]$, which as the target space of a sigma model yields a pure $G/H$-gauge theory.

An intuition for this is as follows. The point here is that for $H$ a central subgroup of $G$, there is an action of $BH$ on $BG$, which from the point of view of gauge theory essentially tensors $H$ bundles into $G$ bundles.  If we topologically gauge $BH$, then we identify any two $G$ bundles related by an $H$ bundle -- which is equivalent to working with principal $(G/H)$ bundles.

A different perspective comes from considering the fiber sequence
\begin{equation}
    \cdots \to BH \to BG \to B(G/H) \xrightarrow{\omega} B^2H=B(BH),
\end{equation}
with $\omega: B(G/H)\to B^2H=B(BH)$ is the map that classifies the central extension. This fibration thus exhibits $BG$ as a higher principal $BH$ (regarded as a 2-group) bundle over $B(G/H)$, where the action of the 2-group $BH$ on $BG$ is the left multiplication described above.

Now, consider the opposite direction.  Since we start with no gauge theory at all,
there is no global Rep$(G)^{[D-3]}$ symmetry present.  However, we can always topologically gauge a trivially-acting symmetry.  Now, topologically gauging a trivially-acting $p$-form symmetry results in a theory with a global $(p+1)$-form symmetry, whose topological operators arise as the ``twisted sector operators'' (defect operators) of the trivially-acting $p$-form symmetry.
Here, topologically gauging a trivially-acting Rep$(G)^{[D-3]}$ symmetry
should result in a theory with a global Rep$(G)^{[D-2]}$ symmetry.  Such a symmetry is precisely the quantum symmetry of a $G$ gauge theory in $D$ dimensions, so we see that the yoga is consistent.

We highlight, however, that these arguments are only at the level of bundles, not connections.

\subsection{${\mathbb R}^{\times}$ gauge theory}

In this section, we briefly consider noncompact gauge theories with gauge group
${\mathbb R}^{\times} \cong {\mathbb R} \times {\mathbb Z}_2$ in $D=2$ spacetime dimensions.

We can relate ${\mathbb R}^{\times}$ gauge theories to other gauge theories in at least
two ways.
\begin{itemize}
\item From the fact that ${\mathbb R}^{\times} / {\mathbb Z}_2 \cong {\mathbb R}$,
in two dimensions, from~(\ref{eq:yoga2}), one has
\begin{equation}
\mbox{${\mathbb R}$ gauge theory} \: \mathrel{\mathop{\rightleftarrows}^{ \mathrm{ {\mathbb Z}_2^{[-1]}  } }_{\mathrm{{\mathbb Z}_2^{[1]}}}}
\:
\mbox{${\mathbb R}^{\times}$ gauge theory},
\end{equation}
where 
on the right we restrict to ${\mathbb R}^{\times}$ gauge theories with a 
${\mathbb Z}_2^{[1]}$ symmetry (i.e.~matter is invariant under ${\mathbb Z}_2 \subset {\mathbb R}^{\times}$). 

On the right, topologically gauging $B {\mathbb Z}_2 = {\mathbb Z}_2^{[1]}$ changes the gauge group
from ${\mathbb R}^{\times}$ to ${\mathbb R}^{\times}/{\mathbb Z}_2 \cong {\mathbb R}$.

On the left, the ${\mathbb R}$ gauge theory has no instantons, hence no theta angles.
To get the ${\mathbb R}^{\times}$ gauge theory, we topologically gauge a trivially-acting
${\mathbb Z}_2^{[-1]}$, which results in a theory with a ${\mathbb Z}_2^{[0]}$ symmetry,
the quantum symmetry of the ${\mathbb R}^{\times}$ gauge theory.

In terms of decomposition, an ${\mathbb R}^{\times}$ gauge theory with a $B {\mathbb Z}_2 = {\mathbb Z}_2^{[1]}$ symmetry in $D=2$ decomposes into $| {\rm Rep}({\mathbb Z}_2) | = 2$ universes, each an ${\mathbb R}^{\times}/{\mathbb Z}_2 \cong {\mathbb R}$ gauge theory.
Topologically gauging the $B {\mathbb Z}_2 = {\mathbb Z}_2^{[1]}$ symmetry selects one of the universes
(depending upon a $B {\mathbb Z}_2$ theta angle), as in \cite{Sharpe:2019ddn}, and the quantum symmetry
for that gauging is ${\mathbb Z}_2^{[-1]}$.

An example is as follows.
Consider a  ${\mathbb R}^{\times}$ gauge theory with a real scalar $\phi$ that transforms under ${\mathbb R}^{\times}$ as $\phi \mapsto \phi + \ln |b|$ for
$b \in {\mathbb R}^{\times}$.
(We are utilizing the projection map ${\mathbb R}^{\times} \rightarrow {\mathbb R}$ given by
$b \mapsto \ln |b|$.)
This is a well-defined action\footnote{
Note that it is important to check that one indeed has an action in the technical sense,
which means for example that the identity in the group leaves the matter invariant.
For example, if under ${\mathbb R}^{\times}$, we defined $\phi \mapsto (\ln |b|) \phi$,
then the identity in ${\mathbb R}^{\times}$ (namely, $1$)
would not send $\phi \mapsto \phi$, but rather,
$\phi \mapsto 0$.  This is not a group action in the technical sense.
} of ${\mathbb R}^{\times}$ (meaning, the identity maps $\phi$ to itself, and the group law on ${\mathbb R}^{\times}$ is respected), with a ${\mathbb Z}_2$ kernel,
hence this theory has a global $B {\mathbb Z}_2 = {\mathbb Z}_2^{[1]}$ symmetry.

Topologically gauging that ${\mathbb Z}_2^{[1]}$ symmetry gives an ${\mathbb R}^{\times}/{\mathbb Z}_2 \cong {\mathbb R}$ gauge theory with a real scalar $\phi$ that transforms under ${\mathbb R}$ as $\phi \mapsto \phi + x$ for $x \in {\mathbb R}$.  Furthermore, from decomposition,
the ${\mathbb R}^{\times}$ gauge theory is equivalent to two copies of this ${\mathbb R}$ gauge theory.

\item From the fact that ${\mathbb R}^{\times} / {\mathbb R} \cong {\mathbb Z}_2$,
in two dimensions, from~(\ref{eq:yoga2}), one has
\begin{equation}
\mbox{${\mathbb Z}_2$ gauge theory} \: \mathrel{\mathop{\rightleftarrows}^{ \mathrm{ {\mathbb R}^{[-1]}  } }_{\mathrm{{\mathbb R}^{[1]}}}}
\:
\mbox{${\mathbb R}^{\times}$ gauge theory},
\end{equation}
where on the right we restrict to ${\mathbb R}^{\times}$ gauge theories with a $B {\mathbb R} = {\mathbb R}^{[1]}$ global symmetry (i.e.~matter is invariant under ${\mathbb R} \subset {\mathbb R}^{\times}$).

On the right, topologically gauging $B {\mathbb R} \cong {\mathbb R}^{[1]}$ changes the gauge group from ${\mathbb R}^{\times}$ to ${\mathbb R}^{\times}/{\mathbb R} \cong {\mathbb Z}_2$,
and then topologically gauging the quantum symmetry, namely Rep$({\mathbb R})^{[-1]}$, returns the original gauge theory.

An example is as follows.  Let $\phi$ be a collection of matter fields on which
${\mathbb R}^{\times}$ acts as $\phi \mapsto (-)^{{\rm sgn}(b)} \phi$ for $b \in {\mathbb R}^{\times}$.  Here, the sign of $b \in {\mathbb R}^{\times}$ completely determines the group action, hence ${\mathbb R} \subset {\mathbb R}^{\times}$ acts trivially, so the theory has
a global $B {\mathbb R} = {\mathbb R}^{[1]}$ symmetry, and as a result, has a decomposition into
Rep$({\mathbb R}) \cong {\mathbb R}$ universes, a continuous family.

The corresponding ${\mathbb Z}_2$ gauge theory has matter fields $\phi$ under which the generator of ${\mathbb Z}_2$ acts as $\phi \mapsto - \phi$, a conventional ${\mathbb Z}_2$ orbifold in other words.

\end{itemize}

Next we consider the decomposition of pure ${\mathbb R}^{\times}$ gauge theory in $D=2$.
Now, in principle, in an abelian gauge theory in $D=2$, if a subgroup $G$ of the gauge group acts trivially, then there is a decomposition into universes indexed by Rep$(G)$,
so here, we would expect a decomposition into universes indexed by
Rep$({\mathbb R}^{\times}) \cong {\mathbb R}^{\times}$.  Let us perform a couple of consistency tests on this expectation.
\begin{itemize}
\item A pure ${\mathbb R}^{\times}$ gauge theory has a $B{\mathbb Z}_2 = {\mathbb Z}_2^{[1]}$ global symmetry, and so decomposes into two copes of a pure ${\mathbb R}$ gauge theory.  Each of those pure ${\mathbb R}$ gauge theories decomposes, from appendix~\ref{sect:r-gauge}, into a family of ${\mathbb R}$ universes, each an invertible field theory,
so that altogether the ${\mathbb R}^{\times}$ gauge theory decomposes into ${\mathbb R} \times {\mathbb Z}_2 \cong {\mathbb R}^{\times}$ universes, each an invertible field theory.
\item A pure ${\mathbb R}^{\times}$ gauge theory has a $B {\mathbb R} = {\mathbb R}^{[1]}$ global symmetry, and so decomposes into Rep$({\mathbb R}) \cong {\mathbb R}$ copies of a pure ${\mathbb R}^{\times}/{\mathbb R} \cong {\mathbb Z}_2$ gauge theory.
Now, a pure ${\mathbb Z}_2$ gauge theory is the same as two-dimensional Dijkgraaf-Witten theory for ${\mathbb Z}_2$, and it
decomposes \cite{Hellerman:2006zs} into two universes, each an invertible field theory,
so that altogether
the ${\mathbb R}^{\times}$ gauge theory decomposes into ${\mathbb Z}_2 \times {\mathbb R} \cong {\mathbb R}^{\times}$ universes, each an invertible field theory, consistent with the computation above.
\end{itemize}
We therefore propose that a pure ${\mathbb R}^{\times}$ gauge theory in $D=2$ decomposes into Rep$({\mathbb R}^{\times}) \cong {\mathbb R}^{\times}$-many universes, each an invertible field theory. This example illustrates that even in noncompact, disconnected groups, topological gauging can eliminate discrete components and yield connected gauge theories.

\subsection{${\mathbb C}^{\times}$ gauge theory}

In this section, we briefly consider noncompact gauge theories with gauge group
${\mathbb C}^{\times} \cong {\mathbb R} \times U(1)$ in $D=2$ spacetime dimensions.

We can relate ${\mathbb C}^{\times}$ gauge theories to other gauge theories in at least two ways.
\begin{itemize}
\item From the fact that ${\mathbb C}^{\times} \cong {\mathbb R} \times U(1)$,
hence ${\mathbb C}^{\times} / U(1) \cong {\mathbb R}$, 
we have that in two dimensions, from~(\ref{eq:yoga2}),
\begin{equation}
\mbox{${\mathbb R}$ gauge theory} \: \mathrel{\mathop{\rightleftarrows}^{ \mathrm{ {\mathbb Z}^{[-1]}  } }_{\mathrm{ U(1)^{[1]}}}}
\:
\mbox{${\mathbb C}^{\times}$ gauge theory},
\end{equation}
where on the left we restrict to ${\mathbb R}$ gauge theories with a ${\mathbb Z}^{[-1]}$ symmetry
and on the right we restrict to ${\mathbb C}^{\times}$ gauge theories with a $U(1)^{[1]}$ symmetry
(i.e.~matter is invariant under $U(1) \subset {\mathbb C}^{\times}$).

On the right, topologically gauging $BU(1) = U(1)^{[1]}$ changes the gauge group from ${\mathbb C}^{\times}$ to ${\mathbb C}^{\times} / U(1) \cong {\mathbb R}$.

On the left, the ${\mathbb R}$ gauge theory has no instantons, hence no theta angles.  To get the ${\mathbb C}^{\times}$ gauge theory, we topologically gauge a trivially-acting ${\mathbb Z}^{[-1]}$, which results in a theory with a Rep$({\mathbb Z})^{[1]} \cong U(1)^{[1]}$ quantum symmetry, the global symmetry of the ${\mathbb C}^{\times}$ gauge theory.

In terms of decomposition, a ${\mathbb C}^{\times}$ gauge theory with a 
$BU(1) = U(1)^{[1]}$ global symmetry in $D=2$ spacetime dimensions decomposes into Rep$(U(1)) \cong {\mathbb Z}$ universes,
each a ${\mathbb C}^{\times}/U(1) \cong {\mathbb R}$ gauge theory.  Topologically gauging that
$BU(1) = U(1)^{[1]}$ selects one of the universes (depending upon a $BU(1)$ theta angle),
as in \cite{Sharpe:2019ddn}, and the quantum symmetry for that gauging is
Rep$(U(1))^{[-1]} \cong {\mathbb Z}^{[-1]}$.

An example is as follows.  Consider a ${\mathbb C}^{\times}$ gauge theory with a real scalar $\phi$
that transforms under ${\mathbb C}^{\times}$ as $\phi \mapsto \phi + \ln |b|$ for $b \in {\mathbb C}^{\times}$.  This is a well-defined action with a $U(1)$ kernel, hence this theory has a global $BU(1) = U(1)^{[1]}$ symmetry.

Topologically gauging that $U(1)^{[1]}$ symmetry gives a ${\mathbb C}^{\times}/U(1) \cong {\mathbb R}$ gauge theory, with a real scalar $\phi$ that transforms under ${\mathbb R}$ as $\phi \mapsto \phi + x$ for $x \in {\mathbb R}$.  Furthermore, from decomposition, the ${\mathbb C}^{\times}$ gauge theory is equivalent to Rep$(U(1)) \cong {\mathbb Z}$ copies of this ${\mathbb R}$ gauge theory.

\item From the fact that ${\mathbb C}^{\times} / {\mathbb R} \cong U(1)$, we also have
\begin{equation}
\mbox{$U(1)$ gauge theory} \: \mathrel{\mathop{\rightleftarrows}^{ \mathrm{ {\mathbb R}^{[-1]}  } }_{\mathrm{ {\mathbb R}^{[1]}}}}
\:
\mbox{${\mathbb C}^{\times}$ gauge theory},
\end{equation}

On the right, topologically gauging $B {\mathbb R} = {\mathbb R}^{[1]}$ changes the gauge group from ${\mathbb C}^{\times}$ to ${\mathbb C}^{\times}/{\mathbb R} \cong U(1)$.
To return to the original theory, we topologically gauge the quantum symmetry, which is 
Rep$({\mathbb R})^{[-1]} \cong {\mathbb R}^{[-1]}$.

In terms of decomposition, a ${\mathbb C}^{\times}$ gauge theory with a $B {\mathbb R} = {\mathbb R}^{[1]}$ global symmetry in $D=2$ spacetime dimensions decomposes into 
Rep$({\mathbb R}) \cong {\mathbb R}$ universes.  Topologically gauging that $B {\mathbb R} = {\mathbb R}^{[1]}$ global symmetry selects one of the universes, as in \cite{Sharpe:2019ddn},
and the quantum symmetry for that gauging is Rep$({\mathbb R})^{[-1]} \cong {\mathbb R}^{[-1]}$.

An example is as follows.  Consider a ${\mathbb C}^{\times}$ gauge theory with a complex scalar $\phi$ that transforms under ${\mathbb C}^{\times}$ as $\phi \mapsto (b / |b|) \phi$ for
$b \in {\mathbb C}^{\times}$.  This is a well-defined action with kernel ${\mathbb R}$, hence this theory has a global $B {\mathbb R} = {\mathbb R}^{[1]}$ symmetry.

Topologically gauging that ${\mathbb R}^{[1]}$ symmetry gives a ${\mathbb C}^{\times}/{\mathbb R} \cong
U(1)$ gauge theory with a complex scalar $\phi$ that transforms under $U(1)$ as $\phi \mapsto \alpha \phi$ for $\alpha \in U(1)$.  Furthermore, from decomposition, the
${\mathbb C}^{\times}$ gauge theory is equivalent to Rep$({\mathbb R}) \cong {\mathbb R}$
copies of this $U(1)$ gauge theory.

\end{itemize}

Next, we consider the decomposition of pure ${\mathbb C}^{\times}$ gauge theory in $D=2$.
As mentioned previously, in principle, in an abelian gauge theory in $D=2$, if a subgroup $G$ of the gauge group acts trivially, then there is a decomposition into universes indexed by Rep$(G)$,
so here, we would expect a decomposition into universes indexed by
Rep$({\mathbb C}^{\times}) \cong {\mathbb Z} \times {\mathbb R}$.  Let us perform a couple of consistency tests on this expectation.
\begin{itemize}
    \item A pure ${\mathbb C}^{\times}$ gauge theory has a $BU(1) = U(1)^{[1]}$ global symmetry, and so decomposes into Rep$(U(1)) \cong {\mathbb Z}$ copies of a pure
    ${\mathbb C}^{\times}/U(1) \cong {\mathbb R}$ gauge theory.
    Each of those pure ${\mathbb R}$ gauge theories decomposes, from appendix~\ref{sect:r-gauge}, into a family of ${\mathbb R}$ universes, each an invertible field theory,
    so that altogether the ${\mathbb C}^{\times}$ gauge theory decomposes into
    ${\mathbb Z} \times {\mathbb R}$ universes, each an invertible field theory.
    \item A pure ${\mathbb C}^{\times}$ gauge theory has a $B {\mathbb R} = {\mathbb R}^{[1]}$ global symmetry, and so decomposes into Rep$({\mathbb R}) \cong {\mathbb R}$ copies of a pure ${\mathbb C}^{\times}/{\mathbb R} \cong U(1)$ gauge theory.
    Now, a pure $U(1)$ gauge theory in $D=2$ spacetime dimensions decomposes
    (see e.g.~\cite{Cherman:2020cvw,Nguyen:2021yld,Nguyen:2021naa},  \cite[section 2]{Komargodski:2020mxz}) into ${\mathbb Z}$ universes, each an invertible field theory, so that altogether the pure ${\mathbb C}^{\times}$ theory decomposes into ${\mathbb Z} \times {\mathbb R}$ universes, each an invertible field theory, consistent with the computation above.
\end{itemize}
We therefore proposal that a pure ${\mathbb C}^{\times}$ gauge theory in $D=2$ spacetime dimensions decomposes into ${\mathbb Z} \times {\mathbb R}$ universes, each an invertible field theory. 

We conclude this section by noting that topological gauging applies naturally to abelian gauge groups built from products such as $\mathbb{C}^\times \cong \mathbb{R} \times U(1)$. The presence of two independent $(-1)$-form symmetries reflects the factorization of the group, and gauging either component yields theories with only partial instanton suppression. In this sense, $\mathbb{C}^\times$ theories interpolate between the structures studied earlier in this section.

\section{Sigma models and GLSM realizations}   \label{sect:sigma}

In this section we will discuss instanton restriction and decomposition from the perspective of a sigma model whose target space is a suitable gerbe, following the spirit of \cite{Pantev:2005zs}.  We will also describe realizations in supersymmetric gauged linear sigma models (GLSMs) \cite{Witten:1993yc} with noncompact gauge groups, and discuss their properties.

\subsection{Instanton restriction, theta angles, and decomposition}

In sigma models, there is an analogue of summing / integrating over theta angles.  It is an old result that the role of theta angles is played, in two-dimensional theories, by the topological term
\begin{equation}
    \int \phi^* B,
\end{equation}
where $B$ is a closed two-form on the target space, and $\phi: \Sigma \rightarrow B$ is the sigma model map.  Summing/integrating over theta angles in a gauge theory corresponds in a sigma model to summing/integrating over choices of $B$.

In orbifolds specifically, choices of discrete torsion are understood as 
as lifts of the orbifold group action to $B$ fields \cite{Sharpe:2000ki,Sharpe:2000wu},
making them
(analogues of) theta angles, as previously discussed in section~\ref{sect:finite-1}.
Thus,
sums over theta angles often involve sums over choices of discrete torsion.  
We have already discussed orbifolds in this paper, but we take a moment to remind the reader of a simple prototypical example:  in a $D_4$ orbifold of a space $X$,
where ${\mathbb Z}_2 \subset D_4$ acts trivially, one has
\cite[section 5.2]{Hellerman:2006zs}
\begin{equation}
    {\rm QFT}\left( [X/D_4] \right) \: = \:
    {\rm QFT}\left( [X/{\mathbb Z}_2 \times {\mathbb Z}_2] \right) \, \coprod \,
    {\rm QFT}\left( [X/{\mathbb Z}_2 \times {\mathbb Z}_2]_{\omega} \right),
\end{equation}
where $\omega$ denotes the nontrivial element of discrete torsion in a 
$D_4/{\mathbb Z}_2 \cong {\mathbb Z}_2 \times {\mathbb Z}_2$ orbifold.

Much as in the case of gauge theories, we can implement a total instanton restriction, here by integrating over $B$ fields instead of theta angles, as was previously suggested in 
\cite[appendix D.1]{Pantev:2023dim}.  Formally, this follows the same pattern as discussed for gauge theories.  Instead of a single instanton degree, there is a vector of instanton degrees, with as many components as dim $H_2(X, {\mathbb Z})$.  The partition function
formally can be written
\begin{equation}
    Z(\vec{\theta}) \: = \: \sum_{\vec{n} \in H_2(X, {\mathbb Z})} Z(\vec{n}) \exp(i \vec{n} \cdot \vec{\theta}),
\end{equation}
where 
\begin{equation}
    \vec{n} \cdot \vec{\theta} \: \propto \:  \int \phi^* B,
\end{equation}
(which references some fixed basis of $H_2(X,{\mathbb Z})$, for example).
By a (multidimensional) Fourier transform, physically interpreted as an integration over $\theta$ angles, one can recover $Z(\vec{n})$ independently, the partition function for fixed
$\vec{n}$.  This is the same idea we discussed in the context of gauge theories in
section~\ref{sect:gauge-1}.

For example, in the case of the $D_4$ orbifold example, as previously reviewed in section~\ref{sect:rev} (see also \cite[section 5.2]{Hellerman:2006zs}), 
summing over choices of discrete torsion has the effect of cancelling out contributions from
nonperturbative sectors (partial traces) that appear in a ${\mathbb Z}_2 \times {\mathbb Z}_2$ orbifold but not in a $D_4$ orbifold.  Integrating over continuous $(-1)$-form symmetries can, for reasons discussed repeatedly elsewhere in this paper, accomplish a total instanton restriction.

\subsection{Application to Gross-Taylor}  \label{sect:app:gt}

It has been argued (see e.g.~\cite{Gross:1993hu}) that two-dimensional pure Yang-Mills theory with gauge group $SU(N)$
is, in the large $N$ limit, equivalent to a string theory, with target space the Riemann surface on which the pure Yang-Mills is defined.
It is also known that two-dimensional pure Yang-Mills theory has a decomposition into invertible field theories, indexed by the irreducible representations of the gauge
group \cite{Nguyen:2021yld,Nguyen:2021naa}.  

The intersection of these two phenomena was discussed in \cite{Pantev:2023dim}.
It was argued there that each universe of the decomposition of two-dimensional pure $SU(N)$ Yang-Mills theory, could be interpreted at large $N$ as a sigma model in which instantons (holomorphic maps) were restricted to a single (typically nonzero) degree, fixed by the representation determining the universe.

Such a total instanton restriction was more extreme than had previously been studied in examples of decomposition, and \cite[appendix D.1]{Pantev:2023dim} expressed reservations about this interpretation, but with the examples of this paper in mind, we can now understand the phenomenon there more concretely, as a special case of the total instanton restrictions discussed here.

Now, this paper has largely focused on restrictions to instanton number zero,
but restrictions to nonzero degree are also possible, see
section~\ref{sect:gauge-1}.  
Some supersymmetric proposals for such variations in nonlinear sigma models were
discussed in \cite[appendix D.2]{Pantev:2023dim}, and essentially look like supersymmetrized Tanizaki-\"Unsal \cite{Tanizaki:2019rbk} restrictions (as in section~\ref{sect:tu:susy}) but with a fixed, classical, background field strength $Y$ instead of a dynamical field.  Those precise actions, as discussed in \cite[appendix D.2]{Pantev:2023dim}, have the property that they break supersymmetry, but are still compatible with a topological twist, and so are still relevant to the Gross-Taylor string.  See \cite[appendix D.2]{Pantev:2023dim} for further details.

\subsection{Sigma models on ${\mathbb Z}$ gerbes and GLSM realizations}
\label{sect:zgerbe}

In this section we will discuss how total instanton restriction can be realized geometrically in terms of sigma models whose target spaces are gerbes, and their GLSM realizations.

We should perhaps begin with some general remarks on geometry.
In previous work e.g.~\cite{Pantev:2005rh,Pantev:2005wj,Pantev:2005zs}, 
we have discussed Deligne-Mumford stacks, realized by gauge theories
in which only a finite subgroup of the gauge group acts trivially on any point of the Higgs branch -- the geometry has only finite stabilizers, in other words.
In this section, the stacks describe geometries in which the stabilizers are not finite groups.  Corresponding stacks still exist, but they are no longer Deligne-Mumford stacks, and less is understood about geometric interpretations of such stacks, although many statements still apply.
(For one example, curiously, more general stacks can have negative dimensions, which is correctly reflected in the central charge computations discussed in \cite[section 2.1, equ'n (2.14)]{Eager:2020rra}.)

Next, we turn to interpretations of maps, which can be discussed independently of whether a stack is Deligne-Mumford.
As discussed in e.g.~\cite[section 3.1]{Pantev:2005zs}, \cite[section 2]{Hellerman:2010fv},
a map from a space $X$ into any gerbe ${\cal G}$ over $M$ is 
equivalent to a map $f: X \rightarrow M$ together with a trivialization of $f^* {\cal G}$.
Now, the reference \cite[section 2]{Hellerman:2010fv} studied maps into ${\mathbb Z}_p$ gerbes
on a space $M$, and argued that such maps must have degrees divisible by $p$.
We can also take a $p \rightarrow \infty$ limit, and in this case, we will see a restriction to maps of degree zero.
To see this, suppose ${\cal G}$ is a ${\mathbb Z}$ bundle over $M$, then a map
$f: X \rightarrow M$ induces a map
\begin{equation}   \label{eq:Zgerbe-constr}
    f^*: \: H^2(M,{\mathbb Z}) \: \longrightarrow \: H^2(X, {\mathbb Z}).
\end{equation}
For example, if $X = {\mathbb P}^1$ and $M = {\mathbb P}^N$, then $f: X \rightarrow M$ is characterized by an integer (its degree), call it $d$, and the map~(\ref{eq:Zgerbe-constr}) is multiplication by $d$.  If we let $n \in H^2(M, {\mathbb Z})$ denote the characteristic class of ${\cal G}$, then the requirement that $f^* {\cal G}$ implies that $d n = 0$, or more simply,
that $d=0$.  In this fashion, we see that maps into ${\mathbb Z}$ gerbes are restricted to degree zero.  For more general spaces, where the `degree' is no longer merely an integer, the characteristic class of the gerbe determines the restriction:  
\begin{itemize}
    \item For freely-generated parts of $H^2(M,{\mathbb Z})$, if the gerbe characteristic class is nonzero, then the map must act with degree zero,
    \item For torsion contributions to $H^2(M,{\mathbb Z})$, if the gerbe characteristic class is nonzero, then the map must have degree that maps the gerbe characteristic class to zero.
    (In other words, the map degree need not vanish, but must have suitable divisibility properties to map the gerbe characteristic class to zero.)
\end{itemize}

In the case of the ${\mathbb Z}$-gerbe, a sigma model with target the ${\mathbb Z}$-gerbe
has a $B {\mathbb Z} = {\mathbb Z}^{[1]}$ symmetry, and so decomposes into a family of universes parametrized by Rep$({\mathbb Z}) \cong U(1)$.  We discussed examples of this form in
sections~\ref{sect:gauge-1}, \ref{sect:ex:u1-r}, and observed there that physically, the elimination of maps of nonzero degree (sigma model instantons) is due to a multiverse interference effect, the fact that each universe carries a slightly different $\theta$ angle (here, $B$ field), so that contributions from maps of nonzero degree cancel out.

Since the ${\mathbb Z}$-gerbe has a global $B {\mathbb Z} = {\mathbb Z}^{[1]}$ symmetry (essentially, rotations of the fibers), and it decomposes into a family of universes parametrized by $U(1)$, we propose that the physical realization of a ${\mathbb Z}$ gerbe be in terms of a ${\mathbb R}$ gauge theory
(as opposed to the $p \rightarrow \infty$ limit of the Tanizaki-\"Unsal construction, for example).  

We can also give more concrete realizations of these gerbes in gauged linear sigma models (GLSMs),
following the same pattern as \cite{Pantev:2005zs}.
There, a prototypical ${\mathbb Z}_p$ gerbe on ${\mathbb P}^n$ was described by the supersymmetric ${\mathbb P}^n$ model in which all fields have\footnote{
The meaningfulness of nonminimal charges in two-dimensional $U(1)$ gauge theories was discussed in \cite{Pantev:2005rh,Pantev:2005wj,Pantev:2005zs}.  Briefly, although the theories are perturbatively identical, they are nonperturbatively distinct, as can be seen in theta angle periodicities on noncompact spacetimes, and from carefully defining fields on compact spacetimes. 
} charge $p$:
\begin{center}
    \begin{tabular}{cc}
    Gauge group & $\phi_{0,\cdots,n}$ \\ \hline
    $U(1)$ & $p$ 
    \end{tabular}
\end{center}
More general ${\mathbb Z}_p$ gerbes on projective spaces, of characteristic class $m \mod p$, were described as a $U(1)^2$ gauge theory with matter as in the table below:
\begin{center}
    \begin{tabular}{ccc}
    Gauge group & $\phi_{0,\cdots,n}$ & $y$ \\ \hline
    $U(1)$, $\lambda$ & $1$ & $m$ \\
    $U(1)$, $\nu$ & $0$ & $p$
    \end{tabular}
\end{center}
where $\lambda, \mu$ label the two $U(1)$'s, and to realize in a GLSM, we take $m < 0$.
In the special case $m=-1$, this describes the same geometry as the previous GLSM, as can be seen by working with a different basis\footnote{
Physical equivalence can be checked by e.g.~comparing symmetries.  For example, if in the second presentation we used $\nu$ instead of $\nu^{1/p}$, the second theory would have two global $B {\mathbb Z}_p$
symmetries, and so could not be in the same universality class of RG flow as the first, which only has one.
} for $U(1)$'s:
\begin{center}
    \begin{tabular}{ccc}
    Gauge group & $\phi_{0,\cdots,n}$ & $y$ \\ \hline
    $\lambda^p \nu$ & $p$ & $0$ \\
    $\nu^{1/p}$ & $0$ & $1$
    \end{tabular}
\end{center}
In this theory, D terms force $y \neq 0$, and it is removed by the $\nu^{1/p}$ rotation and RG flow, so that this is equivalent to the first prototypical GLSM for a ${\mathbb Z}_p$ gerbe
given by
\begin{center}
    \begin{tabular}{cc}
    Gauge group & $\phi_{0,\cdots,n}$ \\ \hline 
    $U(1)$ & $p$.
    \end{tabular}
\end{center}

We propose that
a similar story exists in the present case, for ${\mathbb Z}$ gerbes.
Specifically, first, 
we propose that a supersymmetric ${\mathbb R}$ gauge theory in two dimensions with
$n+1$ chiral superfields of charge $1$ (as in section~\ref{sect:ru1:mincharge}) is the physical realization of a sigma model with target a ${\mathbb Z}$-gerbe of characteristic class $-1 \in H^2({\mathbb P}^n,{\mathbb Z})$.
In other words:
\begin{center}
    \begin{tabular}{cc}
    Gauge group & $\phi_{0,\cdots,n}$ \\ \hline
    ${\mathbb R}$ & $1$
    \end{tabular}
\end{center}
Not only is this closely analogous to the GLSM realization prototypical ${\mathbb Z}_p$ gerbe discussed above, but it also has a $B {\mathbb Z} = {\mathbb Z}^{[1]}$ symmetry, as 
$2 \pi {\mathbb Z} \subset {\mathbb R}$ acts trivially.  This is as expected for a ${\mathbb Z}$ gerbe, as geometrically that symmetry describes `translations' along the fibers.
For reasons discussed earlier in section~\ref{sect:ru1:mincharge}, on a two-dimensional worldsheet, this GLSM decomposes into a family of universes
parametrized by $U(1)$, in which each universe describes an ordinary supersymmetric
${\mathbb P}^n$ model, with gauge group $U(1)$ instead of ${\mathbb R}$.

More generally, given a ${\mathbb Z}$-gerbe on ${\mathbb P}^n$ of characteristic class $m \in H^2({\mathbb P}^n, {\mathbb Z})$, for $m < 0$, we propose that its physical realization is as a $U(1) \times {\mathbb R}$ gauge theory with chiral
superfields $\phi_{0, \cdots, n}$, $y$, with charges
\begin{center}
    \begin{tabular}{ccc}
    Gauge group & $\phi_{0, \cdots, n}$ & $y$ \\ \hline
    $U(1)$, $\lambda$ & $1$ & $m$ \\
    ${\mathbb R}$, $\nu$ & $0$ & $1$,
    \end{tabular}
\end{center}
following the same pattern as GLSMs for general ${\mathbb Z}_p$ gerbes described
in \cite{Pantev:2005zs}.
Again, these theories have a $B {\mathbb Z} = {\mathbb Z}^{[1]}$ symmetry, as expected for a theory interpreted as a sigma model with target a ${\mathbb Z}$ gerbe, and for the same reasons as in section~\ref{sect:ru1:mincharge}, we expect this two-dimensional theory to decompose into a family of universes parametrized by $U(1)$, each of which is a GLSM in which the ${\mathbb R}$ gauge group factor is replaced by $U(1)$.

In the special case $m=-1$, we can perform a field redefinition to relate the GLSM above to the
supersymmetric ${\mathbb P}^n$ model in which the $U(1)$ is replaced by ${\mathbb R}$.
Let us take a moment to walk through those details.  
Given the group $U(1) \times {\mathbb R}$ acting as
\begin{equation}
    \phi \: \mapsto \: \alpha \exp(i b) \phi
\end{equation}
for $\alpha \in U(1)$, $b \in {\mathbb R}$
one can define an action of ${\mathbb R}$ as
\begin{equation}
    \phi \: \mapsto \: \exp\left( i b - i \ln \alpha \right) \phi,
\end{equation}
where $b - \ln(\alpha) \in {\mathbb R}$, and the action is well-defined because
$2 \pi {\mathbb Z}$ acts trivially.
In the GLSM with gauge group $U(1) \times {\mathbb R}$ and $m=-1$ above, we perform a field redefinition\footnote{
As before, physical equivalence can be checked by comparing global symmetries.
For example, if we did not convert the ${\mathbb R}$ symmetry labelled $\nu$ to the
${\mathbb R}/{\mathbb Z} \cong U(1)$ symmetry labelled $\overline{\nu}$, then the theory above would have two global $B {\mathbb Z}$ symmetries instead of one.
} in the path integral to
\begin{center}
    \begin{tabular}{ccc}
    Gauge group & $\phi_{0,\cdots,n}$ & $y$ \\ \hline 
    ${\mathbb R}$, $\lambda \nu$ & $1$ & $0$ \\
    $U(1)$, $\overline{\nu}$ & $0$ & $1$,
    \end{tabular}
\end{center}
in close analogy with the analogous ${\mathbb Z}_p$ gerbe above.
Just as in that ${\mathbb Z}_p$ gerbe example, D-terms require $y \neq 0$, so the effect
of gauging $\overline{\nu}$ is to remove $y$ from the theory in the IR, so that this GLSM is equivalent to the GLSM
\begin{center}
    \begin{tabular}{cc}
    Gauge group & $\phi_{0,\cdots,n}$ \\ \hline 
    ${\mathbb R}$ & $1$.
    \end{tabular}
\end{center}

\subsection{Analysis of (2,2) GLSMs with gauge group ${\mathbb R}$}
\label{sect:qc:r}

In this section we will outline basic properties of two-dimensional (2,2) supersymmetric GLSMs in which a $U(1)$ factor in the gauge group has been replaced by\footnote{
See for example \cite{deWit:1983xe,Inoue:2009np} for previous work on four-dimensional supersymmetric gauge theories
with noncompact gauge groups.
} ${\mathbb R}$.  Such a GLSM has a global $B {\mathbb Z}$ symmetry, and so decomposes into
Rep$({\mathbb Z}) \cong U(1)$ universes, as we shall check by comparing various computations.

Although we will make general statements, we will primarily consider three prototypical examples in turn in each subsection:
\begin{enumerate}
    \item The ordinary supersymmetric ${\mathbb P}^n$ model, with gauge group $U(1)$ and $n+1$ chiral multiplets of charge $1$.
    \item ${\mathbb Z}_p$ gerbes on ${\mathbb P}^n$.  The simplest versions will be described by a $U(1)$ gauge theory with $n+1$ chiral multiplets of charge $p$ \cite{Pantev:2005zs}, a close analogue of the
    supersymmetric ${\mathbb P}^n$ model.  We will also sometimes discuss more general cases which are quotients of line bundles over ${\mathbb P}^n$, in which the line bundle has been over-gauged away, via a $U(1)$ gauge factor under which it has charge $p$.
    \item ${\mathbb Z}$ gerbes on ${\mathbb P}^n$, for which the prototype will have gauge group ${\mathbb R}$ and $n+1$ chiral multiplets of charge $1$, in the sense of
    section~\ref{sect:ru1:mincharge}.
\end{enumerate}
Of course, it is the last case that is of primary interest to us; however, studying the other two cases in parallel will assist in understanding it.

We will outline computations of Coulomb branch relations and related quantities that, at least for a Deligne-Mumford stack, would correspond to quantum cohomology, mirrors, and quantum $K$-theory.  The reader should be cautioned, however, that we are considering ${\mathbb Z}$-gerbes, which are not Deligne-Mumford, so such geometric interpretations may be too quick.
In particular, we are not aware of any mathematics results on quantum cohomology or quantum $K$-theory rings of such gerbes \cite{tsengpriv,herrpriv}, so any mathematical interpretations of the Coulomb branch computations should be considered conjectural. 

\subsubsection{Supersymmetric topological gauging}

Earlier in section~\ref{sect:top-gauge-defn} we defined topological gauging in terms of a local action, but only at the level of bosonic fields. 
Before discussing specific computations, let us take a moment to explicitly demonstrate that topological gauging of $(-1)$-form symmetries can still be performed by a local action in two-dimensional (2,2) supersymmetric theories.

In the bosonic case, for a topologically-gauged $(-1)$-form symmetry in $D=2$ dimensions, we promote the theta angle to a scalar $\phi$ and add a dynamical one-form field $c$ with interaction term
\begin{equation}
    \int d \phi \wedge c
\end{equation}
(omitting the background field, for simplicity).  Here, in (2,2) supersymmetry,
introduce a pair of twisted chiral superfields $T$, $Y$,
\begin{eqnarray}
        Y & = & y - i \theta^+ \overline{\upsilon}_+ - i \overline{\theta}^- \upsilon_- + 
    \theta^+ \overline{\theta}^- (D - i F_{01}) + \cdots,
    \\
    T & = & t - i \theta^+ \overline{\chi}_+ - i \overline{\theta}^- \chi_- + 
    \theta^+ \overline{\theta}^- (r_x - i r_y) + \cdots,
\end{eqnarray}
in the notation of \cite{Witten:1993yc},
where $T$ encodes $\phi$ (as the real part of $t$, of vev $i r + \theta/2\pi$),
and $Y$ is the curvature superfield corresponding to the vector superfield containing $c$
(meaning, $F = d c$).
Then, add the interaction terms
\begin{equation}
    i \int d \theta^+ d \overline{\theta}^- \, TY \: + \: {\rm c.c.}.
\end{equation}
Expanded into components, this has bosonic terms
\begin{equation}
    r D - \frac{\phi}{2\pi} d c + {\rm Im}\, \left( y (r_x - i r_y) \right).
\end{equation}
(We omit fermionic terms for simplicity.)
Integrating out $Y$ sets $r_x = r_y = r = 0$ and gives the constraint $dc = 0$.
This results in the usual topological gauging.

Henceforward for simplicity we will not mention this structure, but we feel it important to mention its existence, to emphasize the fact that topological gauging can be accomplished with a local supersymmetric action, not just a local action, as will be pertinent for the analyses in the remainder of this section.

\subsubsection{Quantum cohomology}

We begin our discussion by describing correlation functions and quantum cohomology ring relations for the three prototypical examples discussed above.
We will also briefly discuss the axial R-symmetry and its nonanomalous subgroups, leaving a more detailed discussion for 
subsection~\ref{sect:mirrors}.
\begin{enumerate}
    \item First, we review the A-twist of the ordinary supersymmetric ${\mathbb P}^n$ model with gauge group $U(1)$.  The nonzero correlation functions are
    \begin{equation}
    \langle \sigma^{n(d+1) + d} \rangle \: = \: \langle \sigma^n \sigma^{d(n+1)} \rangle\: = \: q^d \langle \sigma^n \rangle \: = \: q^d,
\end{equation}
where $\sigma$ is an operator corresponding to a generator of $H^2({\mathbb P}^n, {\mathbb C})$,
and $d \geq 0$.  One can immediately read off the well-known OPE ring
\begin{equation} \label{eq:basicqh}
    \sigma^{n+1} \: = \: q.
\end{equation}
This can also be computed directly along the Coulomb branch, where it emerges as the critical locus equation of a (perturbative) twisted one-loop effective superpotential \cite{Morrison:1994fr}.

The axial R-symmetry $U(1)_A$ acts as
\begin{equation}
    \sigma \: \mapsto \: \sigma \exp(2 i \alpha).
\end{equation}
We see that it generically is not a symmetry of~(\ref{eq:basicqh}), as expected as it is anomalous,
but the nonanomalous subgroup ${\mathbb Z}_{2(n+1)} \subset U(1)_A$ does leave
(\ref{eq:basicqh}) invariant.
\item Next, consider the case of a ${\mathbb Z}_p$ gerbe on ${\mathbb P}^n$ \cite{Pantev:2005zs}.  The simplest version can be described as a $U(1)$ gauge theory with $n+1$ chiral superfields of charge $p$.
The nonzero correlation functions can be described as
\begin{equation}
    \langle \sigma^{n(pd+1) + pd} \rangle \: = \:
    \langle \sigma^n \sigma^{p d(n+1)} \rangle \: = \: q^{pd} \langle \sigma^n \rangle
    \: = \: q^{pd},
\end{equation}
where here we interpret $\sigma$ as the pullback to the gerbe from the generators of
$H^2({\mathbb P}^n,{\mathbb C})$.

One way to think about this, as described in \cite[section 3.1]{Pantev:2005zs}, is as the ${\mathbb P}^n$ model but with a restriction to instantons/vortices of degree divisible by $p$.
As a result, only correlation functions proportional to a positive power of $q^p$ appear.

We also get this structure from decomposition.  This theory is equivalent to a disjoint union of $p$ copies of the ordinary ${\mathbb P}^n$ model, but with the theta angle in each shifted by a $p$th root of unity, which is realized by multiplying $q$ by a $p$th root of unity.  This means
\begin{eqnarray}
    \langle \sigma^{n(d+1) + d} \rangle_{\rm gerbe} & = & \sum_{k=1}^{p} 
    \langle \sigma^{n(d+1) + d} \rangle_k \: = \: \sum_{k=1}^{p}
    \left( \Upsilon^k q \right)^d,
    \\
    & \propto & \left\{ \begin{array}{cl}
    0 & \mbox{$d$ not divisible by $p$} \\
    q^d & d = p m, \: \: \: m \in {\mathbb Z}_{\geq 0}.
    \end{array} \right.
\end{eqnarray}
In the expression above, we have used a subscript $k$ to denote the $k$th universe,
and $\Upsilon$ generates the $p$th roots of unity.  For simplicitly, we are also absorbing proportionality factors, to keep expressions as simple as possible.

From the correlation functions above, one can immediately see the OPE ring relation\footnote{
This is often written as merely $\sigma^{p(n+1)} = q$ instead of $q^{p}$.  We are using this form in order to clarify comparisons to related examples.
}
\begin{equation} \label{eq:qh:gerbe}
    \sigma^{p(n+1)} \: = \: q^{p},
\end{equation}
which can also independently be derived from the Coulomb branch of the GLSM
\cite[section 3.1]{Pantev:2005zs}.  Taking the $p$th roots, we recover
\begin{equation}
    \sigma^{n+1} \: = \: \Upsilon q,
\end{equation}
for $\Upsilon$ a $p$th root of unity.  These are the quantum cohomology relations for each of the universes, and of course the $p$th power is the quantum cohomology relation for the gerbe as a whole.

Next, we consider the axial R-symmetry $U(1)_A$.  It is anomalous, though a direct computation in
\cite{Pantev:2005zs} showed that it has a nonanomalous ${\mathbb Z}_{2p(n+1)}$ subgroup.
We also see that the axial R-symmetry is generically not a symmetry of~(\ref{eq:qh:gerbe}),
as expected.  Now, a ${\mathbb Z}_{2(n+1)} \subset U(1)_A$ does leave each universe invariant.
In addition, other elements of ${\mathbb Z}_{2p(n+1)}$ generate phases which can be
absorbed into $\Upsilon$ as $p$th roots of unity -- meaning that they interchange
universes.  As a result, the entire theory, the sum of the universes, has a
nonanomalous ${\mathbb Z}_{2p(n+1)} \subset U(1)_A$, though only ${\mathbb Z}_{2(n+1)}$ preserves each universe separately.  We will explore this further in the discussion
of mirrors in subsection~\ref{sect:mirrors}.

\item Finally, we turn to the main case of interest:  the analogue of the supersymmetric ${\mathbb P}^n$ model in which the gauge group $U(1)$ is replaced by ${\mathbb R}$.
In the A-twist of this theory, with gauge group ${\mathbb R}$, the only nonzero correlation function is
\begin{equation}
    \langle \sigma^n \rangle \: = \: 1,
\end{equation}
since the ${\mathbb R}$ gauge theory has no instantons/vortices.
This is also consistent with the decomposition into Rep$({\mathbb Z}) \cong U(1)$ universes,
as can be seen as follows.
If we use subscripts to denote whether the gauge group is ${\mathbb R}$ or $U(1)$, then formally
\begin{equation}
    \langle {\cal O}_1 \cdots {\cal O}_k \rangle_{ {\mathbb R} } \: = \:
    \int_0^{2 \pi} d \theta \,
    \langle {\cal O}_1 \cdots {\cal O}_k \rangle_{ U(1) }
\end{equation}
and so any ${\mathbb R}$-gauge theory correlation function proportional to any positive power of $q = \exp(-r + i \theta)$
will vanish, as
\begin{equation}
    \int_0^{2 \pi} d \theta \, \exp(i k \theta) \: = \: 0
\end{equation}
for $k > 0$.  Thus, decomposition predicts that the only correlation function that survives, is the classical correlation function, consistent with expectations for ${\mathbb R}$ gauge theories.

Now, let us compare the Coulomb branch relations arising in this GLSM.  
These were computed in \cite{Morrison:1994fr} just within perturbation theory.
The idea is that although the GLSM describes a geometry on its Higgs branch in the UV,
the theory in the IR flows onto a Coulomb branch, where the geometry has vanished,
and is replaced by a (perturbative) twisted one-loop effective superpotential.

Now, a crucial difference relative to the $U(1)$ gauge theory lies in the interpretation of $\theta$.  In the ${\mathbb R}$ gauge theory, there is no $\theta$ coupling on a boundary-free worldsheet.
Furthermore, 
implicit in our description of the ${\mathbb R}$ gauge theory as a topologically gauged $(-1)$-form symmetry is that
we have effectively integrated over values of $\theta$, so that in principle $\theta$ is not meaningful\footnote{
More precisely, $\theta$ does not appear to be meaningful in a two-dimensional ${\mathbb R}$ gauge theory with a global $B {\mathbb Z}$ symmetry, as we discuss in greater detail in appendix~\ref{app:r:theta}.  
} in the ${\mathbb R}$ gauge theory.  Ordinarily $\theta$ appears in the complexification of the Fayet-Iliopoulos (FI) parameter, an important part of the (2,2) supersymmetric Lagrangian construction.  The fact that $\theta$ does not appear in the supersymmetric ${\mathbb R}$ gauge theory suggests that one might want to describe the gauge field with a different supersymmetric multiplet than the twisted chirals used in \cite{Morrison:1994fr}, for example with multiplets in which the FI parameter is not complexified.

To describe the computation most cleanly, we will therefore perform it universe-by-universe,
using ordinary twisted chirals as in \cite{Morrison:1994fr}.
(As each universe is an ordinary supersymmetric $U(1)$ gauge theory, they have no subtleties of the sort just discussed.  We leave a direct computation in the ${\mathbb R}$ gauge theory for future work.)
Proceeding in this fashion, we 
write $\overline{q} = \Upsilon q$ for some fixed $q$, for $\Upsilon \in U(1)$ parametrizing the universes.
Then, in the universe corresponding to $\Upsilon$, we get the relation 
\begin{equation}
\sigma^{n+1} = \Upsilon q
\end{equation}
which is the ordinary quantum cohomology relation in that universe.  Furthermore, because we integrate over theta angles,
every value of $\Upsilon \in U(1)$ appears.

We turn next to the $U(1)_A$ symmetry.  We believe the $U(1)_A$ symmetry is nonanomalous, ultimately because there are no nontrivial ${\mathbb R}$ bundles, hence no nonperturbative origin for an anomaly.  However, as this will be important for our analysis, and the point is obscure, we shall analyze the $U(1)_A$ more carefully.

To be more thorough, we consider a direct computation of the divergence of the axial R-current.
    This is a perturbative, one-loop computation, and so it is independent of whether the gauge group is $U(1)$ or ${\mathbb R}$. It has the form
     \begin{equation}  \label{eq:chiralanom}
        \partial_{\mu} J^{\mu}_A \: \propto \: F,
    \end{equation}
    where $F$ is the curvature of the ${\mathbb R}$ gauge field. 
    Although the cohomology class $[F] = 0$, as there are no nontrivial ${\mathbb R}$ bundles, 
    $F$ itself need not vanish.
    That said, we can redefine the current slightly to clarify.
    Write $F = dA$ for a globally-defined one-form $A$, and write the chiral anomaly equation~(\ref{eq:chiralanom}) as
    $d * J_A = \alpha d A$ for some constant $\alpha$.  Then, if we define
    \begin{equation} \label{eq:tilde-j-defn}
        \tilde{J}_A \: = \: J_A - \alpha * A,
    \end{equation}
    we have $d^* \tilde{J}_A = 0$, and so $\tilde{J}_A$ is anomaly-free.
    (We expect a similar story for the two-dimensional Konishi anomaly (see e.g.~\cite[equ'n (4.12)]{Hori:2001ax}),
the supersymmetrization of the chiral anomaly.)

To be clear, this argument would not work in a $U(1)$ gauge theory.  The reason typically cited is lack of gauge invariance, but we believe at a more fundamental level, the problem is one of global well-definedness.  On a topologically nontrivial spacetime, in a $U(1)$ gauge theory, $F$ need not be cohomologically trivial, so there is no globally-defined $A$ in general for which $F = dA$.  On a coordinate patch, one can pick $A$ so that $F = dA$, and then define $\tilde{J}_A$ as in~(\ref{eq:tilde-j-defn}), but because $J_A$ is gauge-invariant, the resulting $\tilde{J}_A$ would only be defined on that one patch.  Phrased another way, $J_A$ is a globally-defined one-form but $A$ is not in a $U(1)$ gauge theory, so only if the linear combination $\tilde{J}_A$ were gauge-invariant could one hope to construct a globally-defined one-form.

By contrast, in a ${\mathbb R}$ gauge theory, there exists a globally-defined one-form $A$ such that $F = dA$ everywhere.
Both the $J_A$ and $*A$ terms are globally-defined one-forms, hence the linear combination $\tilde{J}_A$ is also globally-defined.

Now, in the ${\mathbb R}$ gauge theory case, although $\tilde{J}_A$ is globally-defined, it is not gauge-invariant.  However, the operator we require is
\begin{equation}
    \exp\left( i \int \tilde{J}_A \right),
\end{equation}
and the integral of $\tilde{J}_A$ over a closed loop is gauge-invariant under global gauge transformations.
This also is an improvement over a $U(1)$ gauge theory:  there, even for a globally-defined $A$, its integral over a closed loop is only invariant up to a lattice translation under global gauge transformations, whereas in a ${\mathbb R}$ gauge theory, the integral over $A$ over a closed loop is completely invariant under global gauge transformations.  (For local gauge transformations, the integrals are invariant in both cases.)

As a result, because we are working in a ${\mathbb R}$ gauge theory instead of a $U(1)$ gauge theory, we believe we can remedy $J^{\mu}_A$ to create a globally-defined divergence-free current, and corresponding topological operator, hence the axial R-symmetry is anomaly-free.

Now that we have established that the $U(1)_A$ symmetry is anomaly-free, let us interpret it in the quantum cohomology ring.
For reasons already discussed,
${\mathbb Z}_{2(n+1)} \subset U(1)_A$ leaves the quantum cohomology ring of each separate universe invariant, and when we allow for the possibility of the R-symmetry exchanging universes as in ${\mathbb Z}_p$ gerbes, then since $\Upsilon$ is an arbitrary element of $U(1)$ which can absorb any phase, the $U(1)_A$ is unconstrained.  All elements of
$U(1)_A$ are allowed: those elements which do not preserve universes, instead map them into one another.
This correctly reflects the fact that in the ${\mathbb R}$ gauge theory, the entire axial R-symmetry $U(1)_A$ is nonanomalous.

We will also recover a closely-related result from the $p \rightarrow \infty$ limit
 of a Tanizaki-\"Unsal construction
later in section~\ref{sect:tu:susy}, which serves as another consistency check.

For completeness we mention one other possible subtlety, arising from the normalization by the volume of the gauge group.  Ordinarily in a gauge theory, at least at a formal level, one normalizes by the volume of the gauge group (see e.g.~\cite[chapter 15.5]{Weinberg:1996kr}, \cite[section 2.2]{Witten:1991we}).  In the present case, since the volume of ${\mathbb R}$ is infinite, it may be that {\it all} of the classical correlation functions vanish, for example
\begin{equation}
    \langle \sigma^n \rangle \: = \: \frac{1}{{\rm Vol}({\mathbb R})} \: = \: 0.
\end{equation}
In this case, if all classical correlation functions vanish, then the OPE ring is completely undetermined.
That said, we have not seen such a volume factor arise when topologically gauging the $(-1)$-form symmetry, so we are not inclined to believe that this is the entire story.

\end{enumerate}

So far we have discussed projective spaces, but similar ideas hold more generally.
Consider for example an abelian GLSM describing a Fano toric variety.
(For the moment, we assume only $U(1)$ factors in the gauge group.)
For such theories, the quantum cohomology ring was computed 
 in \cite{Morrison:1994fr}.  
 From the critical locus of the one-loop effective twisted superpotential along the Coulomb branch, they found the relations \cite[equ'n (3.45)]{Morrison:1994fr}
\begin{equation} \label{eq:qc:toric}
    \prod_i \left( \sum_b Q_i^b \sigma_b \right)^{Q_i^a} \: = \: q_a,
\end{equation}
where the $a$ index factors of $U(1)$ or ${\mathbb R}$ in the gauge group, the $i$'s index chiral superfields, and $Q_i^a$ is the charge of the $i$th chiral superfield under the $a$th gauge group factor.  These are relations amongst the $\sigma_a$, which parametrize the Coulomb branch, and in the dictionary to quantum cohomology, correspond to the generators of $H^2(X,{\mathbb R})$.

Now, suppose the gauge group has ${\mathbb R}$ factors, but all matter has integral charges,
so that the GLSM has a global $B {\mathbb Z}$ symmetry.
As discussed previously, for any ${\mathbb R}$ factor in the gauge group, there will be a decomposition
into universes indexed by $U(1)$.  If such a gauge factor has index $a$,
the quantum cohomology ring for any given universe will be of the same form as~(\ref{eq:qc:toric}), albeit replacing $q_a$ with
$\Upsilon_a q_a$, as in our discussion of projective spaces, for $\Upsilon_a \in U(1)$, corresponding to the choice of universe.

\subsubsection{Mirrors}  \label{sect:mirrors}

We can apply similar ideas to understand mirrors \cite{Morrison:1995yh,Hori:2000kt} of two-dimensional supersymmetric abelian\footnote{
See for example \cite{Gu:2018fpm,Chen:2018wep,Gu:2019zkw,Gu:2020ivl} for mirrors to nonabelian GLSMs in two dimensions.
} GLSMs with ${\mathbb R}$ factors in the gauge group.  Briefly, much of the mirror computation of \cite{Hori:2000kt} takes place at the level of free chiral superfields, before gauging, so whether the gauge group has a $U(1)$ or an ${\mathbb R}$ is irrelevant for much of the analysis. For example, the periodicity of the twisted
chiral $Y$ arises because it is dual to an angular variable.

As before, we will walk through three prototypical examples systematically.

\begin{enumerate}
    \item First, the ordinary supersymmetric ${\mathbb P}^n$ model, with gauge group $U(1)$ and $n+1$ chiral superfields of charge $1$.  The Toda dual / mirror to this theory is well-known
    (see e.g.~\cite{Hori:2000kt} and references therein) to be a Landau-Ginzburg model with superpotential
    \begin{equation}  \label{eq:toda}
    W \: = \: \exp(-Y_1) \: + \: \cdots \: + \: \exp(-Y_n) \: + \: q \exp\left( + Y_1 + Y_2 + \cdots + Y_n \right),
\end{equation}
We will see variations of this in our next several examples.

The axial R-symmetry $U(1)_A$ acts on the mirror as
(see e.g.~\cite[equ'n (3.30)]{Hori:2000kt}, \cite[section 2]{Gu:2018fpm})
\begin{equation}
    Y_i \: \mapsto \: Y_i \: - \: 2 i \alpha,
\end{equation}
so under the R-symmetry, 
\begin{eqnarray}
    \exp(-Y_i) & \mapsto & \exp(2 i \alpha) \exp(-Y_i),
    \\
    \exp(+Y_1 + \cdots + Y_n) & \mapsto & \exp(-2 i n \alpha) \exp(+ Y_1 + \cdots + Y_n).
\end{eqnarray}
For generic values of $\alpha$, the two terms in the superpotential transform differently, and so we see that (most of the) $U(1)_A$ is not a symmetry perturbatively in the mirror, consistent
with the fact that in the original A-twisted theory, it is anomalous.  However, there is an anomaly-free subgroup.  To see it,
we require that each of the terms transform in the same way under the axial R-symmetry, hence we require
    \begin{equation}
        \exp(2 i \alpha) \: = \: \exp(- 2 i n \alpha )
    \end{equation}
    which implies
    \begin{equation}
        \exp(2 i (n+1) \alpha) \: = \: 1,
    \end{equation}
    recovering the fact that ${\mathbb Z}_{2(n+1)} \subset U(1)_A$ is the anomaly-free subgroup.

\item Next, we consider a ${\mathbb Z}_p$ gerbe on ${\mathbb P}^n$, described as a $U(1)$ gauge theory with $n+1$ chiral multiplets of charge $p$.  As discussed in \cite[section 6.1.1]{Pantev:2005zs}, 
the mirror is a Landau-Ginzburg model with superpotential
   \begin{equation}  \label{eq:toda:gerbe}
    W \: = \: \exp(-Y_1) \: + \: \cdots \: + \: \exp(-Y_n) \: + \: q \Upsilon \exp\left( + Y_1 + Y_2 + \cdots + Y_n \right),
\end{equation}
where $\Upsilon$ is a $p$th root of unity, whose values are summed over in the path integral, and $m \mod p$ the characteristic class of the gerbe.
We interpret each value of $\Upsilon$ as defining a different universe, and for fixed $\Upsilon$, the theory above is clearly a copy of the ordinary Toda dual to
${\mathbb P}^n$, albeit with the phase of $q$ shifted.

The action of the axial R-symmetry is the same as for the ordinary ${\mathbb P}^n$ model above, and as there, for generic $\alpha$, it is broken, as expected.

In this case, if we demand that each term in the superpotential transform the same way, we find a ${\mathbb Z}_{2(n+1)}$ symmetry as above.  
However, one can compute the anomaly-free 
subgroup directly \cite[section 3.1]{Pantev:2005zs}, to find that it is 
${\mathbb Z}_{2p(n+1)}$, larger than ${\mathbb Z}_{2(n+1)}$.
We can see this larger subgroup in the mirror by allowing
the action to rotate universes into one another, providing an analogue of the Witten effect for dyons \cite{Witten:1979ey} (see also \cite[section 4.1.2]{Pantev:2023dim} for a discussion of the Witten effect in the decomposition of two-dimensional pure $U(1)$ Maxwell theory).  Explicitly, this means we only need to require
    \begin{equation}
        \exp(2 i \alpha) \: = \: \Upsilon \exp( - 2 i n \alpha)
    \end{equation}
    for $\Upsilon$ some $p$th root of unity.  Then, we have
    \begin{equation}
        \exp(2 i (n+1) \alpha) \: = \: \Upsilon,
    \end{equation}
    hence 
    \begin{equation}
        \exp(2 i p (n+1) \alpha) \: = \: 1,
    \end{equation}
    so that $\exp(i \alpha) \in {\mathbb Z}_{2 p (n+1)}$.
    This matches the direct computation of the anomaly-free subgroup of the axial R-symmetry in the original A-twisted ${\mathbb Z}_p$ gerbe case  \cite{Pantev:2005zs}.

\item Finally, we turn to the case of interest, the analogue of the supersymmetric ${\mathbb P}^n$ model with gauge group ${\mathbb R}$ instead of $U(1)$.  
Given that this theory decomposes into a family of universes, each a copy of the ordinary
${\mathbb P}^n$ model, parametrized by $U(1)$, we can describe\footnote{
We leave a first-principles derivation of the mirror, following some version of the analysis 
in \cite{Hori:2000kt}, directly from the ${\mathbb R}$ gauge theory, without appealing to decomposition, for future work.
} the mirror as a family of
Landau-Ginzburg models with superpotential
\begin{equation}
    W \: = \: \exp(-Y_1) \: + \: \cdots \: + \: \exp(-Y_n) \: + \: q \Upsilon\exp(+Y_1 + \cdots + Y_n),
\end{equation}
where $\Upsilon \in U(1)$, whose values index the universes.

Following the same logic as in the ${\mathbb Z}_p$ gerbe case, the anomaly-free subgroup of $U(1)_A$ should be given by $\exp(i \alpha)$ such that
\begin{equation}
        \exp(2 i (n+1) \alpha) \: = \: \Upsilon,
    \end{equation}
    for $\Upsilon$ an arbitrary element of $U(1)$.  This is of course no constraint at all, consistent with the fact that all of $U(1)_A$ is nonanomalous in this case.
    A ${\mathbb Z}_{2(n+1)} \subset U(1)_A$ subgroup will preserve each universe; the rest will rotate the universes into one another, just as in the ${\mathbb Z}_p$ gerbe case.

\end{enumerate}

\subsubsection{Quantum $K$-theory}

For completeness, we mention that quantum $K$-theory computations follow the same format.
In principle this is computed by GLSMs on three-dimensional spaces of the form
$S^1 \times \Sigma$, Kaluza-Klein reduced to $\Sigma$, where they can be computed in the same form as Coulomb branch quantum cohomology computations, albeit with a modified twisted superpotential (that takes into account the tower of KK modes) \cite{Bullimore:2014awa,Jockers:2018sfl,Jockers:2019wjh,Jockers:2019lwe,Jockers:2021omw,Ueda:2019qhg,Koroteev:2017nab,Closset:2016arn,Closset:2017zgf,Closset:2018ghr,Closset:2019hyt,Closset:2023vos,Closset:2023bdr,Gu:2020zpg,Gu:2023tcv,Gu:2022yvj,Gu:2023fpw,Huq-Kuruvilla:2025nlf,epiga1}.  Since the Coulomb branch computations are perturbative, the form of the results is essentially independent of whether the gauge group is $U(1)$ or ${\mathbb R}$.
We give here some simple examples, following the same format as in previous subsections.

\begin{enumerate}
    \item First, consider the quantum $K$-theory ring of (a cover of) an ordinary projective space
${\mathbb P}^n$, described by a gauge theory with gauge group ${\mathbb R}$ (instead of $U(1)$) and $n+1$ chiral superfields of charge $1$.  Following e.g.~\cite[section 2.2]{Gu:2020zpg}, and choosing Chern-Simons levels to match ordinary quantum $K$-theory, but with gauge group ${\mathbb R}$, and since the perturbative physics is insensitive to the global form of the gauge group, we have the same twisted one-loop effective superpotential as for the $U(1)$ case, namely
\begin{equation}
    W \: = \: (\ln q) (\ln X) \: + \: (n+1) {\rm Li}_2(X),
\end{equation}
where $X \sim \exp(\sigma)$, from which we derive the relation
\begin{equation}
    (1-X)^{n+1} \: = \: q,
\end{equation}
for a $q$ whose values are restricted, as previously.

\item For our next example, we consider a ${\mathbb Z}_p$ gerbe over ${\mathbb P}^n$,
defined by a $U(1)$ gauge theory with $n+1$ chiral superfields of charge $p$ rather than charge $1$, so that there is a $B {\mathbb Z}_p$ symmetry in the three-dimensional theory,
and a ${\mathbb Z}_p \times B {\mathbb Z}_p$ symmetry in the $S^1$-reduced two-dimensional theory.  Following the same analysis as above, we get the
relation
\begin{equation}
    \left( 1 - X^p \right)^{p(n+1)} \: = \: q^p.
\end{equation}
This was discussed in \cite{Gu:2021yek,Gu:2021beo,Sharpe:2024ujm}, and is interpreted as a doubled or squared decomposition, as discussed in those references.  More invariantly, this should be interpreted as $| {\mathbb Z}_p | | {\rm Rep}({\mathbb Z}_p)| = p^2$ copies of the quantum $K$-theory ring of the underlying space, the first factor from a spontaneously-broken ${\mathbb Z}_p$ symmetry which becomes equivalent to a decomposition deep in the IR, the second factor from the decomposition due to the $B {\mathbb Z}_p$ symmetry in the two-dimensional theory.
\item Finally, consider a ${\mathbb R}$ gauge theory with $n+1$ chiral superfields of charge $1$, so that there is a $B {\mathbb Z}$ symmetry in the three-dimensional theory, and a ${\mathbb Z} \times B {\mathbb Z}$ symmetry in the $S^1$-reduced two-dimensional theory.  Following the same
analysis as above and as in \cite[section 4.1.2]{Sharpe:2024ujm}, since the Coulomb branch analysis is perturbative and so insensitive to whether the gauge group is $U(1)$ or ${\mathbb R}$, we get the
relation
\begin{equation} \label{eq:qk:base}
    \left( 1 - X \right)^{n+1} \: = \: \Upsilon q,
\end{equation}
for $\Upsilon \in U(1)$.  The $\Upsilon$ reflects the Rep$({\mathbb Z}) \cong U(1)$ arising from the decomposition due to the presence of a $B {\mathbb Z}$ symmetry in the two-dimensional theory.  

The role of the ${\mathbb Z}$ zero-form symmetry is more subtle.
First, recall that in a $U(1)$ gauge theory compactified on $S^1$,
the component of the $\sigma$ field arising from the dual of the three-dimensional gauge field is periodic, ultimately due to charge quantization,
see e.g.~\cite[section 2.2]{Aharony:1997bx}, \cite[section 2.2.1]{Closset:2016arn}.
In the conventions of \cite[section 2.1]{Gu:2020zpg}, $\sigma \sim \sigma + k/R$ for $k \in {\mathbb Z}$ and $R$ the radius of the $S^1$.  
As a result, when solving the Bethe ansatz equations for 
$X = \exp(2 \pi i R \sigma)$, the logarithmic branch cuts do not contribute a tower of extra solutions.

In the case of the ${\mathbb R}$ gauge theory, however, because there is no charge quantization, the scalar corresponding to the dual of the gauge field has no periodicity.
Thus, there is a ${\mathbb Z}$ action on solutions of~(\ref{eq:qk:base}), given by
$\sigma \mapsto \sigma + k/R$ for $k \in {\mathbb Z}$.  This ${\mathbb Z}$ action leaves $X$ invariant, and so is a symmetry of the theory, but unlike the case of a $U(1)$ gauge theory, it generates a ${\mathbb Z}$-fold multiplicity in the solutions, corresponding to different logarithmic branch cuts in solutions for $\sigma$.

Altogether, we see that there is a ${\mathbb Z} \times U(1)$ multiplicity in the solutions of~(\ref{eq:qk:base}), or in other words that there are as many copies of the quantum $K$-theory ring relation for ${\mathbb P}^n$ as elements of ${\mathbb Z} \times U(1)$.
This reflects the presence of both a decomposition ($U(1)$) as well as a spontaneously-broken ${\mathbb Z}$ zero-form symmetry, the latter of which is equivalent to a decomposition in a topologically-twisted theory.  In particular, 
because of 
the ${\mathbb Z} \times B {\mathbb Z}$ global symmetry,
we have an analogue of 
the 
squared decomposition discussed for finite group cases in \cite{Gu:2021yek,Gu:2021beo,Sharpe:2024ujm}.
\end{enumerate}

\subsubsection{Supersymmetric localization}

We can also see the decomposition in the exact partition functions given by supersymmetric
localization \cite{Benini:2012ui,Doroud:2012xw}.  Instead of considering three variations on gerbes on projective spaces, we will make some simple general observations.

Briefly, from \cite[equ'n (3.34)]{Benini:2012ui}, the exact partition function of
a two-dimensional (2,2) supersymmetric theory on $S^2$ is 
\begin{equation}   \label{eq:part-fn}
Z_{S^2} \: = \: \frac{1}{|{\cal W}|} \sum_{\mathfrak m} 
\int \left( \prod_j \frac{d \sigma_j}{2\pi} \right) \,
Z_{\rm class}(
\sigma, {\mathfrak m}) \,
Z_{\rm gauge}(\sigma, {\mathfrak m}) \,
\prod_{\Phi} Z_{\Phi}(\sigma, {\mathfrak m}; \tau, {\mathfrak n}) ,
\end{equation}
where \cite[equ'n (3.35)]{Benini:2012ui}
\begin{eqnarray*}
Z_{\rm class}(\sigma,{\mathfrak m}) 
& = & e^{-4 \pi i \xi {\rm Tr}\, \sigma - i \theta
{\rm Tr} \, {\mathfrak m}} \exp\left( 8 \pi i r
{\rm Re}\, \tilde{W}(\sigma/r + i {\mathfrak m}/(2r)) \right), \\
Z_{\rm gauge}(\sigma,{\mathfrak m}) & = & 
\prod_{\alpha \in G} \left( \frac{ |\alpha( 
{\mathfrak m})| }{2} \: + \: i \alpha(\sigma) \right)
\: = \:
\prod_{\alpha > 0} \left( \frac{ \alpha({\mathfrak m})^2}{4} \: + \:
\alpha(\sigma)^2 \right), \\
Z_{\Phi}(\sigma, {\mathfrak m}; \tau, {\mathfrak n}) & = & 
\prod_{\rho \in R_{\Phi}} \frac{
\Gamma\left( \frac{R[\Phi]}{2} \: - \: i \rho(\sigma) \: - \:
i f^a[\Phi] \tau_a \: - \: \frac{
\rho({\mathfrak m}) + f^a[\Phi] n_a }{2} \right)
}{
\Gamma\left( 1 \: - \: \frac{R[\Phi]}{2} \: + \: i \rho(\sigma)
\: + \: i f^a[\Phi] \tau_a \: - \: \frac{
\rho({\mathfrak m}) + f^a[\Phi]n_a }{2} \right)
} ,
\end{eqnarray*}
and we have used the notation of 
\cite{Benini:2012ui}.
From that reference, for a chiral multiplet $\Phi$, $f^a[\Phi]$ encodes its non-R-charges, $R[\Phi]$ its R-charge, and $R_{\Phi}$
denotes the corresponding representation of the gauge group, which has Weyl group ${\cal W}$.
The 
$\tau = (\tau_a)$ and ${\mathfrak n}=(n_a)$ encode twisted masses.  The ${\mathfrak m}$ are elements of the dual weight lattice of the gauge group, living in the Lie algebra of a Cartan torus, and encode possible Chern roots of the gauge bundle
(see \cite[section 2.6]{Sharpe:2014tca} for further details).

The reference \cite[section 2.6]{Sharpe:2014tca} used these exact expressions to compare two-dimensional supersymmetric $G$ gauge theories with center-invariant matter to $G/Z$ gauge theories (for $Z$ a subgroup of the center) with the same matter, and recovered the standard prediction of decomposition 
\begin{equation}
    G \: = \: \coprod_{\lambda} (G/Z)_{\lambda}
\end{equation}
for partition functions, namely that the partition function of the $G$ theory is the sum of the partition functions for the $G/Z$ gauge theories with discrete theta angles $\lambda$.

For our purposes in this paper, we turn to supersymmetric $U(1)$ and ${\mathbb R}$ gauge theories in two dimensions.  
\begin{itemize}
    \item In a $U(1)$ gauge theory, the theta angle $\theta$ appears only in $Z_{\rm class}$, in the overall multiplicative factor,
the Weyl group ${\cal W}$ is of course trivial, and the ${\mathfrak m}$ lie on a one-dimensional lattice, giving possible values of $c_1$ of a bundle on $S^2$.
\item In a ${\mathbb R}$ gauge theory, by contrast, ${\mathfrak m} = 0$ (as there are no nontrivial bundles), and also there is no $\theta$ coupling, but otherwise the form should be unchanged.
\end{itemize}
If we let $Z_{U(1)}(\theta)$ denote the partition function of a supersymmetric $U(1)$ gauge theory,
and $Z_{\mathbb R}$ the partition function of a ${\mathbb R}$ gauge theory with the same matter,
then we can read off immediately from~(\ref{eq:part-fn}) that 
\begin{equation}
    Z_{\mathbb R} \: \propto \: \int_0^{2\pi} d \theta \, Z_{U(1)}(\theta),
\end{equation}
exactly as expected.  The integral over $\theta$ removes all contributions from ${\mathfrak m} \neq 0$.

\subsubsection{Analysis of symmetries of the various ${\mathbb P}^{n}$ models}

In this section, we will summarize the symmetries of the ${\mathbb P}^{n}$ models with gauge groups $U(1)$ and ${\mathbb R}$.

\begin{enumerate}
    \item We begin with the ordinary ${\mathbb P}^{n}$ model with gauge group $U(1)$.

    In the UV, this has (zero-form) global symmetries
    \begin{equation}
        {\mathbb Z}_{2(n+1)}^{[0]} \times U(1)_V^{[0]} \times PSU(n+1)^{[0]},
    \end{equation}
    where the ${\mathbb Z}_{2(n+1)}$ is the nonanomalous subgroup of the axial R-symmetry $U(1)_A$.

    Deep in the IR, as discussed in \cite{Komargodski:2020mxz,Gu:2025gtb},
    the global symmetry becomes 
    \begin{equation}
        {\mathbb Z}_{n+1}^{[0]} \times{\mathbb Z}_{n+1}^{[1]}.
    \end{equation}
    The $U(1)_V \times PSU(n+1)$ are broken along the RG flow,
    and the ${\mathbb Z}_{n+1}^{[1]}$ is emergent in the IR, reflecting the fact that there are $n$ vacua (in effect, a spontaneous symmetry breaking has become a decomposition, deep in the IR).

    \item Next, we consider the GLSM for ${\mathbb Z}_p$ gerbes on ${\mathbb P}^{n}$, given by $U(1)$ GLSMs with $n+1$ chiral multiplets of charge $p$.
    Here, in the UV, the global symmetry is
    \begin{equation}
        {\mathbb Z}_{2 p (n+1)}^{[0]} \times U(1)_V^{[0]} \times PSU(n+1)^{[0]} \times {\mathbb Z}_p^{[1]}.
    \end{equation}
    The ${\mathbb Z}_{2 p (n+1)}^{[0]}$ is the nonanomalous subgroup of the axial
    R-symmetry $U(1)_A$, and because a ${\mathbb Z}_p$ subgroup of the gauge group acts trivially, there is a $B {\mathbb Z}_p = {\mathbb Z}_p^{[1]}$ symmetry.

    Deep in the IR, following in spirit the same ideas as \cite{Gu:2025gtb},
    the global symmetry becomes
    \begin{equation}
        {\mathbb Z}_{n+1}^{[0]} \times {\mathbb Z}_{p (n+1)}^{[1]},
    \end{equation}
    where the ${\mathbb Z}_{p (n+1)}^{[1]}$ arises as a combination of the UV ${\mathbb Z}_p^{[1]}$ and an emergent ${\mathbb Z}_{n+1}^{[1]}$ in the IR, 
    \begin{equation}
        1 \: \longrightarrow \: {\mathbb Z}_p^{[1]} \: \longrightarrow \:
        {\mathbb Z}_{p (n+1)}^{[1]} \: \longrightarrow \:
        {\mathbb Z}_{n+1}^{[1]} \: \longrightarrow \: 1,
    \end{equation}
    reflecting the fact that, deep in the IR, there are $p(n+1)$ universes.

    \item Finally, we turn to the analogue of the ${\mathbb P}^{n}$ model with gauge
    group ${\mathbb R}$.  Here, in the UV, the global symmetry is
    \begin{equation}
        {\mathbb Z}_{2(n+1)}^{[0]} \times U(1)_V^{[0]} \times PSU(n+1)^{[0]} \times {\mathbb Z}^{[1]},
    \end{equation}
    where the ${\mathbb Z}_{2(n+1)}^{[0]} \subset U(1)_A$ preserves separate universes. 
    The axial $U(1)_A$ is nonanomalous, but, $U(1)_A/{\mathbb Z}_{2(n+1)}$ is absorbed by the topologically-gauged $(-1)$-form symmetry, as acting by the (pure anomalous) axial symmetry rotates the theta angle (and moves between universes).
    As a result, although the $U(1)_A$ is completely nonanomalous, we only list
    the ${\mathbb Z}_{2(n+1)}$ subgroup that preserves separate universes.
    The ${\mathbb Z}^{[1]}$ is present because the gauge group ${\mathbb R}$ has a trivially-acting ${\mathbb Z} \subset {\mathbb R}$ subgroup.

    Deep in the IR, the ${\mathbb Z}_{2(n+1)}^{[0]}$ is replaced by a ${\mathbb Z}_{n+1}^{[0]} \times {\mathbb Z}_{n+1}^{[1]}$, just as in \cite{Gu:2025gtb},
    but the emergent ${\mathbb Z}_{n+1}^{[1]}$ combines with the UV ${\mathbb Z}^{[1]}$,
    essentially because the $n$ vacua of the IR of each UV universe rotate into one another under $\theta$ angle rotations.  In other words, the IR one-form symmetry is ${\mathbb Z}_{\rm IR}^{[1]}$, given as the extension
    \begin{equation}
        1 \: \longrightarrow \: {\mathbb Z}_{\rm UV}^{[1]} \: \longrightarrow \: {\mathbb Z}_{\rm IR}^{[1]} \: \longrightarrow \:
        {\mathbb Z}_{n+1}^{[1]} \: \longrightarrow \: 1.
    \end{equation}
    Altogether, we find that the global symmetry deep in the IR of the ${\mathbb R}$ gauge theory is
    \begin{equation}
        {\mathbb Z}_{n+1}^{[0]} \times{\mathbb Z}_{\rm IR}^{[1]}.
    \end{equation}
\end{enumerate}

\section{Gauging ${\mathbb Q}/{\mathbb Z}$ $(-1)$-form symmetries and $S^1_{\mathbb Q}$ gauge theories}
\label{sect:tu}

So far we have considered instanton restriction by topologically gauging continuous $U(1)^{[D-3]}$-form symmetries in $D$ dimensions, which results in a gauge theory with a noncompact
gauge group.  In the case $D=2$, this 
results in a theory that decomposes into a continuous family of universes.

Next, we consider the trick used by Tanizaki-\"Unsal in \cite{Tanizaki:2019rbk} to build a manifestly local theory in which instantons are restricted to degree divisible by a fixed integer $p$.  It can be thought of as gauging a 
${\mathbb Z}_p^{[-1]}$ symmetry, in any spacetime dimension $D$.  We shall argue that, formally, in the limit $p \rightarrow \infty$, one has a decomposition into universes indexed
by ${\mathbb Q}/{\mathbb Z}$ (as opposed to universes indexed by ${\mathbb R}/{\mathbb Z} \cong U(1)$ considered earlier).
(See also \cite{Putrov:2022pua} for general remarks on ${\mathbb Q}/{\mathbb Z}$ as a symmetry, and e.g.~\cite{Choi:2022jqy,Cordova:2022ieu} for other recent occurrences of ${\mathbb Q}/{\mathbb Z}$ symmetries in physics.  The $p \rightarrow \infty$ limit was also considered in \cite[section 3.1]{Lin:2025oml}, where it was identified with a $U(1)^{[-1]}$ topological gauging, instead of the ${\mathbb Q}/{\mathbb Z}$ gauging we will argue here.)

We should also mention that a sigma-model-version of the Tanizaki-\"Unsal construction
\cite{Tanizaki:2019rbk} was discussed in \cite[appendix D.1]{Pantev:2023dim}.
Briefly, it added to a two-dimensional sigma model, a periodic scalar $\widetilde{\varphi}$,
a $U(1)$ gauge field $A$, and interaction terms
\begin{equation}
    \int_{\Sigma} \widetilde{\varphi} \left( \phi^* \omega - p F \right),
\end{equation}
where $\omega$ is the K\"ahler form on the target space and $F$ the curvature of the $U(1)$ gauge field.  For finite $p$, this restricts to maps whose degree\footnote{As determined by the cohomology class of the fixed K\"ahler form $\omega$.} is divisible by $p$.

In passing, note that because the Tanizaki-\"Unsal construction is equivalent to
topologically gauging a $(-1)$-form symmetry, the corresponding quantum symmetry of the gauged theory
is a $(D-1)$-form symmetry, hence the gauged theory decomposes.

\subsection{Review and limits}

First, let us review the construction of \cite{Tanizaki:2019rbk}, in even dimensions
$D = 2k$.  Consider for simplicity a Yang-Mills theory in dimension $D$, of action
$S_{\rm YM}$.  We add to the theory two new fields:
\begin{itemize}
    \item a scalar field $B$ of periodicity $2 \pi$,
    \item a $(D-1)$-form gauge potential $C^{(D-1)}$,
\end{itemize}
and consider the theory with action
\begin{equation}  \label{eq:tu-action}
    S \: = \: S_{\rm YM} \: + \: i \int d^D x \, B \left( 
    \frac{1}{8 \pi^2}{\rm Tr} F^k \: - \: \frac{p}{2\pi} H^{(D)} \right),
\end{equation}
where $H^{(D)}$ is the curvature of $C^{(D-1)}$, so that locally, $H^{(D)} = d C^{(D-1)}$.
For simplicity, we assume conventions so that
\begin{equation}
    \frac{1}{8 \pi^2}  {\rm Tr} F^k \: \in \: {\mathbb Z}.
\end{equation}
Note that the $C^{(D-1)}$ gauge field implies the presence of a global $(D-1)$-form action,
which from general yoga implies a decomposition, as we shall see explicitly.

The equations of motion for $B$ are
\begin{equation}
 \frac{1}{8 \pi^2}{\rm Tr} F^k \: = \: \frac{p}{2\pi} H^{(D)},
\end{equation}
hence
\begin{equation}
    \frac{1}{8 \pi^2} \int d^D X \, {\rm Tr} F^k \: \in \: p {\mathbb Z},
\end{equation}
so that the instanton number is divisible by $p$.  Integrating out $B$, we recover Yang-Mills theory with a restriction on instanton numbers.

The equations of motion for $C^{(D-1)}$ imply that
\begin{equation}
    p \, dB \: = \: 0,
\end{equation}
hence $B$ is locally constant.  However, if the cohomology class of $H^{(D)}$ is nontrivial,
then the path integral measure factor
\begin{equation}
    \exp\left( - i \int d^D x B \left( \frac{p}{2\pi} \right) H^{(D)} \right)
\end{equation}
is not invariant under shifts of $B$ by a period unless
$B \in \{ 0, 2\pi/p, 4 \pi/p, \cdots, 2 \pi (p-1)/p \}$,
hence $B$ is restricted to those values.
Integrating out $C^{(3)}$, we get a sum over theories, which separately have actions of the form
\begin{equation}
    S_m \: = \: S_{\rm YM} \: + \: \frac{2 \pi m}{p}   \frac{1}{8 \pi^2} \int d^D X \, {\rm Tr} F^k,
\end{equation}
 for $m \in \{0, 1, \cdots, p-1\}$.
The second term (arising from the $B {\rm Tr}\, F^k$ interaction) is a theta angle term,
for a theta angle given as a $p$th root of unity.

In this fashion, Tanizaki-\"{U}nsal \cite{Tanizaki:2019rbk} argued that a theory with a restriction to instantons of degree divisible by $p$ is equivalent to a disjoint union of $p$ theories with slightly different theta angles.  The separate theories in the disjoint union all sum over all instantons; however, instantons of the `wrong' degree cancel out between contributions from different universes, as theta angles weight their contributions by phases that cancel out.

For finite $p$, we interpret this as topologically gauging a $(-1)$-form symmetry, specifically
${\mathbb Z}_p^{[-1]} \subset U(1)^{[-1]}$.  The result of the gauging has the quantum
symmetry Rep$({\mathbb Z}_p)^{[1]} \cong {\mathbb Z}_p^{[1]}$, and decomposes into $p$ universes.

Next, consider the limit $p \rightarrow \infty$.
We propose to interpret this limit as topologically gauging $({\mathbb Q}/{\mathbb Z})^{[-1]} \subset U(1)^{[-1]}$, as ${\mathbb Q}/{\mathbb Z}$ is a direct limit of cyclic groups
(see e.g.~\cite[equ'n (2.8)]{Putrov:2022pua}).
Phrased another way, in the limit of the Tanizaki-\"Unsal construction, one can get any rational value of $\theta$ (up to $2 \pi$ factors).
The resulting theory has a Rep$({\mathbb Q}/{\mathbb Z})^{[1]}\cong \hat{Z}$ symmetry\footnote{
For completeness, $({\mathbb Q}/{\mathbb Z})^* \cong \hat{\mathbb Z}$, the profinite integers,
see \cite[examples 2.1.6.2, 2.9.5.3]{profinite}, \cite{Putrov:2022pua}.
},
and decomposes into universes indexed by ${\mathbb Q}/{\mathbb Z}$.

Now, that said, ${\mathbb Q}/{\mathbb Z}$ is almost the same as $U(1)$, so it might be that, due to additional subtleties, the correct interpretation of the
$p \rightarrow \infty$ limit is in terms of topologically gauging a 
$U(1)^{[-1]}$ symmetry.  Indeed, this was assumed in \cite[section 3.1]{Lin:2025oml}.  We leave that determination for future work, and proceed by treating the ${\mathbb Q}/{\mathbb Z}$ gauging as an independent and physically meaningful construction in its own right.

We do not know of a rigorous construction of limits of quantum field theories, 
but, we observe that there are (at least) two interpretations of the $p \rightarrow \infty$ limit
of the Tanizaki-\"Unsal construction, as we outline next.
\begin{itemize}
    \item If the limit is another $U(1)$ gauge theory, then it would appear to be one
    in which all instantons are suppressed, as no instanton degree is divisible by $p$ in the limit\footnote{The conclusion that the limit $p \rightarrow \infty$ implements a restriction to instanton number zero was also previously reached by Y.~Tanizaki and M.~\"Unsal \cite{yuyapriv}.}.
We can understand the total instanton restriction as another example of multiverse interference.
Here, in the limit $p \rightarrow \infty$, we have a universe for every rational theta angle on the
circle.  We have already seen that superimposing every theta angle on the circle results in total instanton restriction; here, we see that it would suffice to have as many universes as rational points,
rather than the entire circle of theta angles.  (This interpretation of the limit was also assumed in \cite[section 3.1]{Lin:2025oml},
\cite[appendix D.1]{Pantev:2023dim}.)

\item There may be another way of understanding the limit, in terms of a modification of the gauge group.
In previous examples, in spacetime dimension $D=2$, we were able to interpret topologically gauging the $(-1)$-form symmetry in terms of enlarging the gauge group.  For example, when we gauged a Rep$({\mathbb Z}) \cong U(1)$ $(-1)$-form symmetry in a $D=2$ $U(1)$ gauge theory, the gauge group was enlarged to
an extension of $U(1)$ by ${\mathbb Z}$, specifically ${\mathbb R}$:
\begin{equation}
    1 \: \longrightarrow \: {\mathbb Z} \: \longrightarrow \: {\mathbb R} \: \longrightarrow \:
    U(1) \: \longrightarrow \: 1.
\end{equation}
In the present case, if we start with a $U(1)$ gauge theory in $D=2$ and apply a limit of Tanizaki-\"Unsal's construction to, effectively, topologically gauge a ${\mathbb Q}/{\mathbb Z}$ $(-1)$-form symmetry, it is natural to formally conjecture that this may be equivalent to an enlargement of the gauge group to an extension of $U(1)$ by $( {\mathbb Q}/{\mathbb Z})^* \cong \hat{\mathbb Z}$, the profinite integers.
Happily, such an extension exists \cite{tonypriv}, \cite[section 2]{burgosv}.  It is known as the adelic solenoid, and is labelled $S^1_{\mathbb Q}$:
\begin{equation}
    1 \: \longrightarrow \: \hat{\mathbb Z} \: \longrightarrow \: S^1_{\mathbb Q} \: \longrightarrow \: U(1) \: \longrightarrow \: 1,
\end{equation}
where, from \cite[lemma 2.2]{burgosv},
\begin{equation}
    S^1_{\mathbb Q} \: = \: \frac{ \hat{\mathbb Z} \times {\mathbb R} }{ {\mathbb Z}}.
\end{equation}

Now, that all said, unlike the interpretation of the limit as a $U(1)$ gauge theory,
in this interpretation, the limit would have nontrivial instantons, as nontrivial principal $S^1_{\mathbb Q}$ bundles can exist.  Briefly \cite{tonypriv}, for any prime number, $S^1_{\mathbb Q}$ contains a factor of a cyclic group of corresponding order, and certainly on a Riemann surface of genus greater than zero, there can be nontrivial ${\mathbb Z}_n$ bundles for any $n$.

It is also conceivable that there might exist another extension, an alternative to $S^1_{\mathbb Q}$, of greater relevance, possibly one with no nontrivial bundles.

In any event, we do not at present understand concretely how to gauge $S^1_{\mathbb Q}$, except of course by defining it as a $p \rightarrow \infty$ limit, nor do we know if it is in fact the relevant extension, but we believe it offers an intriguing mathematical candidate for potentially capturing the structure of the resulting theory. We leave this conjectural interpretation for future work.

\end{itemize}

Finally, we could also consider the opposite limit, namely the case $p=0$.  This case looks naively as if it could also describe a restriction on 
instantons, as $B$ then acts as a Lagrange multiplier.  However, in this case, $B$ is better thought of as an axion -- in effect, in the $p=0$ case, $\theta$ becomes a dynamical field ($B$),
and so we discount this case for purposes of understanding instanton restriction.

\subsection{Supersymmetrization in two dimensions}   \label{sect:tu:susy}

Next, let us turn to two-dimensional supersymmetric gauge theories with
Tanizaki-\"Unsal-type restrictions.  
We will describe a manifestly local, supersymmetric version of the Tanizaki-\"Unsal construction in two-dimensional (2,2) supersymmetry, compute quantum cohomology rings in simple examples, then discuss the limit $p \rightarrow \infty$, in which the vortex / instanton number is restricted to vanish.

In (2,2) supersymmetry, in the case of a $U(1)$ gauge
theory, the Tanizaki-\"Unsal procedure
can be described as adding a twisted chiral $T$ whose scalar part has periodicity $2 \pi$, a vector superfield $V$, and a twisted superpotential
\begin{equation}
    \widetilde{W} \: = \: T \left( \sigma - p Y \right),
\end{equation}
where (locally) $Y = \overline{D}_+ D_- V$ is a twisted chiral,
and $\Sigma$ is the twisted chiral superfield corresponding to the field strength of the $U(1)$ gauge field.  (See also \cite[appendix D.2]{Pantev:2023dim}.)

As a consistency check, let us expand this in components.
If we write\footnote{
We use the same symbol, $\sigma$, for both the twisted chiral superfield as well as its bosonic component, as $\Sigma$ is too easily confused with a summation.
}
\begin{eqnarray}
    \sigma & = & \sigma - i \theta^+ \overline{\lambda}_+ - i \overline{\theta}^- \lambda_- + 
    \theta^+ \overline{\theta}^= ( D - i F_{01} )  +  \cdots,
    \\
    Y & = & y - i \theta^+ \overline{\upsilon}_+ - i \overline{\theta}^- \upsilon_- + 
    \theta^+ \overline{\theta}^- (D_V - i F_{V 01}) + \cdots,
    \\
    T & = & t - i \theta^+ \overline{\chi}_+ - i \overline{\theta}^- \chi_- + 
    \theta^+ \overline{\theta}^- (r_x - i r_y) + \cdots,
\end{eqnarray}
then 
\begin{equation}
    \int d \theta^+ d \overline{\theta}^- \, T \left( \sigma - p Y \right)
    \: = \: t \left(D - i F_{01} - p (D_V - i F_{V 01}) \right)
    \: + \: (r_x - i r_y)\left( \sigma - p y \right) \: + \: \cdots,
\end{equation}
which includes the terms
\begin{equation}
    t \left(F_{01} + p F_{V 01} \right),
\end{equation}
recovering the $D=2$ specialization of~(\ref{eq:tu-action}).

In passing, we should observe that the coupling above has the same form as promoting the FI parameter and theta angle to a (twisted) chiral superfield, which is known in GLSMs to lead to descriptions of $H$ flux, see e.g.~\cite{Adams:2009zg,Blaszczyk:2011ib,Adams:2012sh,Quigley:2011pv,Quigley:2012gq,Melnikov:2012nm,Caldeira:2018ynv}.  However, 
in this particular case\footnote{We would like to thank I.~Melnikov for a discussion of this point.},
the $Y$ equations of motion set $p T = 0$ ($T=0$ locally) and the $T$ equations of motion set
$\Sigma = p Y$, so there is no nontrivial $H$ flux, and the only effect is to restrict
instantons in the 2d theory.

Integrating out $T$ gives the constraint $\Sigma = p Y$,
which 
forces the $U(1)$ vortex number to be a multiple of $p$.
Integrating out $Y$ forces $\exp(i t p) = 1$, so that $\exp(i t)$ is a $p$th root of unity,
for reasons already discussed.

Next, we consider the quantum cohomology ring.  
First, we consider the projective space ${\mathbb P}^n$.
The 
twisted one-loop superpotential is\footnote{
To avoid confusion with summation signs, we are using $\sigma$ to denote both the twisted chiral superfield for the gauge curvature as well as its bosonic component.
}
\begin{equation}
    \widetilde{W} \: = \: \sigma \left( i \tau \: - \: (n+1) \ln\left( \sigma\right) \right)
    \: + \: T \left( \sigma - p Y \right),
\end{equation}
As the product $TY$ is a mass term, we should remove $Y$, $T$ first, and then
compute the ring relations for $\Sigma$ (as the critical locus of $\widetilde{W}$).

For an ordinary ${\mathbb P}^n$, without the extra terms, the ring relation would be
\begin{equation}
    \sigma^{n+1} \: = \: q.
\end{equation}
Here, by contrast:
\begin{itemize}
\item If we integrate out $T$, then $\sigma = p Y$, and the ring relations become
\begin{equation}
    Y^{p(n+1)} \: = \: q,
\end{equation}
which matches those computed in \cite[section 3.1]{Pantev:2005zs}, for a GLSM with gauge field $U(1)$ and $n+1$ chiral superfields of charge $p$.  If we write $q = \tilde{q}^p$, to better reflect the relation between the gerbe theory and the ordinary supersymmetric ${\mathbb P}^n$ model, this implies
\begin{equation} \label{eq:pn:gerbe}
    Y^{n+1} \: = \: \Upsilon \tilde{q},
\end{equation}
for $\Upsilon$ a $p$th root of unity.  This is precise form of the quantum cohomology ring of each universe
of the decomposition (a copy of ${\mathbb P}^n$), indexed by values of $\Upsilon$.
\item If instead we integrate out $Y$, then $\exp(i t p) = 1$,
and the ring relations become
\begin{equation}
    \sigma^{n+1} \: = \: q \exp(T) \: = \: q \Upsilon,
\end{equation}
where $\Upsilon^p = 1$.  This is another presentation of the quantum cohomology ring of the same gerbe discussed in \cite[section 3.2]{Pantev:2005zs}, corresponding to a different GLSM,
and matches the form~(\ref{eq:pn:gerbe}).  Again, each value of $\Upsilon$ corresponds to a universe.
\end{itemize}

More generally, consider a $U(1)^k$ gauge theory with a restriction to
$U(1)$ vortices of vortex number ($c_1$) divisible by $p_a$ for the $a$th $U(1)$ factor.
From the procedure above, we add $2k$ twisted chiral superfields $T_1, \cdots, T_k$,
$Y_1, \cdots, Y_k$, and along the Coulomb branch, the one-loop corrected twisted chiral superpotential 
takes the form
\begin{equation}
    \widetilde{W} \: = \: \sum_{a=1}^k \sigma_a \left( i \tau_a \: - \:
    \sum_i Q_i^a \ln\left( \sum_b Q_i^b \sigma_b \right) \right) \: + \:
    \sum_{a=1}^k T_a \left( \sigma_a - p_a Y_a \right),
\end{equation}
essentially a minor variation of \cite[equ'n (3.36)]{Morrison:1994fr}.
Following \cite{Morrison:1994fr}, and repeating essentially the same reasons as above, the ring relations one computes can be described as either
\begin{equation}
    \prod_i \left( \sum_b Q^b_i Y_b^{p_b} \right)^{Q^a_i} \: = \: q_a
\end{equation}
or
\begin{equation}
    \prod_i \left( \sum_b Q^b_i \sigma_b \right)^{Q^a_i} \: = \: q_a \Upsilon_a
\end{equation}
for $\Upsilon_a$ a $p_a$th root of unity.

So far we have discussed the case of finite $p$, corresponding to gauging 
${\mathbb Z}_p^{[-1]}$.  Next, 
we turn to the limit $p \rightarrow \infty$.  To illustrate the ideas, we specialize to the
projective space ${\mathbb P}^n$. 
In the limit, the relation
\begin{equation}  \label{eq:tu:qc}
    \sigma^{n+1} = q \Upsilon,
\end{equation}
has the same form, albeit with $\Upsilon \in {\mathbb Q}/{\mathbb Z}$, covering the rational points on a circle.  As before, we interpret this as a decomposition, into universes indexed by values of $\Upsilon$, with each universe being a copy of ${\mathbb P}^n$.

If we briefly turn to the axial $U(1)_A$ symmetry of this model, and repeat the analysis of
section~\ref{sect:mirrors} to look for nonanomalous subgroups, we are quickly led to conclude that
the nonanomalous subgroup of $U(1)_A$ is ${\mathbb Q}/{\mathbb Z}$ in this case.
For example, each universe in the mirror is a Landau-Ginzburg model of the usual Toda
form~(\ref{eq:toda}), with $q$ multiplied by a phase $\Upsilon \in {\mathbb Q}/{\mathbb Z}$.
There is a finite subgroup ${\mathbb Z}_{2(n+1)} \subset U(1)_A$ that preserves each universe separately, but much as in the discussion of mirrors to gerbes in section~\ref{sect:mirrors}, there is a larger nonanomalous subgroup of $U(1)_A$ that exchanges the universes, and merely requires that $\exp(i \alpha) \in {\mathbb Q}/{\mathbb Z}$ (in the notation of section~\ref{sect:mirrors}).

The relation~(\ref{eq:tu:qc}) coincides with the form of the Coulomb branch relations of section~\ref{sect:qc:r}, that play a role in understanding the decomposition of the ${\mathbb R}$ gauge theory in that section, with the difference that there, the universes are indexed by $U(1)$ instead of ${\mathbb Q}/{\mathbb Z}$.

So, although the Tanizaki-\"Unsal construction does not (so far as we are aware) reproduce
the corresponding ${\mathbb R}$ gauge theory, nevertheless we see that at least in simple examples, the quantum cohomology rings have similar forms.

\section{Conclusions}

Briefly, in this paper we have discussed examples in which decomposition involves continuous families of universes as well as total instanton restrictions -- typically removing all instanton contributions entirely.  In two dimensions, we have seen in gauge theory examples that these properties are at least sometimes equivalent to making the gauge group noncompact, such as by replacing a $U(1)$ gauge group factor with ${\mathbb R}$.  We have discussed several examples, and commented extensively on how this is described in terms of topologically gauged $(-1)$-form symmetries.
We have also discussed analogous notions in limits of the Tanizaki-\"Unsal construction, which appears to describe a gauged ${\mathbb Q}/{\mathbb Z}$ $(-1)$-form symmetry.

We have focused on the special case of two-dimensional theories, but analogous results
are expected in $D > 2$.
For example, in $D=3$:
\begin{equation}
\mbox{$U(1)$ gauge theory} \: \mathrel{\mathop{\rightleftarrows}^{\mathrm{U(1)^{[0]}}}_{\mathrm{{\mathbb Z}^{[1]}}}}
\:
\mbox{${\mathbb R}$ gauge theory}.
\end{equation}
and in $D=4$:
\begin{equation}
\mbox{$U(1)$ gauge theory} \: \mathrel{\mathop{\rightleftarrows}^{\mathrm{U(1)^{[1]}}}_{\mathrm{{\mathbb Z}^{[1]}}}}
\:
\mbox{${\mathbb R}$ gauge theory}.
\end{equation}
In one direction, gauging the $B {\mathbb Z} = {\mathbb Z}^{[1]}$ is expected to change the gauge group
from ${\mathbb R}$ to ${\mathbb R}/{\mathbb Z} \cong U(1)$, but in the other direction, it is less explicit why topologically gauging a $U(1)^{[D-3]}$ symmetry will remove $U(1)$ vortices in $D > 2$.  We leave this for future work.

Another matter concerns gauge-theoretic interpretations of the
$p \rightarrow \infty$ limit of the Tanizaki-\"Unsal restrictions discussed in
section~\ref{sect:tu}.  We proposed that they be understood as topologically gauging a ${\mathbb Q}/{\mathbb Z}$ $(-1)$-form symmetry, and it would be interesting to explore this further, following
e.g.~\cite{Putrov:2022pua}.
Furthermore, 
in $D=2$ spacetime dimensions, we interpreted the $S^1$-parametrized universes of $U(1)$ gauge theories as ${\mathbb R}$ gauge theories, and we proposed that in the ${\mathbb Q}/{\mathbb Z}$ case, a family of ${\mathbb Q}/{\mathbb Z}$-parametrized universes of $U(1)$ gauge theories could be understood as an $S^1_{\mathbb Q}$ gauge theory, where $S^1_{\mathbb Q}$ is an extension of $U(1)$ by the profinite integers.  It would be interesting to understand that proposal in more detail, which we leave for future work.

Another direction concerns the IR behavior of pure supersymmetric gauge theories with noncompact gauge groups.  Such questions for gauge theories with compact gauge groups were studied in e.g.~\cite{Aharony:2016jki,Gu:2018fpm,Chen:2018wep,Gu:2019zkw,Gu:2020ivl} and references therein.
Naively, the constructions for noncompact gauge groups should be similar, modulo understanding technical issues such as the non-positive-definiteness of the gauge kinetic term.  We leave this for future work.

Yet another interesting direction concerns possible links between JT gravity and decompositions, via arguments  \cite{Iliesiu:2019xuh} that certain two-dimensional gauge theories with noncompact gauge groups (which are conjectured to decompose) also realize JT gravity, which exhibits the closely-related notion of an ensemble.  We refer the reader to section~\ref{sect:gauge-1} and also \cite{Sharpe:2023lfk} for a discussion of the relationship between decompositions and ensembles, and leave a detailed study of possible links via  \cite{Iliesiu:2019xuh} for future work.

\section{Acknowledgements}

We would like to thank C.~Closset, W.~Gu, S.~Hellerman, L.~Herr, J~Knapp, I.~Melnikov, T.~Pantev, W.~Taylor, and H.~H.~Tseng for useful discussions.
E.S.~and X.Y.~were partially supported by NSF grant PHY-2310588.

\appendix

\section{Nonabelian analogues of 1-form symmetries} \label{app:nonabel}

Suppose in a two-dimensional quantum field theory,
that one gauges a nonabelian group $G$ (finite or continuous) that acts trivially.
Then,
\begin{itemize}
\item The 1-form symmetry is described by a {\it set} of topological
local operators, indexed by the conjugacy classes of $G$,
corresponding to twist fields / Gukov-Witten operators.
(In the case $G$ is abelian, that set has a group structure, but in
general it is merely a set.)
\item There exist Wilson lines associated to
representations of $G$, which index the universes.
\end{itemize}
The representations and conjugacy classes are dual to one another, and represent different descriptions of the same space.

\section{${\mathbb R}$ gauge theory in two dimensions}
\label{sect:r-gauge}

In this appendix we collect some results on two-dimensional gauge theories with gauge group ${\mathbb R}$, and to a limited extent, other noncompact gauge groups.

\subsection{Lack of instantons}

One way to get a gauge theory without instantons is if the gauge group is noncompact.
For example, consider a ${\mathbb R}$ gauge theory, as compared to a $U(1)$ gauge theory.
Unlike a $U(1)$ gauge theory,
\begin{itemize}
    \item Charges are not quantized in a ${\mathbb R}$ gauge theory, and
    \item There are no nontrivial bundles in a ${\mathbb R}$ gauge theory, hence no instantons (in two dimensions) or vortices (in higher dimensions).
\end{itemize}

In the main text, we also discuss other methods to eliminate instantons, and relations between those other methods and gauge theories with noncompact gauge groups.

\subsection{Theta angles}  \label{app:r:theta}

Briefly, in a ${\mathbb R}$ gauge theory in two dimensions, at least in theories with
global $B {\mathbb Z}$ symmetries, we claim the theta angle is not meaningful.

On a compact closed worldsheet $\Sigma$, we can see this as follows.
Because there are no nontrivial bundles, because the cohomology class of the curvature $F$ always vanishes, there is no $\theta$ angle coupling in a ${\mathbb R}$ gauge theory, simply because
\begin{equation}
    \theta \int_{\Sigma} F \: = \: \theta \int_{\Sigma} d A \: = \: 0,
\end{equation}
as $F = dA$ for a globally-defined $A$.

In the case of noncompact worldsheets,
the paper \cite[section 1]{Banks:1991mb} reached a similar conclusion from a different line of reasoning.  They noted that although there is no $\theta$ coupling on a two-dimensional spacetime without boundary, there is potentially a boundary coupling if the spacetime is noncompact.  However, a light integrally-charged field can be combined with an irrationally-charged heavy field to screen any value of $\theta$ in such circumstances.  As a result, in such theories, again, $\theta$ does not play a role.

In fact, implicit in the approach we take to two-dimensional ${\mathbb R}$ gauge theories in the main text, at least in those with a global $B {\mathbb Z}$ symmetry, is that the $\theta$ angle is not a meaningful quantity, as the ${\mathbb R}$ gauge theory can be obtained by integrating over theta angles in $U(1)$ theories.

In the case that the worldsheet $\Sigma$ has a boundary,
\begin{equation}
    \theta \int_{\Sigma} F \: = \: \theta \int_{\partial \Sigma} A,
\end{equation}
and so the theta angle term instead appears as a Chan-Paton modification.  This can be interpreted in terms of decomposition along the same lines as in \cite[section 7]{Hellerman:2006zs}.  In any event, this is a different role from its usual one as a phase modification of instanton sectors, and so even in the presence of boundaries, we do not consider a ${\mathbb R}$ gauge theory to have a meaningful theta angle per se.

\subsection{Decomposition}  \label{app:r:decomp}

For later use, we consider the decomposition of two-dimensional pure ${\mathbb R}$ gauge theories (i.e.,
without matter).

In general terms, when one gauges a trivially-acting group $G$
in a two-dimensional theory, the resulting gauge theory decomposes.
If $G$ is in the center of a larger group, the gauge theory decomposes into
$|{\rm Rep}(G)|$ universes \cite{Hellerman:2006zs}.

This yoga was originally discussed for finite groups $G$ in \cite{Hellerman:2006zs},
and was extended to cases in which a compact Lie group acts trivially 
in \cite{Cherman:2020cvw,Nguyen:2021yld,Nguyen:2021naa}.
For example, two-dimensional pure Yang-Mills with gauge group $G$
is equivalent to a disjoint union of universes, where the universes
are in one-to-one correspondence with representations of $G$.

In the case that $G = U(1)$, this can be understood as a refinement and strengthening of
an old statement about superselection sectors.  Recall (see e.g.~\cite{Wick:1970bd}) 
that in electromagnetism, there are superselection sectors corresponding to
fixed total charge.  In the special case of two spacetime dimensions and the pure $U(1)$ Maxwell theory,
those superselection sectors are enhanced to universes (see e.g.~\cite{Cherman:2020cvw},  \cite[section 2]{Komargodski:2020mxz}). Briefly,
since spacelike slices are one-dimensional, a pointlike charge acts as a domain wall.
In the pure Maxwell theory, such domain walls are nondynamical, and so separate universes,
of which there are as many as charges -- ${\mathbb Z}$-fold (countably infinitely) many.

Now, consider the case of a pure ${\mathbb R}$ gauge theory in two dimensions.
In such a gauge theory, charges are not quantized, unlike a
$U(1)$ gauge theory, but instead all continuous values are allowed.
For very similar reasons as in the pure $U(1)$ Maxwell theory,
we expect that in the pure ${\mathbb R}$ Maxwell theory, there is again a decomposition,
but this time, there is a continuum of universes, parametrized by the real numbers.

This is also consistent with the yoga of decomposition in two-dimensional gauge theories.
This says, in essence, that a gauge theory with a trivially-acting central subgroup
$K$ decomposes into universes indexed by Rep$(K)$.  Here, $K$ is the entire gauge group,
so $K = {\mathbb R}$, and Rep$({\mathbb R}) \cong {\mathbb R}$.

In the next subsection we will discuss how this decomposition is reflected in the exact expression for the partition function, and 
we shall also see further evidence elsewhere in this paper.

\subsection{Partition function and states of pure gauge theory}

Next, for later reference, we give the exact expression for the partition function.
For a pure $G$ gauge theory in two dimensions, for $G$ noncompact,
the partition function on a Riemann surface $\Sigma$ has the form (see e.g.~\cite[equ'n (3.8)]{Iliesiu:2019xuh})
\begin{equation}  \label{eq:migdalrusakov:noncompact}
    Z_G \: = \: \int_{{\rm Rep}(G)} dR \, \rho(R)^{ \chi(\Sigma)} \exp\left( - A C_2(R) \right),
\end{equation}
where the integral is over the space\footnote{For a noncompact group, there can be continuous families of unitary irreducible representations.} of unitary irreducible representations
Rep$(G)$,
$\rho(R)$ is the Plancherel measure on 
Rep$(G)$, and $A$ is the area of $\Sigma$.  This is an analogue of the statement that for compact Lie groups, on a Riemann surface $\Sigma$, the partition function is  \cite[section 3.7]{Cordes:1994fc},
\cite{Migdal:1975zg,Drouffe:1978py,Lang:1981rj,Menotti:1981ry,Rusakov:1990rs,Witten:1991we,Witten:1992xu,Blau:1993hj}
\begin{equation} 
    \sum_R \left( \dim R \right)^{\chi(\Sigma)} \exp\left( - A C_2(R) \right),
\end{equation}
where the sum is over irreducible representations $R$ of the (compact) gauge group.

For the pure ${\mathbb R}$ gauge theory on a connected Riemann surface, the partition function is 
\begin{equation}  \label{eq:2d:partfn-R}
    Z_{\mathbb R} \: = \: \int_{{\rm Rep}({\mathbb R})} dr \, \exp\left( - A \frac{r^2}{4\pi^2} \right).
\end{equation} 
Here, since the gauge group ${\mathbb R}$ is noncompact, we integrate over
unitary irreducible representations Rep$({\mathbb R}) \cong {\mathbb R}$.
Furthermore, we have taken the Plancherel measure $\rho(r) = 1$, 
as from \cite[section 7.5]{folland} for abelian groups the Plancherel measure coincides with the Haar measure, and from \cite[section 2.2, prop. 2.21, case 5]{folland} the Haar measure on ${\mathbb R}$ is trivial.

To see the decomposition of the pure ${\mathbb R}$ gauge theory, write
\begin{equation}
    Z_{\mathbb R} \: = \: \int_{{\rm Rep}({\mathbb R})} dr \, Z_{\mathbb R}(r),
\end{equation}
for
\begin{equation}
    Z_{\mathbb R}(r) \: = \: \exp\left( -A \frac{r^2}{4\pi^2} \right),
\end{equation}
a Gaussian centered at the origin.  We interpret $Z_{\mathbb R}(r)$ as the partition function of the invertible field theory
universe associated with $r \in {\rm Rep}({\mathbb R}) \cong {\mathbb R}$, in the decomposition of the pure two-dimensional ${\mathbb R}$ gauge theory, in the same way that for compact groups, universes are associated to irreducible representations.

The expression for the partition function above implicitly encodes information about the states.
First, we review the case of pure $U(1)$ gauge theories in two dimensions,
with partition function
\begin{equation}
    Z_{U(1)}(\theta) \: = \: \sum_{q \in {\mathbb Z}} \exp\left( - A \left( q + \frac{\theta}{2\pi} \right)^2 \right).
\end{equation}
(See e.g.~\cite[section 2]{Komargodski:2020mxz}, \cite[section 4.1.2]{Pantev:2023dim} and references therein.)
As already mentioned, such theories decompose into countably infinitely many invertible field theories.  The invertible field theories are associated to $q$'s in the partition function above, and each is associated with a single state.  (Since two-dimensional pure Maxwell theory has no propagating states, there is not a Fock space of states; instead, it has a Hilbert space, much like a topological field theory, which it becomes in the limit $A \rightarrow 0$.)
If we take the worldsheet to be a square torus, so that the area $A$ is proportional to the
length $T$ of the timelike direction, then in the spirit of finite temperature field theory,
we see that the Hamiltonian 
\begin{equation}
H \: \propto \: \left( q + \frac{\theta}{2\pi} \right)^2.
\end{equation}

In the case of the pure ${\mathbb R}$ gauge theory in two dimensions, it decomposes into uncountably (${\mathbb R}$) many invertible field theories, parametrized by $r$.
Its Hilbert space consists of ${\mathbb R}$-many states, each associated to a single invertible field theory, parametrized by $r$.
For the same reasons as the $U(1)$ case, the Hamiltonian $H \propto r^2$.

In section~\ref{sect:ex:u1-r} we will explore the relationship between $U(1)$ and ${\mathbb R}$ gauge theories in more detail.

One can also consider more general noncompact gauge groups than ${\mathbb R}$.  We do not claim to understand the role of the fact that the gauge kinetic terms have an indefinite metric for more general noncompact groups, or that the volume is infinite, as mentioned in the introduction,
but, given the validity of the expansion~(\ref{eq:migdalrusakov:noncompact}) above, it is natural to conjecture that a two-dimensional
pure gauge theory with noncompact group $G$ should have a decomposition into invertible field theories,
indexed by unitary irreducible representations, following the same format as the partition
function expression~(\ref{eq:migdalrusakov:noncompact})  above, in the same way that pure two-dimensional gauge theories with compact gauge groups decompose \cite{Nguyen:2021yld,Nguyen:2021naa}.
(See also \cite{Margolin:1992rg} for some comments on superselection sectors in gauge theories with noncompact gauge groups that may be relevant here.)
We leave a detailed understanding for future work.

In passing, it was argued in \cite{Iliesiu:2019xuh} that a two-dimensional pure gauge theory with a noncompact gauge group is related to JT gravity.  Given the conjecture above, that two-dimensional pure gauge theories decompose, we are immediately reminded of the distinction between ensembles and decompositions, discussed in section~\ref{sect:gauge-1} and also \cite{Sharpe:2023lfk}.
We leave a study of that relationship for future work.

\section{Gauging $(-1)$-form symmetries in orbifolds}
\label{app:gauge-1orb}

An $G$ orbifold $[X/G]$ in two dimensions with trivially-acting central\footnote{
We specialize to central trivially-acting subgroups for simplicity, but decomposition exists and has been
systematically characterized for general trivially-acting subgroups,
see \cite{Hellerman:2006zs}.
}
$K \subset G$ is equivalent to a disjoint union of $G/K$ orbifolds with choices of
discrete torsion \cite{Hellerman:2006zs}, as reviewed in section~\ref{sect:rev}.

Suppose instead that one starts with an arbitrary two-dimensional orbifold $[X/G]$,  where $G$ need not have a trivially-acting subgroup, and then sums over all choices of discrete torsion.  In this appendix we will construct a covering group $\widetilde{G}$ such that
\begin{equation} \label{eq:tildegdefn}
    {\rm QFT}\left( [X/\widetilde{G}] \right) \: = \: \coprod_{\omega \in H^2(G,U(1))}
    {\rm QFT}\left( [X/G]_{\omega} \right),
\end{equation}
where the sum is over all possible choices of discrete torsion in $G$.
This tells us that in two-dimensional orbifolds, topologically gauging the $(-1)$-form symmetry corresponding to discrete torsion can be interpreted as increasing the size of the gauge group, a pattern we shall see repeated elsewhere in this paper.

We propose that such a $\widetilde{G}$ is given by a Schur covering group,
a central extension of $G$ by the Schur multiplier $M(G)$, 
\begin{equation}
    1 \: \longrightarrow \: M(G) \: \longrightarrow \:\widetilde{G} \: \longrightarrow \: G \: \longrightarrow \: 1.
\end{equation}
Now, 
\begin{equation}
    {\rm Hom}( M(G), U(1) ) \: = \: H^2(G, U(1)).
\end{equation}
The suggestion is that if one sums over compositions of $H^2(G,M(G))$ with homomorphisms
$\rho: M(G) \rightarrow U(1)$, one will be summing over the different choices of discrete torsion.
Note also that $\widetilde{G}$ is not unique unless $G$ is perfect.

For concreteness, we follow \cite[Section 2]{KSchur}. For a finite group $G$, its Schur multiplier $M(G)$ is defined as the group cohomology group $M(G):=H^2(G, U(1))$. 
In other words, $M(G)$ is the group of possible discrete torsions in $G$; 
topologically gauging the $(-1)$-form symmetry will sum over elements of $M(G)$.

A covering group $\widetilde{G}$ of $G$ is a group with an abelian subgroup $A$ such that 
\begin{itemize}
\item $A\leq Z(\widetilde{G})\,\cap\, [\widetilde{G},\widetilde{G}]$, 
\item  $A\cong M(G)$, and 
\item  $G\cong \widetilde{G}/A$, 
\end{itemize}
thus describing an exact sequence
\begin{equation}\label{eq:coveringsequence}
    1\: \longrightarrow \: A \: \longrightarrow \: \widetilde{G} \: \longrightarrow \: G \: \longrightarrow \: 1.
\end{equation}
It can be shown \cite[Theorem 2.1.4]{KSchur} that the transgression map induced by this sequence $\text{Hom}(A, U(1))\to M(G)$ is bijective. In other words, this implies that summing over the compositions
\begin{equation}
    p\circ\Omega:\: G\times G  \: \stackrel{\Omega}{\longrightarrow} \: 
    A  \: \stackrel{p}{\longrightarrow} \: 
    U(1)
\end{equation}
for $p\in \text{Hom}(A, U(1))$ and $\Omega \in Z^2(G,A)$ a 2-cocycle representing the extension (\ref{eq:coveringsequence}) indeed sums over all possible discrete torsion phases $M(G)=H^2(G, U(1))$ of $G$.

Thus, for a given group $G$, we have constructed a covering group $\widetilde{G}$
with trivially-acting subgroup $M(G)$ which, from the decomposition conjecture
\cite{Hellerman:2006zs}, obeys~(\ref{eq:tildegdefn}).

Now, such covering groups $\tilde{G}$ are not unique.
For a given $G$, the covering groups are all \textit{isoclinic} \cite[Section 2.13.8-9]{KSchur}. Two groups $\widetilde{G}_1,\widetilde{G}_2$ are called isoclinic if there exist isomorphisms 
\begin{equation}
\phi:\widetilde{G}_1/Z(\widetilde{G}_1) \: \longrightarrow \: \widetilde{G}_2/Z(\widetilde{G}_2), 
\: \: \:
\psi:[\widetilde{G}_1,\widetilde{G}_1] \: \longrightarrow \: [\widetilde{G}_2,\widetilde{G}_2]
\end{equation}
 such that $[a',b']=\psi[a,b]$, when $\phi(aZ(\widetilde{G}_1))=a'Z(\widetilde{G}_2)$ and $\phi(bZ(\widetilde{G}_1))=b'Z(\widetilde{G}_2)$.

A familiar example of this is given by $G=\Z_2\times\Z_2$, where 
\begin{equation}
M(\Z_2\times\Z_2) \: = \: H^2(\Z_2\times\Z_2, U(1)) \: = \: \Z_2, 
\end{equation}
and two covering groups are
\begin{equation}
    1 \: \longrightarrow \: \Z_2 \: \longrightarrow \: D_4 \: \longrightarrow \: \Z_2\times \Z_2 \: \longrightarrow \: 1,
\end{equation}
\begin{equation}
    1 \: \longrightarrow \: \Z_2 \: \longrightarrow \: Q_8 \: \longrightarrow \: Z_2\times\Z_2 \: \longrightarrow \: 1,
\end{equation}
where $D_4$ and $Q_8$ are not isomorphic but are isoclinic \cite{Hall40}.
The orbifolds $[X/D_4]$ and $[X/Q_8]$, with the same trivially-acting ${\mathbb Z}_2$ center, seem to define the same closed-string CFTs, as discussed in \cite{Hellerman:2006zs}.

We conjecture that isoclinic covering groups $\widetilde{G}$ of $G$ with the same trivially-acting subgroup $M(G)$, embedded as in the construction above, define isomorphic CFTs.  (We note, however, that there are other notions of isomorphism, such as isocategorical groups, and one of those alternative notions may be the better choice to describe physics, by for example capturing open string information.)

When $G$ is perfect, then there is a unique covering group (up to isomorphism), called the universal central extension $\widetilde{G}$ of $G$ \cite[Section 2.10]{KSchur}.

We also highlight that not only are the covering groups generically not unique, but that one could also observe \textit{non}invertible orbifolds as such covering theories, as expected from the form of the decomposition conjecture of such symmetries in \cite{Perez-Lona:2025wup}.

\end{document}